\documentclass[11pt, a4paper]{article}
\pdfoutput=1
\usepackage{jheppub,amsmath,amssymb,slashed,url,bm,textgreek,upgreek}
\usepackage{jheppub}  
\usepackage{tikz,lipsum,lmodern}
\usepackage{slashed}
\usepackage[most]{tcolorbox}
\usepackage{amssymb} 
\usepackage{amsmath}
\usepackage{mathtools}
\usepackage{amsfonts}    
\usepackage{dsfont}
\usepackage{pdfpages}
\usepackage{verbatim}
\usepackage{tensor}
\usepackage{mathrsfs}
\usepackage{textgreek} 
\usepackage[mathscr]{euscript}
\usepackage[normalem]{ulem}
\usepackage{tikz}
\usepackage{makecell}
\usepackage[verbose]{placeins}
\usepackage{subcaption}
\usetikzlibrary{3d, arrows.meta, decorations.pathreplacing, decorations.markings,calc,shapes.misc,decorations.pathmorphing,patterns.meta, math}
\usepackage{pgfplots}
\pgfplotsset{compat=1.18}

\newcommand{\beq}{\begin{equation}}
\newcommand{\eeq}{\end{equation}}
\newcommand{\nn}{\nonumber\\} 
\newcommand{\bea}{\begin{eqnarray}}
\newcommand{\ea}{\end{eqnarray}}
\newcommand{\barr}{\begin{array}}
\newcommand{\earr}{\end{array}}

\DeclareMathOperator{\sech}{sech}

\def\be{\begin{equation}}
\def\ee{\end{equation}}
\def\ba#1\ea{\begin{align}#1\end{align}}
\def\bg#1\eg{\begin{gather}#1\end{gather}}
\def\bm#1\em{\begin{multline}#1\end{multline}}
\def\bmd#1\emd{\begin{multlined}#1\end{multlined}}

\def\d{{\rm d}}
 \def\i{{\rm i}}
\newcommand{\ii}{\mathrm{i}}

\newcommand\U{\text{U}}

\newcommand\SU{\text{SU}}

\def\vp{\chi}

\def\t{{\sf t}}
\def\r{{\sf r}}

\def\wg{\wedge}

\def\no{\nonumber}

\def\({\left(}
\def\){\right)}
\def\[{\left[}
\def\]{\right]}
\def\<{\langle}
\def\>{\rangle}

\def\bea{\begin{eqnarray}}
\def\eea{\end{eqnarray}}

\newcommand{\tr}{\operatorname{tr}}

\newcommand{\zb}{{\bar z}}

\def\nn{\nonumber}

\begin{document}

\global\long\def\aad{(a\tilde{a}+a^{\dagger}\tilde{a}^{\dagger})}%

\global\long\def\ad{{\rm ad}}%

\global\long\def\bij{\langle ij\rangle}%

\global\long\def\df{\coloneqq}%

\global\long\def\bs{b_{\alpha}^{*}}%

\global\long\def\bra{\langle}%

\global\long\def\dd{{\rm d}}%

\global\long\def\dg{{\rm {\rm \dot{\gamma}}}}%

\global\long\def\ddt{\frac{{\rm d^{2}}}{{\rm d}t^{2}}}%

\global\long\def\ddg{\nabla_{\dot{\gamma}}}%

\global\long\def\del{\mathcal{\delta}}%

\global\long\def\Del{\Delta}%

\global\long\def\dtau{\frac{\dd^{2}}{\dd\tau^{2}}}%

\global\long\def\ul{U(\Lambda)}%

\global\long\def\udl{U^{\dagger}(\Lambda)}%

\global\long\def\dl{D(\Lambda)}%

\global\long\def\da{\dagger}%

\global\long\def\id{{\rm id}}%

\global\long\def\ml{\mathcal{L}}%

\global\long\def\mm{\mathcal{\mathcal{M}}}%

\global\long\def\mf{\mathcal{\mathcal{F}}}%

\global\long\def\ket{\rangle}%

\global\long\def\kpp{k^{\prime}}%

\global\long\def\lr{\leftrightarrow}%

\global\long\def\lf{\leftrightarrow}%

\global\long\def\ma{\mathcal{A}}%

\global\long\def\mb{\mathcal{B}}%

\global\long\def\md{\mathcal{D}}%

\global\long\def\mbr{\mathbb{R}}%

\global\long\def\mbz{\mathbb{Z}}%

\global\long\def\mh{\mathcal{\mathcal{H}}}%

\global\long\def\mi{\mathcal{\mathcal{I}}}%

\global\long\def\ms{\mathcal{\mathcal{\mathcal{S}}}}%

\global\long\def\mg{\mathcal{\mathcal{G}}}%

\global\long\def\mfa{\mathcal{\mathfrak{a}}}%

\global\long\def\mfb{\mathcal{\mathfrak{b}}}%

\global\long\def\mfb{\mathcal{\mathfrak{b}}}%

\global\long\def\mfg{\mathcal{\mathfrak{g}}}%

\global\long\def\mj{\mathcal{\mathcal{J}}}%

\global\long\def\mk{\mathcal{K}}%

\global\long\def\mmp{\mathcal{\mathcal{P}}}%

\global\long\def\mn{\mathcal{\mathcal{\mathcal{N}}}}%

\global\long\def\mq{\mathcal{\mathcal{Q}}}%

\global\long\def\mo{\mathcal{O}}%

\global\long\def\qq{\mathcal{\mathcal{\mathcal{\quad}}}}%

\global\long\def\ww{\wedge}%

\global\long\def\ka{\kappa}%

\global\long\def\nn{\nabla}%

\global\long\def\nb{\overline{\nabla}}%

\global\long\def\pathint{\langle x_{f},t_{f}|x_{i},t_{i}\rangle}%

\global\long\def\ppp{p^{\prime}}%

\global\long\def\qpp{q^{\prime}}%

\global\long\def\we{\wedge}%

\global\long\def\pp{\prime}%

\global\long\def\sq{\square}%

\global\long\def\vp{\varphi}%

\global\long\def\ti{\widetilde{}}%

\global\long\def\wg{\widetilde{g}}%

\global\long\def\te{\theta}%

\global\long\def\tr{{\rm Tr}}%

\global\long\def\ta{{\rm \widetilde{\alpha}}}%

\global\long\def\sh{{\rm {\rm sh}}}%

\global\long\def\ch{{\rm ch}}%

\global\long\def\Si{{\rm {\rm \Sigma}}}%

\global\long\def\sch{{\rm {\rm Sch}}}%

\global\long\def\vol{{\rm {\rm {\rm Vol}}}}%

\global\long\def\reg{{\rm {\rm reg}}}%

\global\long\def\zb{{\rm {\rm |0(\beta)\ket}}}%

\title{The Fate of Nucleated Black Holes in de Sitter Quantum Gravity
\vspace{-0.3cm}}

\author{Xiaoyi Shi, Gustavo J. Turiaci and Chih-Hung Wu}
\affiliation{Department of Physics, University of Washington, Seattle, WA 98195, USA}
\emailAdd{xiaoys5@uw.edu, turiaci@uw.edu, chwu29@uw.edu}

\abstract{The Euclidean Nariai geometry has long been proposed as the instanton describing the nucleation of maximal-mass black holes in de Sitter space. We place this interpretation on firmer footing by showing that, once an observer is included, the gravitational path integral produces the imaginary phase required for a transition rate. As a warmup, we revisit the Hawking-Moss instanton and, as a byproduct, find that scalar fields can enhance black-hole nucleation, suggesting a quantum-gravity bound on scalar potentials with de Sitter solutions. We then study the subsequent semiclassical evolution of the nucleated black hole. We show that the previously claimed ``anti-evaporation'' channel is unphysical, arising from a quantum state with singular horizons. In a smooth state, the black hole instead undergoes standard thermal Hawking evaporation. We verify explicit agreement with the no-boundary state and argue that this evaporation is not subject to large quantum-gravity corrections. The nucleated black hole thus evaporates completely back to the maximally-entropic empty de Sitter vacuum, making the full process a Boltzmann fluctuation.}

\maketitle

\newpage

\section{Introduction}

Shortly after the discovery that an entropy can be associated to black-hole horizons, Gibbons and Hawking proposed a similar idea for cosmological horizons \cite{Gibbons:1977mu}. These early observations were promoted to the conjecture that all physics taking place in the static patch can be described by a quantum mechanical Hilbert space $\mathcal{H}$ of dimension 
\bea\label{intro:dSentropy}
\text{dim} \,\mathcal{H} \sim \text{exp}\Big( \frac{A_{\text{dS}}}{4 G_N}\Big),
\ea
where $A_{\text{dS}}$ is the cosmological horizon area. Some early references on this proposal are \cite{Banks:2000fe,Banks:2005bm,Susskind:2021omt,Susskind:2021dfc}. According to it, the static patch of empty de Sitter (dS) represents the state with maximum entropy and any further excitation can only reduce it. This was recently supported by studying the algebra of observables in the presence of an observer \cite{Chandrasekaran:2022cip}.

Since equation \eqref{intro:dSentropy} is valid in the semiclassical regime, it is important to learn how to compute quantum-gravity corrections to it. A reasonable proposal is that the dimensionality of the Hilbert space is computed by the gravitational path integral on closed Euclidean spacetimes with positive curvature~\cite{Gibbons:1976ue}. In this formalism equation \eqref{intro:dSentropy} is the leading contribution arising from the sphere. Perturbative quantum-gravity corrections can then be explicitly evaluated, at least to leading order, see \cite{Anninos:2020hfj}. 
An immediate issue with this interpretation is the fact that the gravitational path integral on the sphere is not manifestly a real and positive state-counting quantity; rather, it can carry a nontrivial complex phase~\cite{Polchinski:1988ua}, see also~\cite{Maldacena:2024spf, Shi:2025amq, Ivo:2025yek, Chen:2025jqm, Ivo:2025xek, Ivo:2025fwe}. This issue was addressed recently by Maldacena in \cite{Maldacena:2024spf}. When an observer is added to the spacetime, the path integral that carries a state-counting interpretation is indeed real and positive \cite{Maldacena:2024spf,Chen:2025jqm}.

The purpose of this paper is to clarify the role of the Nariai geometry~\cite{Nariai} in dS quantum gravity. The Schwarzschild-de Sitter (SdS) geometry admits two distinct physical interpretations. First, it serves as the leading saddle point contributing to the Hartle-Hawking state~\cite{Hartle:1983ai} for universes with spatial topology $S^1 \times S^{D-2}$ \cite{Laflamme:1986bc, Anninos:2012ft, Conti:2014uda, Maldacena:2019cbz, Turiaci:2025xwi}. In this context, the geometry is generically complex, and the Nariai limit is achieved when the size of the $S^1$ becomes large relative to that of the $S^{D-2}$. Second, in the more conventional interpretation that we adopt in this work, it describes a black hole in dS, where the Nariai limit corresponds to the maximal-mass solution. While the former perspective is more natural in the asymptotic future, where the spacetime is asymptotically dS$_D$, the latter is more appropriate inside the static patch, where the geometry reduces to dS$_2 \times S^{D-2}$.

According to \cite{Ginsparg:1982rs, Bousso:1995cc, Bousso:1996au, Volkov:2000ih, Bousso:1998na, Draper:2022xzl, Morvan:2022ybp}, the Euclidean Nariai solution  evaluates the rate of black hole nucleation in dS. A picture was advocated by Susskind~\cite{Susskind:2021omt,Susskind:2021dfc}, who drew an analogy between this process and the Hawking-Moss (HM) instanton \cite{Hawking:1981fz}, describing false vacuum decay across a broad potential barrier (see Figure~\ref{fig:HM_Potential}). According to this picture, the instanton corresponds to a homogeneous configuration where the scalar field sits at the top of the potential barrier. Upon analytic continuation to Lorentzian signature, a quantum instability can initiate a runaway process in which the field rolls down to the true vacuum; however, this dynamics is not visible in the purely Euclidean instanton. The same negative mode responsible for rendering the HM path integral imaginary, appropriate to evaluate a nucleation rate, is also responsible for initiating this runaway behavior.\footnote{This homogeneous, patch-sized excursion is distinct from the case of Coleman-De Luccia (CDL) bubbles~\cite{Coleman:1980aw}, which feature a configuration close to the true vacuum in the interior and close to the false vacuum in the exterior.} In this analogy, the Nariai geometry is similar to the HM instanton in that it represents a patch-sized modification of the geometry. The nucleation process was studied by Bousso and Hawking in the context of primordial black holes (PBHs) created during inflation~\cite{Bousso:1995cc, Bousso:1996au}. 

Our goal is to establish this interpretation of the Nariai geometry using the gravitational path integral. However, this approach encounters two immediate issues. The first concerns the phase of the path integral. For a process to be interpreted as a nucleation rate, the path integral must be imaginary \cite{coleman_1985}. In the absence of an observer, the gravitational path integral evaluated on the Euclidean Nariai geometry $S^2 \times S^{D-2}$ is instead real and positive \cite{Shi:2025amq, Ivo:2025yek}. Building on the results of \cite{Maldacena:2024spf, Chen:2025jqm}, we will show that in the presence of an observer witnessing the process, the path integral can become imaginary, thereby allowing us to evaluate the nucleation rate of Nariai black holes. The rate is exponentially suppressed in the dS entropy. For example, in $D=4$ it is 
\bea
\Gamma \sim \exp\left(-\frac{\pi}{G_N \Lambda}\right).
\ea
The observer involved in this calculation sits at a fixed radial distance, where the gravitational attraction from the black hole balances the cosmological expansion. This choice also fixes the normalization~\cite{Bousso:1996au} of the physical horizon temperatures and the subsequent evaporation process. In the Euclidean description, this is the trajectory that winds once on $S^2$ and is located on $S^{D-2}$. Its worldline saddle has the lowest action and one negative mode, thus giving the leading imaginary contribution. Other possible observer trajectories are discussed in Appendix~\ref{app:ObserverandBetaIntegral}.

We also consider the case in which the dS cosmological constant is generated by a scalar field trapped in a false vacuum $\phi_{\text{fv}}$, with vacuum energy $V(\phi_{\text{fv}})$, in close analogy with the HM process. The nucleation of black holes in the false vacuum is always exponentially suppressed, just as in HM decay. More interestingly, the nucleation rate for dS black holes in the true vacuum $\phi_{\text{tv}}$ can be parametrically enhanced, depending on the shape of the scalar potential. In particular, in $D=4$ and assuming that $V(\phi_{\text{tv}})>0$ we have
\beq
\Gamma \sim O(1),~~~\text{if}~~~V(\phi_{\text{tv}}) \lesssim \frac{2}{3} V(\phi_{\text{fv}}),
\eeq
the semiclassical action suggests an enhanced rate for producing black holes in the true vacuum.  We do not interpret this as an uncontrolled proliferation of such black holes, but rather as a non-trivial quantum-gravity constraint $V(\phi_{\text{tv}}) > \frac{2}{3} V(\phi_{\text{fv}})$ on scalar potentials allowed in quantum gravity.

The second issue we address here is the subsequent evolution of the nucleated Nariai black hole. Since the Nariai black-hole horizon is in equilibrium with the cosmological horizon, one might think that the configuration is stable and will remain so forever. However, this scenario is inconsistent with the proposals for quantum gravity in the dS static patch \cite{Banks:2000fe,Banks:2005bm,Susskind:2021omt,Susskind:2021dfc}, as it would imply that the maximally mixed state corresponding to empty dS is unstable. A more physical scenario is that, upon analytic continuation to Lorentzian signature, a quantum instability will make the black-hole horizon slightly hotter. This triggers an evaporation process that ultimately leaves behind empty dS once the black hole has completely evaporated. This physically unstable mode was identified by Ginsparg and Perry in \cite{Ginsparg:1982rs}. This avoids the need to construct constrained instantons for finite-mass, non-Nariai black holes in dS \cite{Bousso:1998na, Draper:2022xzl, Morvan:2022ybp}, where a smooth Euclidean section is unavailable. In our picture, the Euclidean saddle that mediates the nucleation is instead the smooth exact Nariai geometry, and the subsequent departure from Nariai is driven by its Lorentzian instability.

Bousso and Hawking appealed to the no-boundary prescription to argue that near-Nariai black holes should evaporate at late times, although the thermal character of this process was never made explicit. At the same time, their analysis exhibited an anti-evaporation branch~\cite{Bousso:1997wi, Nojiri:1998ue, Nojiri:1998ph}, which is often treated as unphysical or as a mode that could only be excited at early times~\cite{Bousso:1997wi, Niemeyer:2000nq}. Such an anti-evaporating branch runs directly counter to the thermodynamic picture in which the pair-created black hole relaxes back to empty dS space. As Hawking himself remarked, ``I regard anti-evaporation as a pathology''~\cite{Elizalde:1999dw}. Indeed, we demonstrate in this paper that this branch is unphysical and arises entirely from a choice of state with a singular horizon. Instead, a physical choice of state that is smooth across the horizon correctly reproduces the evaporation flux expected from thermodynamics~\cite{Hawking:1975vcx, Hartle:1976tp}. Our results align with those obtained via alternative methods in \cite{Markovic:1991ua} and  \cite{Tadaki:1990aa, Tadaki:1990cg, Choudhury:2004ph, Easson:2025ekn}. Finally, we analyze the no-boundary construction of the state and its time evolution, explicitly reproducing the standard thermal Hawking flux.  After nucleation, the Nariai black hole evaporates in a time scale of order the Page time~\cite{Page:1993df, Page:1993wv, Page:2013dx}. This time scale was computed in \cite{Bousso:1996au} under the assumption of an ordinary thermal spectrum. Our analysis provides a justification for precisely this thermal behavior.

To derive the evaporation history of the near-Nariai black hole, we use the approach of Bousso and Hawking \cite{Bousso:1997wi}, which we also reformulate in a more modern language. There they use a 2d dilaton-gravity model along the temporal and radial directions to describe fluctuations in the metric and matter fields inside the static patch. Regarding the leading-order classical solution, we show that their approach is completely equivalent to dS$_2$ Jackiw-Teitelboim (JT) gravity~\cite{Teitelboim:1983ux, Jackiw:1984je} with a linear dilaton potential on a background with a slightly inhomogeneous dilaton profile \cite{Maldacena:2019cbz,Cotler:2019nbi}. The dilaton $\upphi$ in this description is proportional to the difference between the transverse $S^2$ area and its Nariai value, which is small inside the static patch. For the quantum analysis, it is important to keep the leading correction quadratic in $\upphi$ to the strict JT limit inherited from the four-dimensional picture. The resulting action then takes the form in the near-Nariai regime
\beq
S = \frac{1}{2}\int \d^2 x \sqrt{-g} \left( \upphi R + 2 \upphi - \varepsilon \upphi^2 \right) + S_{\text{matter}}, 
\eeq
where $\varepsilon\ll 1$ parametrizes the departure from the exact Nariai point and controls the splitting of horizon temperatures in~\eqref{eq:Nariaitemperature}. To describe the thermal evaporation of the near-Nariai black hole, one must include the quadratic term and go beyond strict JT limit. In JT gravity, the deviation from the Nariai limit is encoded in the entropy difference between the horizons, measured by $\upphi$, while the 2d metric remains rigid dS$_2$. To leading order, the matter fields do not couple to the dilaton, so the black-hole and cosmological horizons remain in exact thermal equilibrium. It is the $\upphi^2$ term that backreacts on the geometry, creating a difference in surface gravities between the two horizons and driving the evaporation process.

The same lesson appears in the Euclidean sphere partition function. JT gravity on the sphere comes with a triplet of exact dilaton zero-modes \cite{Maldacena:2019cbz,Mahajan:2021nsd,Cotler:2024xzz}. Even though the quadratic correction to the dilaton potential is small, it provides a non-trivial weight in the path integral for these three modes, and the path integral with $\varepsilon>0$ is finite \cite{Ivo:2025yek}. Similarly, we find that in order to obtain a well-behaved solution in Lorentzian signature, this same deformation is required. This calculation also implies, upon analytic continuation, that there are no large quantum corrections to the evaporation history of a near-Nariai black hole, since all metric fluctuations are heavily suppressed in the action. Such corrections might be expected in light of the recent progress on near-extremal black holes evaporation~\cite{Ghosh:2019rcj,Iliesiu:2020qvm,Iliesiu:2022onk,Brown:2024ajk}, and our conclusion is consistent with the thermodynamic arguments of~\cite{Aalsma:2025lcb}.

The paper is organized as follows. In Section~\ref{sec:bigpicture}, we use the HM instanton to clarify the relation between negative modes, path-integral phases, and decay rates, and then apply the same logic to Euclidean Nariai, showing that an observer supplies the phase needed for black-hole nucleation. In Section~\ref{sec:anatomyNariai}, we derive the near-Nariai effective theory from 4d Einstein gravity and identify the perturbation away from the exact Nariai limit. In Section~\ref{sec:Antievaporation}, we include quantum matter effects and determine the Lorentzian evolution of the nucleated black hole, showing that a smooth horizon gives the standard thermal Hawking flux, consistent with the no-boundary construction, and that the Bousso-Hawking anti-evaporation branch is a singular-state artifact. Section~\ref{sec:discussion} contains a summary and outlook, while technical details are deferred to the appendices.

\section{The Nariai geometry and the gravitational path integral} \label{sec:bigpicture}

We begin by discussing the HM instanton from the perspective of the gravitational path integral, since it provides the cleanest example in which the role of negative modes and path-integral phases can be stated sharply. We then return to the Nariai geometry and explain why the same logic naturally leads to its interpretation as the instanton for black-hole nucleation in dS. This will set up the Lorentzian evolution question addressed later.

\subsection{Hawking-Moss and the gravitational path integral}\label{sec:Hawking-Moss}

The basic setting of \cite{Hawking:1981fz}, see also \cite{Weinberg:2006pc}, is a theory of $D$-dimensional gravity coupled to a scalar field $\phi$,
\beq\label{eq:EH+scalar}
I = - \frac{1}{16 \pi G_N} \int \d^D x\,\sqrt{g}\left[ -R + (\partial \phi)^2 + 2V(\phi) \right],
\eeq
with scalar potential $V(\phi)$. This is precisely the type of system that appears in many cosmological applications, including inflation.

\begin{figure}[t!]
    \centering
    \begin{tikzpicture}[scale=1.0]
        \draw[->,thick] (-0.3,0) -- (6.5,0) node[right] {$\phi$};
        \draw[->,thick] (0,-0.2) -- (0,3.5) node[above] {$V(\phi)$};
        \draw[thick,smooth] plot coordinates {
            (0.3,2.7)
            (1.0,1.4)
            (1.7,0.9)
            (2.5,1.4)
            (3.2,2.7)
            (3.8,2.9)
            (4.4,2.3)
            (5.0,1.0)
            (5.6,0.35)
            (6.1,0.6)
        };
        \fill (1.7,0.9) circle (1.7pt) node[below=5pt] {$\phi_{\text{fv}}$};
        \fill (3.67,2.92) circle (1.7pt) node[above=5pt] {$\phi_{\text{top}}$};
        \fill (5.65,0.35) circle (1.7pt) node[below=5pt] {$\phi_{\text{tv}}$};
    \end{tikzpicture}
    \caption{Schematic scalar potential relevant for the Hawking--Moss transition.}
    \label{fig:HM_Potential}
\end{figure}
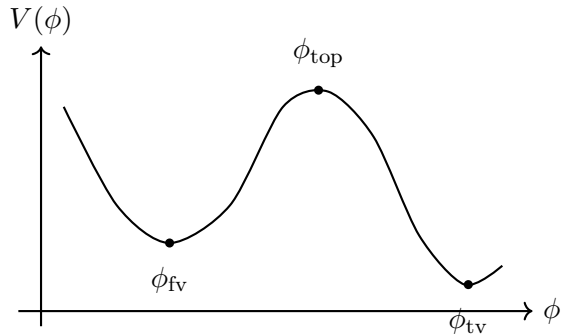

The HM instanton is relevant when the potential has three distinguished stationary points: a false vacuum $\phi_{\text{fv}}$, which is a local minimum; a true vacuum $\phi_{\text{tv}}$, which is the global minimum; and an unstable critical point $\phi_{\text{top}}$ at the top of the barrier separating them. For simplicity we assume that $V(\phi)>0$ throughout so that the relevant homogeneous Euclidean saddles are compact dS spheres.\footnote{Relaxing this assumption leads to qualitatively different situations, including decays to Minkowski or AdS vacua and cases in which the HM interpretation becomes more subtle.} A schematic example is shown in Figure~\ref{fig:HM_Potential}.

Place the system in the homogeneous false-vacuum configuration 
\beq
\phi(x)=\phi_{\text{fv}},~~~~~\d s^2 = \frac{(D-2)(D-1)}{2 V(\phi_{\text{fv}})}\left( -\d t^2 + \cosh^2 t \, \d \Omega_{D-1}^2\right).
\eeq
 The geometry is dS space with effective cosmological constant $\Lambda_{\text{fv}} = V(\phi_{\text{fv}})$. In Euclidean signature $\tau = \i t$ is a round sphere $S^D$, with effective radius
\beq
{\sf R}_{\text{fv}}^2 = \frac{(D-2)(D-1)}{2\Lambda_{\text{fv}}}.
\eeq
The true vacuum is likewise homogeneous, with $\phi(x)=\phi_{\text{tv}}$ and
\beq
\Lambda_{\text{tv}} = V(\phi_{\text{tv}}) < \Lambda_{\text{fv}},~~~~~{\sf R}_{\text{tv}} > {\sf R}_{\text{fv}}
\eeq
How does a universe prepared in the false vacuum make the transition to the true vacuum?

The answer depends sensitively on the shape of the barrier. When the potential between the two vacua is sufficiently steep, the dominant process is the nucleation of a localized bubble whose interior is close to the true vacuum. In that regime the wall separating the two phases is thin, and the decay is described by the CDL instanton \cite{Coleman:1980aw}. This bubble regime was analyzed from the gravitational path integral viewpoint in \cite{Ivo:2025fwe}. The physics changes when the barrier becomes broad and shallow. As $|V''(\phi_{\text{top}})|$ is decreased, the bubble thickens, and in dS space its size can eventually become comparable to a Hubble patch. In that regime the relevant saddle is the HM instanton \cite{Hawking:1981fz}: a completely homogeneous Euclidean solution with
\beq
\phi(x)=\phi_{\text{top}},~~~~~\d s^2 = \frac{(D-2)(D-1)}{2 V(\phi_{\text{top}})}\left( \d \tau^2 + \cos^2 \tau\,  \d \Omega_{D-1}^2\right)
\eeq
throughout the entire $S^D$. The instanton therefore computes the probability for the system to fluctuate homogeneously from the false vacuum up to the top of the barrier. Once one returns to Lorentzian signature, the configuration perched at the top is unstable, and quantum fluctuations drive it away from $\phi_{\text{top}}$. The subsequent evolution can be approximated by starting at $\phi = \phi_{\text{top}} + \delta \phi_{\text{quantum}}$ and evolving the equations classically, with $\delta \phi_{\text{quantum}}$ being of the order of the size of quantum fluctuations of the unstable mode.

Let us analyze the phase of the path integral over the HM instanton. The first simplification, compared to the CDL setup, is that the background scalar profile is homogeneous $\phi(x)=\phi_0$ (this could be either the false vacuum value or the HM one). Because $\partial\phi=0$ on the saddle, the term $\int (\partial\phi)^2$ contributes only at quadratic order in fluctuations. Likewise, since the background satisfies $V'(\phi_0)=0$, the expansion of $\sqrt{g}\,V(\phi)$ begins quadratically in the scalar fluctuation. As a result, the scalar and metric fluctuations decouple at quadratic order. The metric sector sees only the same quadratic structure as pure gravity with cosmological constant $\Lambda = V(\phi_0)$. As a result, the phase of the saddle can be read off as the product of a pure-gravity contribution and a scalar contribution. The answer takes the form
\beq
Z = \ii^{D+2}\,\ii^{-N_\phi}\,|Z|.
\eeq
The gravity factor $\ii^{D+2}$ comes from the contour rotation required to define the Gaussian integral over the conformal mode \cite{Polchinski:1988ua}, see also \cite{Maldacena:2024spf,Shi:2025amq,Ivo:2025yek}. The second factor depends on the scalar sector: $N_\phi$ is the number of negative modes of the quadratic scalar operator. Therefore the relative phase between the HM saddle and the false vacuum is
\beq
\frac{Z_{\text{HM}}}{Z_{\text{fv}}} = \ii^{-\Delta N_\phi}\,\left|\frac{Z_{\text{HM}}}{Z_{\text{fv}}}\right|,
\eeq
where $\Delta N_\phi$ is the change in the number of scalar negative modes between the two saddles. If the HM  saddle differs by exactly one scalar negative mode $\Delta N_\phi=1$, then the ratio has the phase appropriate to an instability, and it is natural to interpret its absolute value
\beq
\Gamma = \left|\frac{Z_{\text{HM}}}{Z_{\text{fv}}}\right| \sim \exp\left(-\Delta S_{\rm dS}\right)
\eeq
as the decay rate of the false vacuum, with $\Delta S_{\rm dS}=S_{\rm dS}(\phi_{\rm fv})-S_{\rm dS}(\phi_{\rm top})$.

The counting of $N_\phi$ is straightforward. Expanding the scalar about a homogeneous background $\phi(x)=\phi_0$, the quadratic fluctuation operator is
\beq
\Delta_\phi = -\nabla^2 + V''(\phi_0),
\eeq
so $N_\phi$ is simply the number of negative eigenvalues of $\Delta_\phi$. When $\phi_0$ is either $\phi_{\text{fv}}$ or $\phi_{\text{tv}}$, one has $V''(\phi_0)>0$, and since $-\nabla^2$ is non-negative there are no unstable scalar modes:
\beq
N_\phi = 0~~~\text{for}~~~\phi(x)=\phi_{\text{fv}}~~\text{or}~~\phi_{\text{tv}}
\eeq
At the top of the barrier the situation changes because $V''(\phi_{\text{top}})<0$. The constant fluctuation $\delta\phi = \text{const.}$ is then automatically a negative mode, so the HM saddle always has at least one unstable direction. This is the $\ell=0$ scalar harmonic on $S^D$. A pertinent question is whether there are additional non-homogeneous negative modes. For those, the Lichnerowicz-Obata theorem implies \cite{Shi:2025amq}
\beq
-\nabla^2 \geq \frac{2D}{(D-1)(D-2)}\,\Lambda_{\text{top}}
= \frac{2D}{(D-1)(D-2)}\,V(\phi_{\text{top}})
\eeq
on all non-constant modes. The lower bound is saturated by the $\ell=1$ harmonic on $S^D$. It follows that the first inhomogeneous fluctuation is still stable provided
\beq
\frac{|V''(\phi_{\text{top}})|}{V(\phi_{\text{top}})} \leq \frac{2D}{(D-1)(D-2)}.
\eeq
In that range one has exactly one negative scalar mode,
\beq
N_\phi = 1,~~~\text{for}~~~\phi(x)=\phi_{\text{top}},
\eeq
which is precisely what is needed for the HM saddle to describe a decay process.

If the barrier is steeper than this bound, the story changes again. The $\ell=1$ scalar harmonics on the sphere then become unstable, contributing an additional $D+1$ negative modes. In that case one finds $N_\phi = D+2$, and the ratio $Z_{\text{HM}}/Z_{\text{fv}}$ becomes real and positive. At first sight this seems troubling: the would-be decay saddle no longer carries the phase appropriate for an instability.

The resolution is simple and physically natural. The issue is that the HM saddle has ceased to be the dominant contribution to vacuum decay. As derived in \cite{Jensen:1983ac}, when $|V''(\phi_{\text{top}})|$ is increased beyond a critical value, the CDL bubble solution comes into existence. The details in \cite{Jensen:1983ac} apply to $D=4$ but a straightforward generalization to $D$ dimensions shows that such a bubble exists whenever\footnote{Linearize the Euclidean scalar equation near the top of the barrier on the dS four-sphere,
$
\delta\phi''-(D-1)\tan\tau\,\delta\phi' = {\sf R}_{\text{top}}^2 V''(\phi_{\text{top}})\,\delta\phi  .
$
The lowest nontrivial mode satisfying the regularity conditions at $\tau=\pm \pi/2$ is $\delta\phi\propto \sin(\tau)$, which exists precisely when $-V''(\phi_{\text{top}})=D/{\sf R}_{\text{top}}^2$. For $|V''(\phi_{\text{top}})|<D/{\sf R}_{\text{top}}^2$, a configuration released near the top is always an overshoot, so no nontrivial CDL bounce exists. For $|V''(\phi_{\text{top}})|>D/{\sf R}_{\text{top}}^2$, one finds both overshoot and undershoot solutions, and continuity then guarantees a CDL solution. At the threshold $|V''(\phi_{\text{top}})|=D/{\sf R}_{\text{top}}^2$, the CDL bounce degenerates with  HM. Mathematically this is the same eigenvalue crossing, although physically one statement concerns the local stability of the HM saddle while the other concerns the existence of a distinct bounce solution.} 
\beq
|V''(\phi_{\text{top}})| \geq D/{\sf R}_{\text{top}}^2,~~~\text{or equivalently}~~~
\frac{|V''(\phi_{\text{top}})|}{V(\phi_{\text{top}})} \geq \frac{2D}{(D-1)(D-2)},
\eeq
which is exactly the same threshold at which the HM phase stops being imaginary. In other words, the apparent failure of the HM interpretation occurs precisely when a more appropriate thick-bubble saddle becomes available and takes over the decay channel. As $|V''(\phi_{\text{top}})|$ increases, the bubble becomes thinner and the CDL approximations apply.

There is one final subtlety. Even if the ratio $Z_{\text{HM}}/Z_{\text{fv}}$ has the correct phase, one must still decide how to fix the overall $\ii^{D+2}$ phase of the vacuum partition function itself, both for the false and the true one. For homogeneous saddles this is again a question in pure gravity, since the metric and scalar sectors decouple. As proposed by Maldacena \cite{Maldacena:2024spf} and further discussed in \cite{Chen:2025jqm}, this phase can be fixed by including an observer. The same prescription should therefore be applied to the HM saddle. With that understood, the path integral gives a consistent description of the decay of unstable dS space in the presence of the observer.

\subsection{Black-hole nucleation in de Sitter}\label{sec:Nariai-GPI-decay}

We now return to the Nariai geometry and use the same logic to argue that it should be interpreted as the saddle for black-hole nucleation in dS.\footnote{A related story for black holes in AdS is being considered in \cite{WIP}.} Consider now Einstein gravity with a positive cosmological constant
\beq
I = \frac{1}{16 \pi G_N} \int \d^D x \, \sqrt{-g} (R- 2\Lambda).
\eeq
For now, we focus on the pure-gravity sector with a positive cosmological constant, leaving possible matter fields turned off in the classical solutions. The maximally symmetric vacuum is de Sitter space, 
\beq
\d s^2 = {\sf R}^2 \left( -\d t^2 + \cosh^2 t \, \d \Omega_{D-1}^2\right),~~{\sf R}^2 =\frac{(D-2)(D-1)}{2\Lambda}.
\eeq
The Lorentzian solutions relevant for our discussion are the Schwarzschild-de Sitter (SdS) geometries,
\beq
\d s^2 = - f(r)\, \d t^2 + \frac{\d r^2}{f(r)} + r^2 \d \Omega_{D-2}^2,
\qquad
f(r) = 1 - \frac{2\mu}{r^{D-3}} - \frac{r^2}{{\sf R}^2}.
\label{eq:SdS_metric_section}
\eeq
For an appropriate range of the mass parameter $\mu$, the function $f(r)$ has two positive roots,
\beq
f(r_b)=f(r_c)=0,
\qquad
r_b<r_c,
\eeq
where $r_b$ is the black-hole horizon and $r_c$ is the cosmological horizon. In the Lorentzian spacetime, the remaining roots are negative or complex and therefore unphysical. The static patch lies between the two horizons $r_b<r<r_c$. As the mass $\mu$ increases, the two horizons move toward each other, and in the limiting configuration they coincide. This is the Nariai limit, with horizon radius $r_b=r_c=r_N = {\sf R} \sqrt{(D-3)/(D-1)}$.

The Euclidean continuation makes it immediately clear why this limit is special. Setting $t=-\ii \tau$ gives
\beq\label{eq:4Dmetric}
\d s_E^2 = f(r)\, \d \tau^2 + \frac{\d r^2}{f(r)} + r^2 \d \Omega_{D-2}^2.
\eeq
Regularity at a horizon $r=r_h$ requires the Euclidean time circle to have period
$
\beta_h = \frac{4\pi}{|f'(r_h)|},
$
with $r_h=r_b$ or $r_c$. For a generic SdS black hole, the two horizons have different surface gravities, so no single choice of Euclidean period can remove both conical singularities. In this sense the Euclidean continuation of a generic dS black hole is necessarily singular. The exceptional case is the Nariai limit which has a double root,
$
f(r_N)=0,
$
$
f'(r_N)=0,
$
with $r_b,r_c \to r_N$.  Although the radii are degenerate, the proper distance between the two horizons remains finite, so the region between them does not collapse. As we review in Section~\ref{sec:nearNariai}, the resulting Lorentzian geometry in the static patch factorizes into 2d dS space times a round sphere, while the Euclidean continuation becomes the smooth compact manifold
\beq
S^2 \times S^{D-2}.
\eeq
A convenient form of the Euclidean metric is
\beq
\d s_E^2 = \frac{D-2}{2\Lambda}\left[\sin^2\chi\, \d \tau^2 + \d \chi^2 + (D-3)\d \Omega_{D-2}^2\right],
\qquad 0\leq \tau <2\pi.
\label{eq:EuclideanNariai}
\eeq
Thus Euclidean Nariai is the unique smooth compact saddle within the dS black hole family, and for this reason it is the natural candidate for black-hole nucleation in dS.

This saddle has a natural dynamical interpretation. Ginsparg and Perry showed that Euclidean Nariai has a negative mode \cite{Ginsparg:1982rs}, suggesting that it should be viewed as the instanton mediating black-hole nucleation in dS space. In that interpretation the path integral on $S^2 \times S^{D-2}$ computes the probability for empty dS to fluctuate into a maximal black hole. At leading semiclassical order one expects a nucleation rate of the form
\beq
\Gamma_{\rm Nariai}
\;\sim\;
\left|\frac{Z_{S^2\times S^{D-2}}}{Z_{S^D}}\right|
\;\sim\;
\exp\!\left[-\bigl(I_{\rm Nariai}-I_{\rm dS}\bigr)\right].
\label{eq:nucleationrate}
\eeq
Once the system nucleates a Nariai configuration, quantum fluctuations will inevitably push it slightly away from the exact degenerate point, just like in the HM instanton quantum fluctuations push it slightly down the potential hill. The resulting Lorentzian solution is then near-Nariai rather than exactly Nariai and after some short initial time it can be treated semiclassically. This initiates the subsequent evaporation of the black hole through a net thermal flux between the black-hole horizon and the cosmological horizon. This will be the topic of Section~\ref{sec:Antievaporation}. In this sense the Nariai instanton should be viewed as the beginning of a decay channel. Unlike HM, it does not lead to a new vacuum, but rather to a black hole configuration that eventually relaxes back to empty dS.

This brings us to the phase of the gravitational path integral. For the Nariai saddle to admit the nucleation interpretation just described, the corresponding amplitude should be imaginary. However, the one-loop analysis of \cite{Shi:2025amq,Ivo:2025yek} found that the pure gravitational answer on Euclidean Nariai is real and positive. More concretely, for the empty saddles one finds
\beq
Z_{S^D} = \ii^{D+2}\, |Z_{S^D}|,
\qquad
Z_{S^2\times S^{D-2}} = |Z_{S^2\times S^{D-2}}|.
\label{eq:phasesempty}
\eeq
These results raise two related problems. First, the dS partition function is not always real or positive, in tension with an interpretation as a count of dS degrees of freedom. Second, the fact that the Nariai path integral is real is precisely what obstructs a direct nucleation-rate interpretation of the empty Nariai path integral.

A prescription for extracting real and positive density of states from the sphere saddle is introduced in \cite{Maldacena:2024spf}, which requires the inclusion of an observer.\footnote{ It was proposed in \cite{Maldacena:2024spf} that the observer prescription can be realized by a lukewarm black hole, namely a magnetically charged black hole in thermal equilibrium with the cosmological horizon; this realization was further checked in \cite{Chen:2025jqm}. In Appendix~\ref{sec:symmetries}, we explain from symmetry considerations why such a lukewarm black hole can play the role of an observer, and why this interpretation breaks down upon taking the Nariai limit.} Semiclassically, the observer is modeled by the geodesic of a massive particle, and the corresponding worldline fluctuation operator carries its own zero and negative modes. These negative modes modify the phase of the bare gravitational saddle by a factor of $(-\ii)^{D-1}$. Furthermore, one must carefully distinguish the bare gravitational path integral from the actual state-counting observable. In the former, the final integral over $\beta$, the proper length of the Euclidean time circle, is taken along the real steepest-descent contour.\footnote{At one loop, the collective variable $\beta$ contains the homogeneous $\ell=0$ trace mode together with other tensor components. Since the conformal factor problem typically requires a Wick rotation of the trace field, one might worry that the $\beta$ integral inherits such a rotation. This is not the case, because the quadratic action for the homogeneous $\ell=0$ mode is positive definite before Wick rotation, unlike the higher $\ell$ modes. Its steepest-descent contour must therefore be rotated back relative to the standard contour for the full trace field. As a result, in the gravitational path integral the $\beta$ integral runs along the real axis.} By contrast, to obtain the state-counting observable one must perform an inverse Laplace transform to project onto the microcanonical density of states. With the addition of a factor of $\ii$ from this contour rotation, the resulting quantity is real and positive, and admits the interpretation of a genuine density of states of dS static patch as seen by the observer,
\begin{equation}
    \frac{Z_{\rm dS}}{|Z_{\rm dS}|}=\ii\,\cdot\, \frac{Z_{S^D}}{|Z_{S^D}|}\, \frac{Z_{\rm particle}}{|Z_{\rm particle}|}=\ii\,\cdot\,\ii^{D+2}\,\cdot\,(-\ii)^{D-1}=1.
\end{equation}

We now implement the same prescription for the Nariai saddle. A new feature relative to the round sphere case is that the Nariai saddle is a product $S^2\times S^{D-2}$, so a massive observer's worldline can wind not only around the thermal circle in the $S^2$ factor, but also along a non-trivial closed circle on $S^{D-2}$. We label the different observer channels by the winding numbers $(n_1,n_2)$, the explicit parameterization of these worldlines is given in Appendix~\ref{app:ObserverandBetaIntegral}. The channel $(1,0)$ is the standard observer, which is static in the dS$_2$ static patch and localized on $S^{D-2}$. Channels with $n_2\neq 0$ continue to observers rotating on the $S^{D-2}$ factor in Lorentzian signature.

To accommodate this angular momentum on the $S^{D-2}$ factor, we introduce a Euclidean twist parameter $\eta_I$ for each Cartan generator of the rotation group $SO(D-1)$. The observer-refined path integral takes the schematic form
\begin{equation}
    Z_{\rm obs}=\int \d\beta\, \prod_{I=1}^{r}\d\eta_I \; Z_N(\beta,\vec\eta),
    \qquad \vec\eta=(\eta_1,\dots,\eta_r).
\label{eq:BetaEtaIntegral}
\end{equation}
where $r=\Bigl\lfloor \frac{D-1}{2}\Bigr\rfloor$ is the rank of $SO(D-1)$. With $\beta$ and $\vec\eta$ held fixed, $Z_N(\beta,\vec\eta)$ denotes the total contribution from the gravitational path integral on the Nariai background, the observer worldline, and the associated measure factors. $\beta$ is the proper length of the Euclidean time circle at the observer location, while the twists $\eta_I$ are defined through the identification 
\begin{equation}
    (\tau,\phi_I)\sim (\tau+\beta,\phi_I+\eta_I), \qquad I=1,\dots,r,
\end{equation}
where $\phi_I$ are azimuthal angles associated with a choice of Cartan generators. 
The $(\beta,\vec\eta)$ integrals are evaluated semiclassically about the relevant saddle. In the probe limit where $m\,G_N/r_N\ll 1$, the gravitational sector dominates and the saddle is pinned near the symmetric Nariai point $\beta=\beta_N$ and ${\vec\eta}_N=0$, with an $O(m\,G_N /r_N)$ correction proportional to $n_2/n_1$. The discrete labels $(n_1,n_2)$ enter the on-shell answer only through the worldline action evaluated on the bare Euclidean Nariai geometry.

The phase of the observer-refined path integral is fixed by the choice of observer channel,
\begin{equation}
    \frac{Z_{\rm obs}}{|Z_{\rm obs}|}= \frac{Z_{S^2\times S^{D-2}}}{|Z_{S^2\times S^{D-2}}|} \frac{Z_{\rm particle}}{|Z_{\rm particle}|}=1\,\cdot\, \frac{Z_{\rm particle}}{|Z_{\rm particle}|},
\end{equation}
The particle path integral in the $(n_1,n_2)$ channel carries the phase $(-\ii)^{N_{\rm neg}}$, where $N_{\rm neg}$ is the number of negative modes given in \eqref{eq:NegativeModesParticle}.

Before turning to the observer sector relevant for black-hole nucleation, we first analyze the contour of the $\beta$ integral. A natural attempt would be to carry over the sphere analysis~\cite{Maldacena:2024spf,Chen:2025jqm}, according to which the final $\beta$ integral in the gravitational path integral differs from the inverse-Laplace-transform contour by an additional factor of $\ii$. If so, the phase for the Nariai saddle entering the nucleation rate would be
\begin{equation}
    \frac{Z_{\rm Nariai}}{|Z_{\rm Nariai}|}\stackrel{?}{=}\ii\,\cdot\,\frac{Z_{\rm obs}}{|Z_{\rm obs}|}.
\end{equation}
We will show, however, that this direct extrapolation is not correct. In the Nariai case, the unstable gravitational fluctuation can mix nontrivially with $\beta$. As a result, the relevant steepest-descent contours cannot be inferred by direct analogy with the sphere. Instead, they must be determined from the Nariai one-loop fluctuations and then compared with the inverse-Laplace-transform contour that extracts the density of states.\footnote{In the probe limit $mG_N/r_N\ll 1$, the worldline fluctuations are parametrically subleading compared to the graviton fluctuations. Accordingly, the steepest-descent contour is fully determined by the gravitational sector. Corrections from backreaction enter at order $m\,G_N/r_N$, see Appendix~F of~\cite{Chen:2025jqm}.}

\paragraph{The homogeneous $(A,B)$ sector.}
To illustrate this point, we integrate out all higher partial waves and truncate the one-loop path integral to the homogeneous sector. These are the only metric fluctuations that preserve the $SO(3)\times SO(D-1)$ symmetry of the Nariai background
\begin{equation}
    \delta g_{\mu\nu} = A\,g^{(2)}_{ab} + B\,g^{(D-2)}_{ij}\,,
\label{eq:HomogPert}
\end{equation}
where $g^{(2)}_{ab}$ and $g^{(D-2)}_{ij}$ are the two blocks of the round metrics on $S^2$ and $S^{D-2}$. Geometrically, $A$ and $B$ are the rescaling of the $S^2$ and $S^{D-2}$ radii,
\begin{equation}
    A = 2\,\frac{\delta r_2}{r_2},
    \qquad
    B = 2\,\frac{\delta r_N}{r_N},
\label{eq:ABrescale}
\end{equation}
with $r_2$ and $r_N=\sqrt{D-3}\,r_2$ the radii of the two spheres.

Substituting \eqref{eq:HomogPert} into the Einstein action and expanding to quadratic order gives
\begin{equation}
S^{(2)}[A,B]=\frac{\Lambda V_{S^2\times S^{D-2}}}{32\pi G_N}
    \left[4AB+(D-4)B^2\right].
\label{eq:QuadActionAB}
\end{equation}
This $2\times 2$ quadratic form is indefinite for $D>4$. The
unstable direction is the negative mode in the transverse-traceless sector identified in \cite{Ginsparg:1982rs}, which parametrizes the relative scaling of $S^2$ and $S^{D-2}$. After absorbing the modulus of the one-loop determinant of all non-homogeneous modes into $|Z'|$ and the phase from the Wick rotation of the higher conformal-factor partial waves into a factor $\ii$,\footnote{In the conformal-factor sector, all higher partial waves have negative kinetic action, while the homogeneous component has positive action. Wick-rotating each negative direction by $+\pi/2$ produces a formal phase $(-\ii)^{\infty-1}$. The divergent $\ii^\infty$ from wick rotating the whole conformal sector $h$ is absorbed into local counterterms, leaving a factor of $\ii$ and positive quadratic action for every mode in conformal factor~\cite{Shi:2025amq}.} the gravitational partition function reads
\begin{equation}
   Z_{\rm grav}=\ii\, |Z'| \int \d A\,\d B\, \exp\left(-\frac{\Lambda V_{S^2\times S^{D-2}}}{32\pi G_N}
    \left[4AB+(D-4)B^2\right]\right).
\end{equation}
For $D>4$, performing the Gaussian integral over $B$ along the real axis gives a single integral over the homogeneous $S^2$-scale mode $A$,
\begin{equation}
    Z_{\rm grav} = \ii\,\bigl|Z''\bigr|
    \int \d A\;
    \exp\!\left(\,
        \frac{\Lambda\,V_{S^2\times S^{D-2}}}{32\pi G_N}
        \cdot\frac{4}{D-4}\,A^2
    \right),
\label{eq:ZgravA}
\end{equation}
where $|Z''|$ absorbs the Gaussian normalization from the $B$-integration into $|Z'|$. The exponent in~\eqref{eq:ZgravA} has a positive sign, so the action is unbounded below as a function of $A$, indicating the scale of $S^2$ is not stabilized near the saddle. This reflects the Ginsparg-Perry instability of the Nariai saddle repackaged in the $S^2$ scale after $S^{D-2}$ has been integrated out. Convergence of the $A$-integral demands a contour rotation onto the imaginary axis. We fix the orientation as
\begin{equation}
    \mathcal{C}_A \;=\; e^{-\i\pi/2}\,\mathbb{R},
\label{eq:Acontour}
\end{equation}
which renders the rotated Gaussian convergent and makes $Z_{\rm grav}$ real and positive. Notice that integrating $B$ over the real axis and $A$ over the imaginary one is akin to the Lorentzian path integral introduced in \cite{Marolf:2022ybi}. 

The identifications~\eqref{eq:ABrescale} have a natural thermodynamic interpretation. The simplest $(1,0)$ observer worldline $\gamma$ is the equatorial circle of $S^2$, located at a fixed point $y_0\in S^{D-2}$. Its proper length is $\beta = 2\pi r_2$, so a variation of the $S^2$ radius shifts the inverse temperature as
\begin{equation}
    \delta\beta \;=\; 2\pi\,\delta r_2 \;=\; \pi r_2\,A.
\label{eq:dBetaA}
\end{equation}
The entropy for each horizon is $S_{\rm hor} = \Omega_{D-2}\,r_N^{D-2}/(4G_N)$, and the first law $\delta E = 2 T\,\delta S_{\rm hor}$ at the Nariai temperature relates the energy fluctuation to the radius of the transverse sphere,
\begin{equation}
    \delta E \;\propto\; \delta r_N \;\propto\; B.
\label{eq:dEB}
\end{equation}
Hence the integration over the homogeneous modes $A$ and $B$ plays exactly the same role as the two dimensional $(\beta,E)$ model of~\cite{Chen:2025jqm}. Following their prescription, we perform the $A$ ($\beta$) integral first to impose the Hamiltonian constraint on $B$ ($E$). The key is that the integration contour \eqref{eq:Acontour} for $A$ turns the cross term $4AB$ to $-4\,\ii\,\tilde AB$, the Fourier transform of a delta function on $B$. Explicitly,
\begin{equation}
\begin{aligned}
&~\ii \,|Z'| \int_{\mathcal{C}_B} \d B\,\int_{\mathcal{C}_A} \d A\, \exp\left(-\frac{\Lambda V_{S^2\times S^{D-2}}}{32\pi G_N}
    \left[4AB+(D-4)B^2\right]\right)\\
   =&  ~\ii \,|Z'''| \int_{\mathcal{C}_B} \d B\,(-\ii)\,\delta(B)\, \exp\left(-\frac{\Lambda V_{S^2\times S^{D-2}}}{32\pi G_N}
    \left[(D-4)B^2\right]\right)
    \\
    =&\,\,~ |Z'''|.
\end{aligned}
\end{equation}
The contour $\mathcal{C}_A$ in~\eqref{eq:Acontour} thus plays a dual role. It is simultaneously the steepest-descent contour of the bare gravitational path integral (forced by the Ginsparg-Perry instability) and the inverse-Laplace-transform contour through which the $A$-integral imposes the Hamiltonian constraint on $B$.
No additional factor of $\ii$ is incurred, in contrast to the sphere and lukewarm analyses of~\cite{Maldacena:2024spf,Chen:2025jqm}, where the bare gravitational $\beta$-contour is the real axis and rotation to the inverse-Laplace-transform contour produces a $-\ii$.

In $D=4$, however, the quadratic action~\eqref{eq:QuadActionAB} reduces to $4AB$, reflecting the fact that the two factors of $S^2\times S^2$ are on equal footing in Euclidean signature. Despite this symmetry, the contour prescription $\mathcal{C}_A=e^{-i\pi/2}\mathbb{R}$ and $\mathcal{C}_B=\mathbb{R}$ remains consistent, as the $A$-integral renders the path integral convergent and simultaneously imposes the Hamiltonian constraint $\delta(B)$, provided $A$ is identified as the inverse-temperature modulus.

Through the identification \eqref{eq:dBetaA}, the contour for $\beta$ in \eqref{eq:BetaEtaIntegral} is fixed to be 
\begin{equation}
    \mathcal{C}_\beta \;=\; \beta_0 + e^{-\i\pi/2}\,\mathbb{R}.
\label{eq:CBetaTransition}
\end{equation}
where $\beta_0$ is the on shell value. We have so far analyzed the $\beta$-contour within the reduced model for homogeneous modes. However, $\beta$ as the proper length can receive contributions from higher partial waves of the metric fluctuations. A complete treatment of them could in principle refine this contour prescription. We leave a full analysis for future work.

For the Cartan twists $\eta_I$, the corresponding metric perturbation is the off-diagonal mode $\delta g_{\tau\phi_I}\propto \eta_I\sin^2\chi$. These modes are orthogonal to both the conformal factor and the Ginsparg-Perry negative mode, so their quadratic action is positive definite. The steepest-descent contour for each $\eta_I$ is thus taken along the real axis $\mathcal{C}_{\eta_I}=\mathbb{R}$.

With the contours for the observer-refined gravitational path integral \eqref{eq:BetaEtaIntegral} in hand, we notice that it is also the integral that projects to the microcanonical observable at $E=0$ and $\vec J=0$,
\begin{equation}
    Z_{\text{Nariai}}(E,\vec J)
    \;=\;
    \int_{\beta_0 - \i\infty}^{\beta_0 + \i\infty}
        \!\d\beta\;e^{\beta E}
    \;\prod_{I=1}^{r}\int_0^{2\pi}\!\frac{\d\eta_I}{2\pi}\;
        e^{-\i\eta_I J_I}
    \;Z_N(\beta,\vec\eta).
\label{eq:RhoEJ}
\end{equation}
The phase of this microcanonical observable is therefore determined by the observer channel. In the $(1,0)$ channel, the Euclidean trajectory is a great circle on $S^2$ and is localized on $S^{D-2}$. Its worldline action carries a single negative mode contributing a factor of $-\ii$~\cite{Maldacena:2024spf}. Combining this with the real-positive bare gravitational partition function on Nariai, we obtain
\begin{equation}
 \frac{Z_{\rm Nariai}}{|Z_{\rm Nariai}|}=\,\frac{Z_{\rm obs}}{|Z_{\rm obs}|}=\frac{Z_{\rm particle}}{|Z_{\rm particle}|}\;\;\;\Rightarrow\;\;\; Z_{\text{Nariai}}=-\ii\,|Z_{\text{Nariai}}|\quad \text{for (1,0) channel.} 
\end{equation}
Thus, in the $(1,0)$ channel, the Nariai path integral in the presence of the observer is purely imaginary, supporting its interpretation as a nucleation rate. The dependence on the observer trajectory is not an ambiguity of the Nariai geometry. Rather, specifying the observer is part of specifying the static-patch quantity computed from the closed Euclidean saddle. The $(1,0)$ observer has a particularly natural Lorentzian interpretation. After Lorentzian continuation, it describes the observer sitting at the center of the dS$_2$ static patch. Its unique static geodesic also picks out a normalization of the horizon temperatures, which is the natural Nariai analogue of the Bousso-Hawking normalization. Moreover, among the nontrivial worldline saddles that admit a timelike Lorentzian continuation, the $(1,0)$ saddle has the smallest Euclidean action, while its single negative mode supplies the factor of $-\ii$. It therefore gives the leading imaginary contribution associated with black hole nucleation in the standard static patch. Other winding sectors either do not continue to timelike observers or, when they do, have larger Euclidean actions and are semiclassically subleading. Their phases are determined by their respective negative mode spectra in Appendix~\ref{app:ObserverandBetaIntegral}. Such sectors may contribute to other quantities, but their detailed interpretations are not needed for the leading nucleation rate considered here.

The conclusion of this analysis is that the empty Euclidean Nariai gravitational path integral naturally fits a norm interpretation and is accordingly real, see \cite{Shi:2025amq}. The observer-dressed Nariai saddle we study here instead plays the role of the instanton controlling black-hole nucleation in dS and is accordingly imaginary. Once this distinction is kept in mind, the path integral interpretation of Nariai runs in close parallel with the HM discussion above, with the only essential difference being that the nucleated configuration eventually evaporates back to empty dS with the observer. It therefore has the interpretation of a rare Boltzmann fluctuation, in which empty dS fluctuates into a lower-entropy black-hole configuration that subsequently evaporates back to the original state.\footnote{In the case of charged black hole nucleation~\cite{Bousso:1996pn}, the
evaporation endpoint may be a lukewarm black hole rather than the
maximally entropic empty dS state, unless there is charged matter that
allows the black hole to discharge. See \cite{Montero:2019ekk} for a related discussion.}

\subsection{Combining the Hawking-Moss process with the Nariai solution}
\label{sec:HMplusNariai}

Let us return to the theory \eqref{eq:EH+scalar} with the potential illustrated in Figure~\ref{fig:HM_Potential}. For simplicity, we consider the case $D=4$, although the generalization to arbitrary dimensions is straightforward. If the system is initially in the false vacuum $\phi_{\text{fv}}$, several processes can occur, each governed by a different transition rate:

\paragraph{Black-hole nucleation in the false vacuum:} A Nariai black hole can nucleate within the false vacuum. As we will see in the following sections, this black hole subsequently evaporates, leaving the system in the false vacuum once again. Consequently, this process does not mediate a decay to the true vacuum. The rate is
\beq
    \Gamma_{\text{fv}\to\text{Nariai}_{\text{fv}}} \sim \exp\left(-\frac{\pi}{G_N V(\phi_{\text{fv}})}\right),
\eeq
and is therefore always exponentially suppressed. 

\paragraph{Hawking-Moss decay:} The false vacuum can transition to the homogeneous dS state at the top of the potential barrier, $\phi_{\text{top}}$, which subsequently rolls down to the true vacuum. The rate for this process is 
\beq
    \Gamma_{\text{fv}\to\text{top}} \sim \exp\left(-\frac{3\pi}{G_N} \left[ \frac{1}{V(\phi_{\text{fv}})} - \frac{1}{V(\phi_{\text{top}})} \right] \right).
\eeq
This is the standard HM instanton.\footnote{It is also possible to nucleate a Nariai black hole at $\phi_{\text{top}}$ that concurrently evaporates and rolls down toward the empty dS true vacuum. However, one can readily verify that this process is always exponentially subleading compared to the HM transition.} Because $V(\phi_{\text{fv}}) < V(\phi_{\text{top}})$, the HM instanton is inherently exponentially suppressed. The relative probability of this decay versus Nariai nucleation depends on the height of the potential barrier. If $V(\phi_{\text{fv}}) < V(\phi_{\text{top}}) < \frac{3}{2} V(\phi_{\text{fv}})$, the HM-mediated decay to the true vacuum is more probable than Nariai nucleation within the false vacuum. Conversely, if $V(\phi_{\text{top}}) > \frac{3}{2} V(\phi_{\text{fv}})$, Nariai nucleation in the false vacuum becomes the dominant process. This does not preclude eventual decay to the true vacuum; the black hole will evaporate back into the false vacuum, allowing for a subsequent HM tunneling to $\phi_{\text{top}}$.

\paragraph{Black-hole nucleation in the true vacuum:} The false vacuum can also nucleate a Nariai black hole directly with $\phi=\phi_{\text{tv}}$, which subsequently evaporates into the empty dS true vacuum. This constitutes a decay channel that is genuinely distinct from the HM instanton. Its rate is 
\beq
    \Gamma_{\text{fv}\to\text{Nariai}_{\text{tv}}} \sim \exp\left(-\frac{\pi}{G_N} \left[ \frac{3}{V(\phi_{\text{fv}})} - \frac{2}{V(\phi_{\text{tv}})} \right] \right).
\eeq
It is expected that black-hole nucleation is exponentially suppressed unless the cosmological constant is comparable to the Planck scale \cite{Bousso:1995cc}. However, this process reveals a loophole. If the potential of the true vacuum is sufficiently low compared to that of the false vacuum, the nucleation rate becomes unsuppressed:
\beq
    V(\phi_{\text{tv}}) \lesssim \frac{2}{3} V(\phi_{\text{fv}}) \quad \implies \quad \Gamma_{\text{fv} \to \text{Nariai}_{\text{tv}}} \sim \mathcal{O}(1).
\eeq
The implications of this observation for quantum gravity remain to be fully understood. One possibility is that this particular process does not contribute to the gravitational path integral. Alternatively, we interpret it as suggesting a new consistency bound on scalar potentials,
\beq
V(\phi_{\text{tv}}) > \frac{2}{3} V(\phi_{\text{fv}}),
\eeq
which must be satisfied by any well-defined theory of quantum gravity. It would be interesting to study similar black hole nucleation processes with potential that have both dS vacua as well as flat or AdS, which we leave for future work.

\section{Anatomy of the near-Nariai geometry: classical analysis} 
\label{sec:anatomyNariai}

In this section, we review the near-Nariai regime of 4d dS black holes and its description by a 2d dilaton-gravity theory approximated by JT gravity \cite{Maldacena:2016upp, Jensen:2016pah, Engelsoy:2016xyb, Mertens:2022irh, Turiaci:2024cad}. We do this carefully in order to expose the residual diffeomorphisms and the SL$(2,\mathbb{R})$ redundancies of the nearly dS$_2$ factor. These will later be essential for identifying which ``horizon dynamics'' is physical and which is pure gauge in the evaporation problem. 

\subsection{The near-Nariai throat and two-dimensional dilaton gravity} \label{sec:nearNariai}

From now on we specialize the discussion to $D=4$ Einstein gravity with positive cosmological constant $\Lambda$. We will specify the matter content later. We consider the metric \eqref{eq:4Dmetric} for generic mass. In 4d we have the empty dS radius ${\sf R} = \sqrt{3/\Lambda}$ and the black-hole mass is given by $\mu = G_N M$. As $M$ increases, the black-hole horizon grows and the cosmological horizon shrinks until they coincide, $r_b=r_c=r_N$, which in $D=4$ is at
\be \label{eq:Nariai_radius_mass}
r_N=\frac{1}{\sqrt{\Lambda}},
\qquad
G_N M_N=\frac{1}{3\sqrt{\Lambda}}.
\ee
The mass cannot be increased further, since the geometry develops a naked singularity.

The purpose of this section is to study the evaporation of a near-Nariai black hole after nucleation. Consider now a dS black hole with mass close to, but smaller than, the Nariai value. We parametrize the horizon radii as
\be
r_{b,c}=r_N\Big(1\mp\epsilon+O(\epsilon^2)\Big),
\ee
where $\epsilon\ll1$ in the near-Nariai limit. Although the radii are close, the proper distance between them remains finite as $\epsilon \to 0$, and there is still a large region of spacetime between the horizons \cite{Ginsparg:1982rs}. Introduce coordinates ${\sf t}$ and $\chi\in(0,\pi)$ by
\be
t=\frac{1}{\epsilon\sqrt{\Lambda}}{\sf t},
\qquad
r=\frac{1}{\sqrt{\Lambda}}\bigg[1-\epsilon\cos\chi-\frac{1}{6}\epsilon^2\bigg],
\ee
which give the near-Nariai metric
\be \label{eq:nearNariai}
\d s^2
=\frac{1}{\Lambda}\bigg[
-\Big(1+\frac{2}{3}\epsilon\cos\chi\Big)\sin^2\chi\,\d{\sf t}^2
+\Big(1-\frac{2}{3}\epsilon\cos\chi\Big)\d\chi^2
+\Big(1-2\epsilon\cos\chi\Big)\d\Omega_2^2
\bigg].
\ee
This is the form usually quoted in \cite{Ginsparg:1982rs} and \cite{Bousso:1997wi}. The connection to JT gravity becomes more direct after the coordinate change
$
\chi\to\chi+\epsilon\,(\frac{2}{3}\sin\chi\,\ln(\sin\chi)+\frac{1}{6}\sin\chi),
$ giving
\begin{equation}
\d s^2=\frac{1}{\Lambda}\Big[\Phi^{-1/2}(\chi)\tilde{\Phi}(\chi)\left(-\sin^2\chi\,\d{\sf t}^2+\d\chi^2\right)
+\Phi(\chi)\,\d\Omega_2^2\Big],
\end{equation}
with
\begin{equation}
\tilde{\Phi}(\chi)=1+\frac{4\epsilon}{3}\cos\chi\log\sin\chi,~~~~\Phi(\chi)=  1- 2 \epsilon \cos \chi.
\end{equation}
For $\epsilon=0$, the geometry is exactly $\text{dS}_2 \times S^2$ with curvature set by $\Lambda$, since $\Phi(\chi)=\tilde{\Phi}(\chi)=1$.\footnote{Even in the Nariai limit, if one looks at the asymptotic future-expanding patch the geometry is very far from dS$_2$ as emphasized in \cite{Kolanowski:2019xpq}. This region is important when the SdS is interpreted as preparing the wavefunction on an $S^1 \times S^{D-2}$ universe.} At finite $\epsilon>0$, the metric still contains an exact $\text{dS}_2$ factor, but its overall scale, set by $\tilde{\Phi}$, and the size of the transverse sphere, set by $\Phi$, vary nontrivially with $\chi$. The $\text{dS}_2$ symmetries are therefore explicitly broken unless $\epsilon=0$. This produces a nearly-dS$_2$ geometry \cite{Maldacena:2019cbz,Cotler:2019nbi}. 

 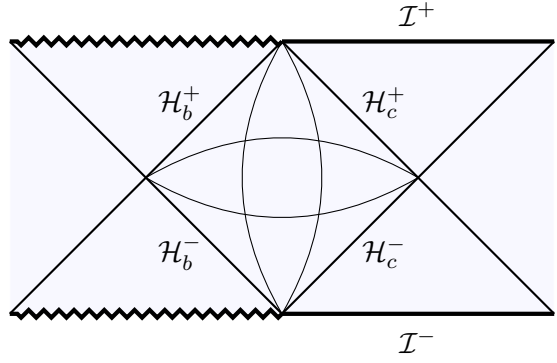
\begin{figure}
\begin{center}
\begin{tikzpicture}[scale=0.9]
\draw[white, fill=blue!10,nearly transparent] (-6,2) -- (-6,-2) -- (2,-2) -- (2,2) -- (-6,2);
\draw[thick] (-2,2) -- (2,-2);
\draw[thick] (-2,-2) -- (2,2);
\draw[thick] (-6,2) -- (-2,-2);
\draw[thick] (-6,-2) -- (-2,2);
\draw[ultra thick,decoration = {zigzag,segment length = 2mm, amplitude = 0.5mm},decorate] (-2,2) -- (-6,2);
\draw[ultra thick,decoration = {zigzag,segment length = 2mm, amplitude = 0.5mm},decorate]  (-2,-2) --  (-6,-2);
\draw[ultra thick] (-2,2)--(2,2);
\draw (0,2.4) node  {$\mathcal{I}^+$};
\draw[ultra thick] (-2,-2)--(2,-2);
\draw (0,-2.4) node  {$\mathcal{I}^-$};
\draw (-0.5,1.1) node {$\mathcal{H}_c^+$};
\draw (-0.5,-1.1) node {$\mathcal{H}_c^-$};
\draw (-3.5,1.1) node {$\mathcal{H}_b^+$};
\draw (-3.5,-1.1) node {$\mathcal{H}_b^-$};
\draw (-2,-2) to[bend right] (-2,2);
\draw (-2,-2) to[bend left] (-2,2);
\draw (-4,0) to[bend left] (0,0);
\draw (-4,0) to[bend right] (0,0);
\end{tikzpicture}
\end{center}
\vspace{-0.1cm}
\caption{Penrose diagram for a black hole in dS with mass $0<M<M_N$. We indicate the coordinates used in the text and the location of the horizons. The vertical lines have constant $\r$ or $\chi$. The horizontal lines have constant $\t$. $\mathcal{H}_c^\pm$ are located at $\chi=\pi$ or $\r=\infty$, while $\mathcal{H}_b^\pm$ are located at $\chi=0$ or $\r=-\infty$.}
\label{Penrosefd}
\end{figure}

It will also be convenient to a shifted radial coordinate
$
\r=\ln\tan\frac{\chi}{2}.
$
The metric takes the form
\be \label{eq:Nariaistatic}
\d s^2
=\frac{1}{\Lambda}\left[\Phi^{-1/2}(\r)\, \tilde{\Phi}(\r) \left(\frac{-\d\t^2+\d\r^2}{\cosh^2 \r}\right)+\Phi(\r)\, \d\Omega_2^2\right],
\qquad \r\in(-\infty,\infty),
\ee
where the same functions $\tilde{\Phi}(\chi)$ and $\Phi(\chi)$, now written in terms of $\r$, are
\be\label{eq:PhitPhiNDSD}
\tilde{\Phi}(\r) = 1 + \frac{4 \epsilon}{3} \tanh \r \, \log \cosh\r ,
\qquad
\Phi(\r)=1+2\epsilon \tanh\r.
\ee
The limits $\r\to-\infty$ and $\r\to+\infty$ still correspond to the black hole and cosmological horizons of the static patch, namely $\chi=0$ and $\chi=\pi$. We illustrate the two coordinate systems in Figure~\ref{Penrosefd}. 

The static-patch coordinates are the most convenient starting point for our Lorentzian quantum-state analysis in Section~\ref{sec:Antievaporation}. They are naturally adapted to the near-horizon Kruskal charts used to impose Hadamard regularity at each Killing horizon. In this patch, regularity and flux conditions are transparent and admit a direct interpretation in terms of what a static observer measures. For this reason, later refinements in the literature \cite{Nojiri:1998ue, Nojiri:1998ph} formulate the near-Nariai evolution problem directly in the static patch. By contrast, in \cite{Bousso:1997wi} the perturbation analysis was phrased in global coordinates, which are better suited to applications to the Hartle-Hawking wavefunction. In Section~\ref{sec:noboundary}, we relate their variables to our static-patch description.

The surface gravity associated with the Killing vector $\xi=\partial_t$ is
$
\kappa_h[\partial_t]=\frac{1}{2}\bigl|f'(r_h)\bigr|
$
and vanishes at a double root, similar to the horizon of an extremal black hole. In the absence of an asymptotic boundary, the normalization of $\xi$ is arbitrary. The temperature measured by a local observer at location $r$ is $T_{\rm loc}(r)=\frac{\kappa[\xi]}{2\pi\sqrt{-\xi^2(r)}}=\frac{\kappa[\partial_t]}{2\pi\sqrt{f(r)}}$ \cite{Tolman:1930zza, Tolman:1930ona}. 
Following Bousso and Hawking \cite{Bousso:1996au}, we fix the normalization using the unique static geodesic observer at $r_O$, defined by $f'(r_O)=0$.\footnote{For complementary studies of static sphere observers and geodesic probes in SdS and Nariai geometries, see~\cite{Faruk:2023uzs, Aalsma:2026kqj}.} This is the same physical static-patch observer that appeared in the observer-refined path integral of Section~\ref{sec:bigpicture}, where its worldline was needed to assign the nucleation saddle the appropriate decay phase. As $r_b\to r_c$, we have $f(r_O)\sim O(\epsilon^2)$ and the horizon temperatures remain finite as $\epsilon\to 0$, so that
\be \label{eq:Nariaitemperature}
T_{c,b}=\frac{\sqrt{\Lambda}}{2\pi}\Big(1\mp \frac{2}{3}\epsilon\Big)+O(\epsilon^2).
\ee
The Nariai value is the intrinsic Gibbons-Hawking temperature of the $\text{dS}_2$ horizon. Thus, near Nariai the system approaches an unstable thermal equilibrium between the two horizons, rather than a ``zero-temperature'' limit.\footnote{In this spirit, \cite{Aalsma:2025lcb} uses the Bousso-Hawking normalization to argue that the static-patch thermodynamic description need not suffer large quantum-gravity corrections as in the near-extremal limit \cite{Preskill:1991tb}. In Section~\ref{sec:discussion}, we will give a more explicit explanation of this fact.}

\subsection{Near-Nariai perturbations and horizon diagnostics}
\label{sec:classicalnearNariai}

As preparation for the study of quantum effects, we first analyze classical perturbations around the Nariai solution. We rewrite the 4d theory as a 2d dilaton-gravity theory coupled to matter in the near-horizon region, see also \cite{Castro:2022cuo}. The relevant 4d modes in the 2d description are
\be \label{eq:spherical_ansatz_gauged}
\d s^2_{4\mathrm{D}}=\frac{1}{\Lambda}\left[\Phi^{-1/2}(x) \, g_{ab}(x) \d x^a \d x^b + \Phi(x) \, D \mu^I D \mu^I \right].
\ee
In this ansatz, $g_{ab}$ ($a,b=0,1$) is a Lorentzian metric for the 2d sector. $\mu^I$ ($I=1,2,3$) are embedding coordinates for $S^2$ satisfying $\mu^I \mu^I = 1$, and $D \mu^I = \d \mu^I + \, \epsilon^{IJK} A^J(x) \mu^K$ is the gauge-covariant derivative. The 1-forms $A^I = A^I_a(x) \d x^a$ are the three $SU(2)$ gauge fields living on the 2d spacetime.\footnote{If $A^I=0$ we can reproduce the round metric used earlier with $\mu^1 = \sin \theta \cos \varphi$, $\mu^2=\sin \theta \sin \varphi$, and $\mu^3=\cos\theta$.} The dilaton $\Phi(x)$ controls the area of the transverse $S^2$ in units of the dS radius. In other words,
\be
\Phi(x)\equiv e^{-2\phi(x)}=\Lambda\,r^2(x).
\label{eq:dilaton_def}
\ee
We have also introduced $\phi(x)$ because the one-loop effective theory in Section~\ref{sec:Antievaporation} is most naturally written in terms of this variable. Starting from the Einstein action with a cosmological constant we can integrate it over the sphere to produce an effective 2d dilaton-gravity action
\be \label{eq:2Daction}
S = \frac{1}{4 G_N \Lambda} \int \d^2 x \sqrt{-g}\Big[ \Phi R - U(\Phi)- \frac{1}{4} Z(\Phi) F_{ab}^I F^{Iab}\Big] + S_{\text{matter}},
\ee
where $F^I = \d A^I +\frac{1}{2} \epsilon^{IJK} A^J \wedge A^K$ is the $\SU(2)$ field strength. The dilaton potentials are 
\be
U(\Phi) = 2 \Phi^{1/2}-\frac{2}{\Phi^{1/2}} ,~~~~Z(\Phi) = \frac{2}{3} \Phi^{5/2}.
\ee
This action could be supplemented with other possible matter gauge fields present in 4d. For a review of classical aspects of such dilaton-gravity theories, see \cite{Grumiller:2002nm,Mertens:2022irh}. The matter action $S_{\text{matter}}$ universally includes Kaluza-Klein (KK) modes of the 4d metric, which appear in the 2d theory as massive matter with a specific spectrum of masses arising from partial waves on $S^2$. Likewise, 4d matter fields and their KK modes produce additional 2d matter fields. In this paper we will focus on the massless modes
\be\label{eq:2D_matter_action}
S_{\text{matter}}= - \frac{1}{2} \sum_{i=1}^N \int \d^2 x \sqrt{-g} \,\Phi (\nabla f_i)^2 + S_{\text{KK}},
\ee
which arise as $\ell=0$ modes of massless 4d fields. All other (massive) KK modes from the metric or matter are part of the remainder $S_{\text{KK}}$. When we analyze the evaporation rate we will ignore this sector. This can be justified by looking at the greybody factors in the black hole geometry, see Appendix \ref{app:greybody}. 

\medskip

From now on we will focus on solutions without angular momentum, i.e. with $A^I=0$. We can easily reproduce the Nariai solution from this formalism. In the 4d language, the size of the transverse sphere is independent of position in the static patch. In the 2d dilaton-gravity language, this is a solution with an exactly constant dilaton. Set $\Phi=\Phi_0$ independent of position. The equation of motion from varying the 2d metric gives $U(\Phi_0)=0$, hence $\Phi_0=1$. The equation from varying the dilaton gives $R=U'(\Phi_0)=2$, which implies the 2d geometry is a constant positive curvature dS$_2$. The solution therefore exactly matches \eqref{eq:nearNariai} with $\epsilon=0$. In our conventions $\Phi_0=1$ corresponds to $r_N^2=1/\Lambda$.

The purpose of the rest of this section is to study classical perturbations around the Nariai background. In the 2d dilaton-gravity language, this corresponds to the expansion
\beq
\Phi(x)= \Phi_0 + 2 G_N \Lambda\,  \upphi(x),
\eeq
where $\upphi$ is an inhomogeneous small correction to the constant-dilaton background. This drastically simplifies the dilaton potential. For small perturbations around Nariai we can approximate $\frac{S_0}{2\pi}U(\upphi) \approx 2 \upphi$ and the 2d dilaton-gravity becomes JT gravity. This regime also simplifies the matter sector since the coupling of $f_i$ to $\upphi$ is a subleading correction. The 2d matter theory is well-approximated by a collection of free massless scalars on the dS$_2$ throat, effectively a 2d conformal field theory (CFT). The full theory reduces to
\bea \label{eq:JTaction}
S[g_{ab},\upphi,f_i]
&\approx &\frac{S_0}{4\pi} \int \d^2x\sqrt{-g}  R  + \frac{1}{2}\int \d^2x\sqrt{-g}\, \upphi (R -2) \nonumber\\
&&~~-\frac{S_0}{24\pi} \int \d^2x\,\sqrt{-g} F^I_{ab} F^{Iab}  - \frac{1}{2} \sum_{i=1}^N \int \d^2 x \sqrt{-g} \, (\nabla f_i)^2,
\ea
where the first term is topological and $S_0=\pi/G_N\Lambda$ is the Bekenstein-Hawking entropy of the Nariai horizon. This is a positive cosmological constant version \cite{Maldacena:2019cbz,Cotler:2019nbi} of the theory that appears in most analyses of nearly-AdS$_2$ gravity and its applications to near-extremal black holes, quantum chaos, black hole evaporation, and more \cite{Mertens:2022irh}. The varying dilaton softly breaks the dS$_2$ conformal isometries of the exact Nariai solution, just like the inflaton breaks the symmetries of dS$_4$ during inflation. We will see later that understanding the evaporation requires keeping the $\upphi^2$ correction to the dilaton potential.

\subsubsection*{Classical equations of motion}
We will study perturbations around Nariai, such that $\upphi$ is of order $\epsilon\ll 1$. The equations of motion of the 2d dilaton-gravity theory are
\be \label{eq:dilgravEoMclass}
 g_{ab} \Box \Phi-\nabla_a \nabla_b \Phi + \frac{1}{2} g_{ab} U(\Phi)= 0,
~~~~R=U'(\Phi).
\ee
For now, we focus on the classical near-Nariai dynamics. Quantum matter effects will be incorporated in Section~\ref{sec:Antievaporation}. To reproduce this dynamics directly in 4d we write the following ansatz:
\be \label{eq:metricansatz} 
\d s^2
=\frac{1}{\Lambda} \Big[e^{\phi(\t,\r)}\underbrace{e^{2\rho(\t,\r)}(-\d\t^2+\d\r^2)}_{\text{2d dilaton-gravity metric}}+ e^{-2\phi(\t,\r)} \d\Omega_2^2\Big],
\ee
and expand around Nariai
\be \label{eq:linearperturbation}
\rho(\t,\r)=\underbrace{- \log \cosh \r}_{=\rho_0(\r)}+\epsilon\,\mathcal{R}(\t,\r),
\qquad
\phi(\t,\r)=\phi_0-\epsilon\,\mathcal{S}(\t,\r),
\ee
$\mathcal{R}$ captures order $\epsilon$ corrections to the dS metric. $\mathcal{S}$ is proportional to the JT dilaton $\upphi$. The 4d Einstein equations, in the small $\epsilon$ regime, in the temporal and radial directions, become
\be \label{eq:nearNariai_tt}
\partial_\r^2 \mathcal{S}+\tanh \r \partial_\r \mathcal{S}+\sech^2 \r \mathcal{S}=0,
\ee
\be \label{eq:nearNariai_rr}
\partial_\t^2 \mathcal{S}+\tanh \r \partial_\r \mathcal{S}-\sech^2 \r \mathcal{S}=0,
\ee
\be \label{eq:nearNariai_tr}
\partial_\r\partial_\t \mathcal{S}+\tanh \r\partial_\t \mathcal{S}=0,
\ee
which are equivalent to the 2d JT equations $g_{ab} \Box \upphi - \nabla_a \nabla_b \upphi +  g_{ab} \upphi=0$ with the identification $\upphi$ proportional to $\epsilon \mathcal{S}$. The angular component of 4d Einstein equations is
\be \label{eq:classicaldilatoneq}
(\partial_\r^2-\partial_\t^2)\mathcal{R}+2\sech^2 \r (\mathcal{R}-\mathcal{S})=0.
\ee
This equation can be reproduced from the quadratic correction to the dilaton potential $U(\upphi) \approx 2\upphi - 2\pi \upphi^2/S_0$. If we work strictly with JT gravity with the $\upphi$ correction, we get the same equation without the $\mathcal{S}$ term.

Next, we solve these equations perturbatively about the Nariai background. After quotienting by residual diffeomorphisms, the physical solutions reduce to a one-parameter family, which precisely reproduces the near-Nariai sector of SdS geometries.

\subsubsection*{Perturbations around Nariai and JT gravity}
Solving equations  \eqref{eq:nearNariai_tt}--\eqref{eq:nearNariai_tr} we get:
\be \label{eq:S0}
\mathcal{S}(\t,\r)=C_{-1} e^{-\t}\sech \r + C_0 \tanh \r + C_{1} e^{\t}\sech \r,
\ee
parametrized by three real constants $C_{-1},C_0,C_1$. This can be simplified using the isometries of dS$_2$. Working in the Nariai static patch and measuring lengths in units of the dS$_2$ radius,
$
\d s^2=\sech^2\r(-\d\t^2+\d\r^2),
$
a convenient basis of $\mathrm{SO}(1,2)$ Killing vectors is 
\be 
\ell_0=\partial_\t,
\qquad
\ell_+=e^{\t}\big(\sinh \r\,\partial_\t+\cosh \r\,\partial_\r\big),
\qquad
\ell_-=e^{-\t}\big(\sinh \r\,\partial_\t-\cosh \r\,\partial_\r\big).
\ee
These are isometries of dS$_2$, and they would also be isometries of the full metric in the exact Nariai limit $\epsilon=0$. At finite $\epsilon$, however, the size of the transverse sphere depends on $\t$ and $\r$ through $\mathcal{S}(\t,\r)$, and is therefore affected by the coordinate transformations generated by $\ell_n$. Acting on the $C_0$ mode with rigid dS$_2$ isometries generates the $C_{1}$ and $C_{-1}$ modes:
\be
\delta_{\ell_0}(\tanh \r)=0,
\quad
\delta_{\ell_+}(\tanh \r)=e^{\t}\sech \r,
\quad
\delta_{\ell_-}(\tanh \r)=-e^{-\t}\sech \r.
\label{eq:Killing_generates_modes}
\ee
 Thus, modulo coordinate transformations, the most general solution is $\mathcal{S}=C_{0} \tanh \r$. Choosing normalization $C_0=1$ matches the radial dependence of the $S^2$ found in \eqref{eq:Nariaistatic} as the near-Nariai limit of SdS. It also matches the dilaton solution from the 2d dilaton-gravity perspective in the JT regime.\footnote{We can reproduce the same result for dS using embedding coordinates \cite{Maldacena:2016upp}. dS$_2$ with $R=2$ can be represented as the hyperboloid in $\mathbb{R}^{1,2}$ satisfying
$
-X_1^2+X_2^2+X_0^2=1.
$
The isometry group $\mathrm{SO}(1,2)$ acts linearly on $X^I$ with $I=0,1,2$. This is related to our coordinates above via
$
X_+\equiv X_2+X_1=\sech\r\,e^{\t}$, $
X_-\equiv X_2-X_1=\sech\r\,e^{-\t}$, $X_0=\tanh\r.
$
The general three-parameter solution \eqref{eq:S0} is simply 
$
\mathcal{S}(\t,\r)=C_I X^I
=(C_1 e^{\t}+C_{-1} e^{-\t})\sech \r + C_{0} \tanh \r.
$
Because $\mathrm{SO}(1,2)$ acts linearly on $(X_1,X_2,X_0)$, it mixes the coefficients $(C_1,C_{-1},C_{0})$, and this freedom can be used to set $C_1=C_{-1}=0$.}

Given $\mathcal{S}$, the dilaton equation \eqref{eq:classicaldilatoneq} determines $\mathcal{R}$. Substituting \eqref{eq:S0} into \eqref{eq:classicaldilatoneq} gives the inhomogeneous equation
\be
\cosh^2 \r (\partial_\r^2-\partial_\t^2)\mathcal{R}+2\mathcal{R}
=2C_{0}\tanh \r,
\ee
whose general solution can be written as
$
\mathcal{R}(\t,\r)=\mathcal{R}_{p}(\t,\r)+\mathcal{R}_{h}(\t,\r),
$
with particular solution
\bea \label{eq:R0p}
\mathcal{R}_{p}(\t,\r)
&=&
\frac{2}{3}C_{0}\tanh \r \big(\log \cosh \r\big).
\eea
The near-Nariai SdS solution \eqref{eq:PhitPhiNDSD} can be reproduced with the particular solution \eqref{eq:R0p} with normalization $C_0=1$.  The remaining freedom $\mathcal{R}_{h}$ solves the homogeneous equation 
$
\cosh^2 \r (\partial_\r^2-\partial_\t^2)\mathcal{R}_{h}+2\mathcal{R}_{h}=0,
$
which is the P\"oschl-Teller Schr\"odinger problem for each temporal Fourier mode. The most general solution can be found using the observation that if $\tilde{\mathcal{R}}$ solves the 2d wave equation $(\partial_\t^2 -\partial_\r^2)\tilde{\mathcal{R}}=0$, then
$\mathcal{R}_{h}=(\partial_\r - \tanh \r) \tilde{\mathcal{R}}$
solves the homogeneous equation. The most general solution of the 2d wave equation is $\tilde{\mathcal{R}}=F(\t+\r)+ G(\t-\r)$, for arbitrary functions $F$ and $G$, and therefore
\bea \label{eq:R0h_separated}
\mathcal{R}_{h}(\t,\r)&=&F'(\t+\r)-G'(\t-\r) - \tanh\r \Big[F(\t+\r) + G(\t-\r) \Big].
\ea
All these solutions can be removed by a diffeomorphism and are therefore unphysical. Consider the 2d sector of the metric, which can always be put in conformal gauge and written in null coordinates:
\be
\d s^2_{2\mathrm{d}}=
e^{2\rho(\t,\r)}(-\d\t^2+\d\r^2)=
-e^{2\rho(u,v)}\d u \d v ,
\qquad
u=\t-\r,\quad v=\t+\r,
\ee
This does not completely fix the gauge, since conformal transformations preserve this form of the metric. The fact that $\rho(\t,\r) = \rho_0(\r) + \mathcal{O}(\epsilon)$ fixes order one conformal transformations. Order $\epsilon$ conformal transformations remain unfixed:
\be\label{eq:residual_diffeo_uv}
u \ \mapsto\ u'=u+\epsilon g(u),
\qquad
v \ \mapsto\ v'=v+\epsilon f(v).
\ee
They preserve the conformal form of the metric and its order one form. This residual diffeomorphism leaves $\mathcal{S}$ unchanged, since its spacetime dependence is already order $\epsilon$. It does induce an inhomogeneous shift of $\mathcal{R}$
\be \label{eq:R_shift}
\mathcal{R}(u,v)\ \longrightarrow\ \mathcal{R}(u,v)
+\frac{1}{2}\bigg[ f'(v)+g'(u)-\tanh \r\,\big(f(v)-g(u)\big) \bigg].
\ee
This shifts the homogeneous solution by $F(v)\to F(v)+ \frac{1}{2}f(v)$ and $G(u)\to G(u)-\frac{1}{2}g(u)$ and this freedom allows us to set $\mathcal{R}_{h}(\t,\r)=0$. For strictly JT gravity the differential equation for $\mathcal{R}$ \eqref{eq:classicaldilatoneq} does not have the $\mathcal{S}$ term and therefore $\mathcal{R}=\mathcal{R}_h = 0$ up to diffeomorphisms. This is the statement that the geometry is rigid dS$_2$ in JT gravity.

To summarize, the family of linearized solutions of Einstein's equations around the Nariai geometry, modulo diffeomorphisms, is one-dimensional. We can set all modes to zero except $C_0$ both in $\mathcal{R}$ and $\mathcal{S}$. Of course, this one-parameter family is nothing but the SdS solution expanded near the Nariai regime, with the parameter measuring the deviation from the Nariai mass. 

\subsubsection*{Location of the horizons}

The 2d metric is dS$_2$ up to small near-Nariai corrections. The horizon location is not determined by the 2d metric but by the dilaton. In nearly-AdS$_2$, the dilaton selects the boundary clock and thus the horizon location \cite{Mertens:2022irh}. In dS$_2$, it instead determines which portion of the asymptotic future expands to asymptotic dS$_4$ and which contracts into the black-hole interior and singularity. 

Following \cite{Bousso:1997wi}, in 4d spherical symmetry one characterizes apparent horizons through the null expansions of the outgoing and ingoing null congruences orthogonal to the $S^2$. These are proportional to $\partial_u\Phi$ and $\partial_v\Phi$. A marginally trapped surface satisfies $\theta_+ \propto \partial_v\Phi=0$ or $\theta_- \propto \partial_u\Phi=0$, depending on the branch. Equivalently, the gradient of $\Phi$ is null,
\be \label{eq:gradPhi_null_22}
(\nabla\Phi)^2 \equiv g^{ab}\partial_a\Phi\,\partial_b\Phi = 0.
\ee
Motivated by this, we take \eqref{eq:gradPhi_null_22} as the general definition of the apparent horizon in 2d dilaton-gravity. In the near-Nariai expansion, $\Phi$ is a large homogeneous background plus a small inhomogeneous fluctuation, so the derivative acts only on the varying piece. In terms of the JT dilaton, this becomes $(\nabla \upphi)^2=0$, or equivalently, in our notation, $(\nabla \mathcal{S})^2=0$.

For the most general time-dependent dilaton solution \eqref{eq:S0}, since $(\nabla \mathcal{S})^2 \propto \partial_u\mathcal{S}\,\partial_v\mathcal{S}$, the horizon is determined by either $\partial_u\mathcal{S}=0$ or $\partial_v\mathcal{S}=0$. These give
$
C_0 + C_1 e^{u_h} - C_{-1} e^{-u_h} = 0$, and 
$C_0 - C_1 e^{v_h} + C_{-1} e^{-v_h} = 0.
$
Evaluating the dilaton there, one finds
\be
\mathcal{S}|_{\text{horizon}} = \pm \sqrt{C_0^2 + 4C_1C_{-1}} = \pm \sqrt{C_I C^I},
\ee
with the plus (minus) sign corresponding to the cosmological (black-hole) horizon. This is time independent and manifestly SL$(2,\mathbb{R})$ invariant, being the norm of the embedding-space vector $C_I$.
\section{The evaporation of pair-created dS black holes}
\label{sec:Antievaporation}

In this section, we incorporate one-loop quantum effects in the near-Nariai throat, study the fate of pair-created dS black holes from the no-boundary prescription, and rule out the classic ``anti-evaporation" story. Working in the double expansion in $\epsilon$ and $\hbar$, we first show that a horizon-smooth Hadamard state~\cite{Kay:1988mu,Radzikowski:1996pa} reproduces the expected thermal Hawking flux in the static patch. We then connect this Lorentzian analysis to the Euclidean ``no-boundary" preparation by Wick rotating global conformal time and projecting onto a single static diamond, demonstrating that the resulting late-time profile is compatible with the thermal flux that would be predicted by a smooth choice of quantum state. Finally, we derive the semiclassical equations governing the dilaton and identify the would-be growing mode that has been interpreted as anti-evaporation, explaining how its physical status depends crucially on gauge choices and on the admissibility of the quantum state.

\subsection{The one-loop effective theory}
\label{sec:onelooptheory}

We begin by deriving the one-loop effective action $\Gamma_{\text{1-loop}}$ that arises from integrating out the matter fields. As explained in Appendix~\ref{app:greybody} we ignore massive KK modes, see also \cite{Brady:1996za,Kanti:2005ja, Crispino:2013pya, Kanti:2014dxa, deCesare:2025ccs}, but here we retain the coupling of massless fields to the dilaton. The central ingredient is the trace anomaly of the renormalized stress tensor, derived in \cite{Mukhanov:1994ax, Bousso:1997cg, Kummer:1998dc, Kummer:1999zy, Balbinot:2000iy, Fabbri:2003vy, Hofmann:2004kk}. The trace anomaly was computed for a general matter theory with action $S_{\text{matter}} = \frac{1}{2} \int \d^2x\sqrt{-g} e^{-2 \varphi(\phi)} (\nabla f)^2$ and with a path integral measure derived from the ultralocal inner product $\langle \delta f_1| \delta f_2\rangle = \int\d^2x \sqrt{-g} e^{-2\psi(\phi)} \delta f_1 \delta f_2$. They get
\be \label{eq:traceanomaly1}
\langle T^a{}_a \rangle =\frac{N\hbar}{24 \pi}\Big(R-6 (\nabla \varphi)^2+4 \Box \varphi + 2 \Box \psi\Big).
\ee
This result is derived using zeta-function regularization of the matter functional determinant appearing in the one-loop approximation. We can extract $\varphi(\phi)$ and $\psi(\phi)$ from the 4d action and ultralocal measure, and get 
\be \label{eq:traceanomaly}
\varphi(\phi)=\phi,~\psi(\phi)=\frac{\phi}{2},~~~~\Rightarrow~~~~\langle T^a{}_a \rangle =\frac{N\hbar}{24 \pi}\Big(R-6 (\nabla \phi)^2+5 \Box \phi\Big).
\ee
From now on we restore $\hbar$ to keep track of matter quantum effects. The term proportional to $R$ in \eqref{eq:traceanomaly} is the usual trace anomaly of a 2d CFT~\cite{Christensen:1977jc}. The additional ``dilaton anomaly" terms reflect the non-conformal nature of the matter fields. 

The basic equations governing the coupled
$(g_{ab},\Phi)$ system together with the matter fields one-loop action are
\be \label{eq:metricEoM}
\frac{1}{2G_N \Lambda} \Big[  g_{ab} \Box \Phi-\nabla_a \nabla_b \Phi + \frac{1}{2} g_{ab} U(\Phi)\Big]= \langle T_{ab} \rangle,
\ee
\be \label{eq:dilatonEOM}
\frac{1}{4 G_N \Lambda}\Big[R-U'(\Phi)\Big] 
=-\frac{1}{\sqrt{-g}}\frac{\delta \Gamma_{\text{1-loop}}}{\delta \Phi}.
\ee
We define the covariant renormalized stress tensor from the one-loop effective action by
\be \label{eq:def_stresstensor}
-\frac{2}{\sqrt{-g}}\,\frac{\delta \Gamma_{\text{1-loop}}}{\delta g^{ab}} \equiv \langle T_{ab}\rangle.
\ee
The anomaly fixes the \emph{anomaly-induced} part $\Gamma_{\text{anom}}$, but only up to Weyl-invariant terms $\Gamma_{\text{W}}$ and local counterterms $\Gamma_{\text{ct}}$:
\be \label{eq:fulloneloop}
\Gamma_{\text{1-loop}}=\Gamma_{\text{anom}}+\Gamma_{\text{W}}+\Gamma_{\text{ct}}.
\ee
Hence, $\Gamma_{\text{anom}}$ is viewed as a particular solution of the functional differential equation~\eqref{eq:def_stresstensor} reproducing~\eqref{eq:traceanomaly}
\be \label{eq:oneloopanom}
\Gamma_{\text{anom}} =-\frac{N\hbar}{96 \pi}\int \d^2 x \sqrt{-g}\bigg(R\frac{1}{\Box}R -12(\nabla\phi)^2\frac{1}{\Box}R+10\phi R
\bigg).
\ee
The action that reproduces \eqref{eq:traceanomaly1} for arbitrary $\varphi$ and $\psi$ can be found in~\cite{Kummer:1999zy}. The first term is precisely the Polyakov action associated with the trace anomaly of the 2d CFT~\cite{Polyakov:1981rd}. The additional dilaton-dependent pieces are required to reproduce the dilaton anomaly in~\eqref{eq:traceanomaly}, and thus encode the higher-dimensional origin of the 2d matter theory. We present the most general prescription for dealing with $\Gamma_{\text{anom}}$ below.

In the near-Nariai regime, derivatives of the dilaton are of order $\epsilon$, so $(\nabla\varphi)^2\sim O(\epsilon^2)$ is parametrically subleading. Moreover, the local $\phi R$ term can be absorbed into a small renormalization of the dS$_2$ length in the leading JT description. Thus, at the order relevant for the Hawking flux in the controlled near-Nariai expansion, the Polyakov term captures the dominant contribution. We nevertheless keep the full anomaly-induced action since this is the natural one-loop effective action of the 4d $s$-wave sector. It also makes the connection with the original Bousso-Hawking treatment manifest~\cite{Bousso:1997wi} and gives a framework that remains meaningful beyond the strict near-Nariai/CFT approximation, for example in the finite-$\epsilon$ SdS family or in the crossover regime $\epsilon\sim\hbar$, where the hierarchy underlying the leading Polyakov description can break down.\footnote{This distinction is important beyond the near-Nariai throat. For an ordinary Schwarzschild black hole, where one has $\Phi=r^2$ varying over the entire exterior region, and the resulting 2d theory is \emph{not} well approximated by a minimally coupled CFT on the 2d metric. See \cite{Wu:2023uyb} for a similar ``anti-evaporation'' problem in the Schwarzschild background, whose resolution depends crucially on the correct one-loop effective theory and is therefore fundamentally different from the near-Nariai case. See also~\cite{Tomasevic:2025kqy, Tomasevic:2025clf} for applications to the quantum cloaking of naked singularities in critical gravitational collapse.}

The term $\Gamma_{\text{W}}$ denotes Weyl-invariant non-local functionals characterizing the most general solution of~\eqref{eq:def_stresstensor}, which is not fixed by the trace anomaly and in general can carry state-dependent information.\footnote{It remains an open question to determine $\Gamma_{\text{W}}$ in closed form, since this requires evaluating the finite part of the one-loop determinant, which is distinct from extracting the UV trace anomaly. See efforts in~\cite{Hofmann:2004kk, Hofmann:2005yv} based on covariant perturbation theory. }  Our working assumption is that, in the near-Nariai throat where conformal symmetry is approximately restored~\cite{Fabbri:2003vy}, one may fix the reference vacuum by matching to the corresponding locally inertial/conformal vacuum and choose the subtraction scheme so that $\Gamma_{\rm W}$ gives no contribution to $\langle T_{ab}\rangle$ in that state, i.e., its vacuum contribution is absorbed into the choice of state, in close analogy with normal ordering.

A structural difference from an ordinary 2d CFT remains important, however: in the presence of dilaton coupling, diffeomorphism invariance constrains the combined $(g_{ab},\phi)$ system, so the stress tensor satisfies the modified conservation law
\be \label{eq:conservation}
\nabla^a \langle T_{ab} \rangle-\frac{1}{\sqrt{-g}}\frac{\delta \Gamma_{\text{1-loop}}}{\delta \phi} \nabla_b \phi=0,
\ee
which may be derived either by reducing the four-dimensional relation $\nabla^\mu\langle T^{(4)}_{\mu\nu}\rangle=0$, or directly from diffeomorphism invariance of the covariant functional $\Gamma_{\text{1-loop}}[g,\phi]$.

This is the same one-loop effective theory used by Bousso and Hawking~\cite{Bousso:1997wi}, up to a choice of the local $\phi R$ term in the effective action (which they parametrize by a constant $w$). In two dimensions, such a term shifts the trace by a total-derivative contribution proportional to $\Box\phi$.\footnote{At the order relevant here, varying $w$ only shifts local $O(\hbar)$ background data and does not affect the gauge-invariant near-Nariai observables extracted at $O(\epsilon\hbar)$. That is, this difference is not what drives the pathological anti-evaporation channel that we will review in Section~\ref{sec:antievaporation}.} The point clarified in later work is that the coefficient of this $\Box\phi$ term is not a genuine ambiguity of the anomaly itself: once the dilaton coupling, path-integral measure, and renormalization prescription are fixed, it is fixed as well~\cite{Kummer:1999zy}. What remains is only the standard freedom to add finite local counterterms $\Gamma_{\text{ct}}$ to the effective action. Under mild assumptions, these are purely local and geometric, hence state-independent~\cite{Wald:1977up, Wald:1978pj, Wald:1978ce}.

To evaluate the effective action (and its associated stress tensor) it is convenient to introduce two auxiliary fields $\chi_1$ and $\chi_2$, whose role is to make the action local. More precisely,
\be \label{eq:Gamma_ext}
\Gamma_{\rm anom}[g,\phi]
=
\operatorname*{ext}_{\chi_1,\chi_2}
\Big(
\Gamma_{\chi_1}[g,\phi,\chi_1]
+\Gamma_{\chi_2}[g,\phi,\chi_2]
+\Gamma_{\phi}[g,\phi]
\Big),
\ee
where  each of the three terms in the RHS is local
\bea
\Gamma_{\chi_1}&=& N\hbar \int \d^2 x \sqrt{-g}\,\bigg[
\frac{1}{2} (\nabla \chi_1)^2+ \chi_1 \big(\lambda_1 R + \lambda_2 (\nabla \phi)^2\big)
\bigg],
\\
\Gamma_{\chi_2}&=& N\hbar \int \d^2 x \sqrt{-g}\,\bigg[
-\frac{1}{2} (\nabla \chi_2)^2+ \chi_2 \big(\mu_1 R + \mu_2 (\nabla \phi)^2\big)
\bigg],
\\
\Gamma_{\phi} & =&-\frac{5N\hbar}{48\pi}\int \d^2x \sqrt{-g}\,\phi R.
\eea
$\lambda_{1,2}$ and $\mu_{1,2}$ are parameters to be fixed by matching to \eqref{eq:oneloopanom}. The extremization equations are
\be \label{eq:aux}
\Box \chi_1= \lambda_1 R+ \lambda_2 (\nabla \phi)^2, \qquad \Box \chi_2=- \mu_1 R-\mu_2 (\nabla \phi)^2.
\ee
Reproducing $\Gamma_{\rm anom}$ fixes the coefficients through
\be \label{eq:constraint1}
\lambda_1^2 -\mu_1^2=-\frac{1}{48 \pi}, \qquad
\lambda_1 \lambda_2-\mu_1 \mu_2=\frac{1}{8 \pi}, \qquad
\lambda_2^2 -\mu_2^2=0.
\ee
Any set of parameters satisfying these constraints will produce the correct effective action. After extremizing over $\chi_{1,2}$ and imposing~\eqref{eq:constraint1}, the resulting $\langle T_{ab} \rangle$ should contain no extra parameters. This extremization is taken with fixed boundary conditions, which encode the choice of quantum state relative to a chosen normal-ordering prescription. Equivalently, in solving \eqref{eq:aux}, the homogeneous parts of $\chi_1$ and $\chi_2$ specify the state-dependent sector of the stress tensor once the normal-ordering has been fixed.

We can write the renormalized stress tensor appearing on the RHS of
\eqref{eq:metricEoM} as
\be
\langle T_{ab} \rangle =
\langle T^{(\chi_1)}_{ab} \rangle
+\langle T^{(\chi_2)}_{ab} \rangle
+\langle T^{(\phi)}_{ab} \rangle,
\ee
where
\bea
\langle T^{(\chi_1)}_{ab} \rangle
&=&
N\hbar \bigg[
-\nabla_a \chi_1 \nabla_b \chi_1
+\frac{1}{2} g_{ab} (\nabla \chi_1)^2
+2\lambda_1\big( \nabla_a \nabla_b \chi_1- g_{ab} \Box \chi_1\big)
\no\\
&&\hspace{2.3em}
-2 \lambda_2 \chi_1 \bigg( \nabla_a \phi \nabla_b \phi-\frac{1}{2} g_{ab} (\nabla \phi)^2 \bigg)
\bigg],
\\
\langle T^{(\chi_2)}_{ab} \rangle
&=&
N\hbar \bigg[
\nabla_a \chi_2 \nabla_b \chi_2
-\frac{1}{2} g_{ab} (\nabla \chi_2)^2
+2\mu_1\big( \nabla_a \nabla_b \chi_2- g_{ab} \Box \chi_2\big)
\no\\
&&\hspace{2.3em}
-2 \mu_2 \chi_2 \bigg( \nabla_a \phi \nabla_b \phi-\frac{1}{2} g_{ab} (\nabla \phi)^2 \bigg)
\bigg],
\\
\langle T^{(\phi)}_{ab} \rangle
&=&
-\frac{5N\hbar}{24 \pi}\big(\nabla_a \nabla_b \phi- g_{ab} \Box \phi\big).
\eea
Similarly, the one-loop contribution to the dilaton equation \eqref{eq:dilatonEOM} is
\be
\frac{\delta \Gamma_{\text{1-loop}}}{\delta \phi}
= N\hbar \bigg[
-2 \lambda_2\big(\chi_1 \Box \phi+ \nabla_a \chi_1 \nabla^a \phi\big)
-2 \mu_2\big(\chi_2 \Box \phi+ \nabla_a \chi_2 \nabla^a \phi\big)
-\frac{5R}{48 \pi}
\bigg].
\ee
One can verify directly that these expressions reproduce the trace anomaly~\eqref{eq:traceanomaly} and satisfy the modified conservation law~\eqref{eq:conservation}. We emphasize that the auxiliary field formulation keeps the state data manifest while cleanly separating the universal anomaly-induced UV structure from the potentially subtle IR/state-dependent sector. Correspondingly, it is useful to regard the renormalized stress tensor on a curved spacetime as admitting the general decomposition
$
\langle T_{ab}\rangle
=\langle :T_{ab}:\rangle_{\text{state}}+\langle T_{ab}\rangle_{\text{geo}},
$
where $\langle T_{ab}\rangle_{\text{geo}}$ denotes the local, geometric, state-independent contribution (up to finite covariant counterterms) fixed by the short-distance structure encoded in the trace anomaly, whereas the normal-ordered $\langle :T_{ab}:\rangle_{\text{state}}$ contains the residual dependence on the quantum state.\footnote{At the formal level, one may make the distinction between the geometric and state-dependent sectors more explicit by rewriting the anomaly-induced non-local pieces involving $\Box^{-1}$ in terms of the Weyl-invariant metric density $g^{ab}\sqrt{-g}$, thereby splitting it into a Weyl-invariant piece and a local piece. That is, $S_{\rm non-loc}=S_{\rm W}+S_{\rm loc}$. This was established in~\cite{Karakhanian:1994gs, Jackiw:1995qh, Navarro-Salas:1995lmi}, and more discussion can be found in~\cite{Fabbri:2005mw, Wu:2023uyb, Tomasevic:2025clf}. In such a decomposition, the local part reproduces the anomaly-controlled geometric contribution to $\langle T_{ab}\rangle$, while the Weyl-invariant part carries the residual state-dependent information. In the present work, however, we will not need to implement this reorganization explicitly: once the stress tensor is evaluated on the background spacetime, the corresponding split becomes manifest.} 

Note that the quantity that sources the semiclassical equations is always the full covariant $\langle T_{ab} \rangle$. But it is important to separate two logically distinct choices for $\langle :T_{ab}:\rangle_{\text{state}}$. The first is the \emph{physical state}, which in our problem must be chosen so that the covariant stress tensor is smooth across the relevant horizons in Kruskal patches (a Hadamard state). The second is the \emph{normal-ordering prescription} adapted to the choice of vacuum of an observer, which in the present Nariai problem is the one naturally associated with the static-patch coordinates $(u,v)$ used by the observer who measures the flux. In the auxiliary-field formalism, the distinction is encoded by the homogeneous solutions of $\chi_i$. Because $\chi_i$ is a scalar, a change of conformal frame shifts the homogeneous sector by the corresponding chiral Jacobians, so changing the normal-ordering prescription is equivalent to shifting the homogeneous auxiliary-field solution. The correct procedure is therefore to keep the static-patch prescription adapted to the observer, while determining the state by
imposing smoothness across the Kruskal horizons.

\subsection{The thermal Hawking radiation of the near-Nariai black hole}
\label{sec:thermalHawking}

In this subsection, we work throughout in the static patch of the (quantum) Nariai/near-Nariai throat and show how the physical thermal Hawking flux is reproduced within the one-loop effective action formalism. In the near-Nariai regime, it is useful to organize the renormalized stress tensor in a simultaneous expansion in $\epsilon$ and $\hbar$:
\be \label{eq:expansion}
\langle T_{\mu\nu}\rangle =\underbrace{O(\hbar)}_{\text{static backreaction on Nariai}} +\,\,\epsilon\,\underbrace{O(\hbar)}_{\text{quantum Hawking flux}} +\cdots ,
\ee
The first piece corresponds to the static backreaction on exact Nariai background. For the Nariai geometry there is
a distinguished Hadamard state whose stress tensor produces a static, symmetric
backreaction on both horizons without driving time dependence, so the background
remains of ``Nariai type." Turning on $\epsilon$ moves us away from this
equilibrium geometry, where genuinely interesting dynamics such as quantum Hawking
flux appears once one-loop effects are included.\footnote{There is a crossover regime where the classical deviation from Nariai becomes comparable to quantum effects, $\epsilon\sim\hbar$. Higher-order terms such as $\epsilon^2$, $\hbar^2$, and dilaton-coupling contributions that break the simple conformal approximation may all enter at the same order as the Hawking flux sitting at $O(\epsilon \hbar)$. The evaporation history may then receive corrections not captured by the leading one-loop near-Nariai treatment. We do not expect these corrections to be parametrically enhanced in the static-patch problem; the point is instead that the clean expansion hierarchy is lost. We leave this crossover regime for future work.}

\subsection*{The quantum Nariai spacetime}

We begin with the exactly degenerate limit $\epsilon=0$, where the classical throat is the Nariai geometry. It turns out that at $O(\hbar)$, the semiclassical equations admit a \emph{static} backreacted solution in which the geometry remains of Nariai form but with quantum-shifted constant radii (equivalently, quantum-shifted 2d curvature and $S^2$ size).

Plugging the metric ansatz~\eqref{eq:metricansatz} into the semiclassical equations and solving the $O(\epsilon^0)$ system, we take
\be
\Lambda \d s^2_{(2)} = e^{2\rho_0(\r)}(-\d\t^2+\d\r^2),\qquad e^{2\rho_0(\r)}\propto \sech^2 \r,\qquad
e^{-2 \phi_0}=\text{const},
\ee
so that the two horizons sit at $\r\to\pm\infty$. At this order, $\rho_0$ and $\phi_0$ include
the $O(\hbar)$ backreaction, as discussed around \eqref{eq:expansion}. In particular, $\phi_0$ remains a constant, while $\rho_0(\r)$ is fixed by
a constant 2d curvature,
\be
R_0=-2 e^{-2 \rho_0(\r)} \,\partial_\r^2 \rho_0(\r),
\ee
with $R_0$ and $\phi_0$ shifted by $O(\hbar)$ effects. Evaluating the auxiliary field equations~\eqref{eq:aux} on the conformal gauge ansatz
gives
\bea
(\partial_{\sf t}^2-\partial_\r^2)\chi_1({\sf t},\r)
+2\lambda_1(\partial_{\sf t}^2-\partial_\r^2)\rho({\sf t},\r)
-\lambda_2\!\left[(\partial_{\sf t}\phi)^2-(\partial_\r\phi)^2\right]&=&0,\\
(\partial_{\sf t}^2-\partial_\r^2)\chi_2({\sf t},\r)
-2\mu_1(\partial_{\sf t}^2-\partial_\r^2)\rho({\sf t},\r)
+\mu_2\!\left[(\partial_{\sf t}\phi)^2-(\partial_\r\phi)^2\right]&=&0.
\eea
At $O(\epsilon^0)$ we have $\phi=\phi_0=\text{const}$, so the $\phi$-dependent terms vanish (and more generally they start only at $O(\epsilon^2)$ in the near-Nariai expansion). We may therefore write
\be
\chi_1(\t,\r)=-2\lambda_1\rho({\sf t},\r)+\eta({\sf t},\r),\qquad
\chi_2(\t,\r)= 2\mu_1\rho({\sf t},\r)+\eta({\sf t},\r),
\ee
where $\eta$ obeys the homogeneous wave equation
\be
(\partial_{\sf t}^2-\partial_\r^2)\eta({\sf t},\r)=0.
\ee
The homogeneous field $\eta$ is precisely the state-dependent datum: at the present order we denote it
by $\eta_0({\sf t},\r)$ (i.e., $\eta=\eta_0+\epsilon\,\eta_1+\cdots)$ in the parameterization used above, and we will see that the $O(\epsilon^0)$ constraints fix $\eta_0$ uniquely.

A controlled semiclassical near-Nariai regime requires two independent small parameters: the geometric deformation $\epsilon \ll 1$ and the ratio of matter degrees of freedom to horizon entropy, $N/S_0\propto O(\hbar) \ll 1$. A convenient formal limit is therefore
\be
N\to\infty,\qquad S_0\to\infty,\qquad \frac{N}{S_0}\ \text{fixed and small},\qquad
 S_0 \equiv \frac{\pi}{G_N \hbar \Lambda}.
\ee
which is analogous to a matter-dominated double-scaling limit. In this regime, the one-loop matter backreaction remains finite but perturbative, while graviton loops are suppressed. The large-$N$ description should be viewed mainly as a device that makes the semiclassical matter backreaction effectively classical, whereas the actual perturbative control is governed by the smallness of $N/S_0$. We present a realistic model justifying this assumption based on a single massless charged fermion in a 4d charged dS black hole background in Appendix~\ref{app:fermions}.

The $O(\epsilon^0)$ equations reduce to
\bea \label{eq:Nariaitt}
2 e^{2 \rho_0}\sinh{\phi_0   }
&=&\frac{N }{12 S_0}\,\Big(2\partial_\r^2\rho_0-(\partial_\r\rho_0)^2+6\partial_\r^2\eta_0\Big),\\ \label{eq:Nariairr}
2 e^{2 \rho_0}\sinh{\phi_0}
&=&\frac{N }{12 S_0}\,\Big((\partial_\r\rho_0)^2-6\partial_{\sf t}^2\eta_0\Big),\\
\label{eq:Nariaitr}
\partial_{\sf t}\partial_\r\eta_0&=&0,\\
2 \cosh{\phi_0}&=&-2 e^{-2 \rho_0-2 \phi_0}\partial_\r^2 \rho_0-\frac{5 N  }{12 S_0}e^{-2 \rho_0}\partial_\r^2 \rho_0
\eea
Combining~\eqref{eq:Nariaitt} and~\eqref{eq:Nariairr} eliminates $\eta_0$ (using $(\partial_{\sf t}^2-\partial_\r^2)\eta_0=0$)
and yields an equation that implies the 2d Ricci scalar $R=-2e^{-2\rho_0}\partial_\r^2\rho_0$ is a constant we denote as $R_0$. The equation becomes an algebraic relation fixing $R_0$ in terms of $\phi_0$. Independently, the dilaton equation gives a second algebraic relation between $R_0$ and $\phi_0$. The two equations are
\be \label{eq:R0sigma0_relation}
R_0
=-\frac{4 8 S_0 \sinh{\phi_0}}{N}= \frac{48  \cosh{\phi_0}}{24 e^{-2 \phi_0}+5 \frac{N}{S_0}}.
\ee
Solving these two relations, we select the solution for $\phi_0$ that reduces to the
classical Nariai value $e^{-2\phi_0}=1$ as $\hbar\to 0$. Expanding that physical solution
consistently to $O(\hbar)$ gives the unique static backreaction
\be \label{eq:staticbackreaction}
R_0=2-\frac{7 N }{12 S_0},\qquad
\phi_0=-\frac{N }{24 S_0},
\ee
in precise agreement with the equilibrium \emph{quantum Nariai} solution found by Bousso-Hawking (the analog of Eq.~(4.1) in~\cite{Bousso:1997wi}, but here written in static patch variables). This result can be reproduced in JT language by doing a field redefinition of the dilaton that removes the $\int \phi R$ coupling originating from $\Gamma_{\text{1-loop}}$, leading to a change in both the background dilaton value and the dS$_2$ radius.

We now discuss the physical role of the homogeneous solution $\eta_0$ in the Nariai spacetime. Although $R_0$ and $\phi_0$ are determined without specifying $\eta_0$, the stress tensor itself does depend on $\eta_0$. As emphasized near the end of Section~\ref{sec:onelooptheory}, $\eta_0$ plays two logically distinct roles: it encodes the physical state, but it also depends on the observer-adapted normal-ordering prescription used to display that same state.

The physical requirement is that $\langle T_{ab}\rangle$ be smooth across each horizon, namely that its covariant components remain finite in horizon-regular Kruskal coordinates. Since the static patch has two distinct horizons, this condition must be imposed patchwise. The static null coordinates $u=\t-\r, v=\t+\r$ are well adapted to the Killing flow and to the flux measured by the observer, but they are not themselves locally inertial at the horizons. 

Near the black-hole horizon $\mathcal{H}_b^\pm$ we use
\be \label{eq:KruskalBH}
\tilde U_b=-e^{-\kappa_b u},\qquad \tilde V_b=e^{\kappa_b v},
\ee
so that $\mathcal H_b^+$ is $\tilde U_b=0$ with $\tilde V_b>0$, and $\mathcal H_b^-$ is $\tilde V_b=0$
with $\tilde U_b<0$. Near the cosmological horizon $\mathcal{H}_c^\pm$ we use
\be \label{eq:KruskalCH}
\tilde U_c=e^{\kappa_c u},\qquad \tilde V_c=-e^{-\kappa_c v},
\ee
so that $\mathcal H_c^+$ is $\tilde V_c=0$ with $\tilde U_c>0$, and $\mathcal H_c^-$ is $\tilde U_c=0$
with $\tilde V_c<0$. The constants $\kappa_b$ and $\kappa_c$ are the dimensionless surface gravities of the black-hole and cosmological horizons, respectively, up to overall normalization; such a normalization only rescales the Kruskal coordinates and plays no role in the regularity analysis. 

Regularity must therefore be tested in $(\tilde U,\tilde V)$ rather than in $(u,v)$. For example, near the future black-hole horizon $\mathcal{H}_b^+$
\be
\langle T_{\tilde U_b\tilde U_b}\rangle
=
\left(\frac{du}{d\tilde U_b}\right)^2\langle T_{uu}\rangle
=
\frac{1}{\kappa_b^2\tilde U_b^{\,2}}\langle T_{uu}\rangle,
\ee
and similarly near $\mathcal H_c^+$ one has
$\langle T_{\tilde V_c\tilde V_c}\rangle
=
\kappa_c^{-2}\tilde V_c^{-2}\langle T_{vv}\rangle$.
Thus a nonzero constant piece in $\langle T_{uu}\rangle$ or $\langle T_{vv}\rangle$ would still lead to a divergence in the Kruskal frame. This is why the physical state must be fixed by horizon smoothness in Kruskal coordinates, even though the flux is naturally described in the static-patch coordinates $(u,v)$.

Now consider the exact Nariai case. We see that~\eqref{eq:Nariaitr} imposes $\partial_{\sf t}\partial_\r\eta_0=0$, together with the wave equation
$(\partial_{\sf t}^2-\partial_\r^2)\eta_0=0$, the general solution is
\be \label{eq:Nariaieta0}
\eta_0({\sf t},\r)=A({\sf t}^2+\r^2)+B\t+C\r+D. 
\ee
Only the coefficient $A$ is relevant, since the stress tensor depends on $\partial_{\sf t}^2\eta_0$ and $\partial_\r^2\eta_0$. Substituting the
general solution back into the~\eqref{eq:Nariaitt} and~\eqref{eq:Nariairr} fixes $A=\frac{1}{12}$ uniquely. With this choice, the renormalized 2d stress
tensor at $O(\epsilon^0)$ is static and depends only on $\r$,
\be \label{eq:Nariaistress}
\langle T_{\t \t}\rangle=-\frac{N\hbar}{24\pi}\sech^2 \r,\qquad
\langle T_{\r \r}\rangle=\frac{N\hbar}{24\pi}\sech^2 \r,\qquad
\langle T_{\t \r}\rangle=0.
\ee
Since $\langle T_{\t \r} \rangle=0$, there is no net  quantum flux, and the full stress tensor vanishes at the two horizons while remaining regular in the corresponding Kruskal frames. This state is viewed as a Hartle-Hawking equilibrium state~\cite{Hartle:1976tp} on the
static patch: it incorporates only vacuum polarization on the quantum Nariai background and
provides an unambiguous reference point for the subsequent near-Nariai analysis.\footnote{Indeed, in the \emph{exact} Nariai geometry this Hartle-Hawking state admits a unique Euclidean preparation: the Euclidean continuation is smooth and compact with topology $S^2\times S^2$, with a single periodicity fixed by regularity at the horizons. The resulting path integral therefore selects a distinguished regular, time-translation-invariant state on the Lorentzian static patch.}

In terms of the $(u,v)$ coordinates, \eqref{eq:Nariaistress} gives $\langle T_{uu}\rangle=\langle T_{vv}\rangle=0$. This simply means that so far we have been using the horizon-smooth Kruskal representative. To display the same physical state in the normal-ordering prescription adapted to the static-patch observer, we shift the homogeneous solution,
\be \label{eq:Nariaishift}
\eta^{\text{(static)}}_0 = \eta^{\text{(Kruskal)}}_0+\delta \eta_0.
\ee
The physical state is defined by positive frequency in smooth Kruskal coordinates, then the natural representative in that frame has vanishing homogeneous part $\eta^{\text{(Kruskal)}}_0=0$. Since the auxiliary fields are scalars, rewriting the horizon-smooth Kruskal representative in the static-patch coordinates induces a homogeneous piece with\footnote{This shift of representative is precisely the Schwarzian term that relates the two normal-ordering prescriptions for the null stress tensor,
\be
:T_{uu}:_{(u)}=\left(\frac{dU}{du}\right)^2:T_{UU}:_{(U)}-\frac{N\hbar}{24\pi}\{U,u\},
\qquad
\{U,u\}\equiv \frac{U'''}{U'}-\frac32\left(\frac{U''}{U'}\right)^2,
\ee
and similarly for $v$ and $V$. Equivalently, one may derive the same shift entirely in the auxiliary-field formalism by requiring that the localized auxiliary field transform as a scalar under $U=U(u)$, $V=V(v)$, which induces the homogeneous piece $\delta\eta_0$ with $\partial_u^2\delta\eta_0=-\frac16\{U,u\}$ and $\partial_v^2\delta\eta_0=-\frac16\{V,v\}$.}
\be
\partial_u^2\delta\eta_0=\partial_v^2\delta\eta_0=\frac{1}{12},
\ee
for exact Nariai, this is precisely
\be
\delta \eta_0 (u,v)=\frac{1}{24}(u^2+v^2)+\cdots,
\ee
the null components of the one-loop stress tensor pick up the state-dependent contributions
\be
\delta\langle T_{uu}\rangle
=
\frac{N\hbar}{4\pi}\,\partial_u^2\delta\eta_0,
\qquad
\delta\langle T_{vv}\rangle
=
\frac{N\hbar}{4\pi}\,\partial_v^2\delta\eta_0.
\ee
Restoring physical units,
\be \label{eq:Nariaiflux}
\Lambda \langle T_{uu} \rangle
=
\Lambda \langle T_{vv} \rangle
=
\frac{\pi N\hbar}{12}T_N^2,
\qquad
T_N=\frac{\sqrt{\Lambda}}{2\pi}.
\ee
Thus exact Nariai exhibits the expected equal left- and right-moving thermal bath at the Nariai temperature. Since the two null components are equal, there is still no net transport,
\be
\langle T_{\t\r}\rangle
=
\langle T_{vv}\rangle-\langle T_{uu}\rangle
=0,
\ee
as expected. This is the precise sense in which exact Nariai is the dS analog of a Hartle-Hawking equilibrium state.

\subsection*{The thermal Hawking flux of the near-Nariai black hole}

We now switch on the near-Nariai deformation with the small non-degeneracy parameter $\epsilon \ll 1$ to study the physical thermal Hawking flux in a clean Lorentzian framework. Since the classical near-Nariai mode structure at $O(\epsilon)$ has already been analyzed in Section~\ref{sec:classicalnearNariai}, here
we focus on the \emph{one-loop backreaction} at $O(\epsilon\hbar)$ and on the associated constraints from
horizon-smoothness of the quantum state.

We therefore keep the leading quadratic dilaton correction beyond the strict JT limit discussed in Section~\ref{sec:classicalnearNariai}, since it is what makes the 2d geometry sensitive to the order-$\epsilon$ splitting between the black-hole and cosmological horizons. Let us begin with the order $\epsilon$ classical solution with $\hbar=0$. The areal-radius perturbation contains the three dS$_2$ linear modes \eqref{eq:S0} and the associated solution $\mathcal{R}(\t,\r)$ of the perturbation on the conformal factor given in \eqref{eq:R0p} plus contributions that can be accounted for by a conformal transformation in $(\t,\r)$. The $(C_1,C_{-1},C_{0})$ is parameterizing the SL$(2,\mathbb{R})$ orbit of the background at the linearized level, as discussed in Section~\ref{sec:classicalnearNariai}. The goal of this section is to turn on $\hbar \neq 0 $ and to further expand 
\beq
\mathcal{S}(\t,\r) = \mathcal{S}_0(\t,\r) +  \mathcal{S}_1(\t,\r) ,~~~~\mathcal{R}(\t,\r) = \mathcal{R}_0(\t,\r) +  \mathcal{R}_1(\t,\r),
\eeq
where $(\mathcal{S}_1, \mathcal{R}_1)$ correspond to the perturbative one-loop correction organized by $N/S_0 \propto \hbar$. We first need to make a choice of the precise form of the classical solution we want to use. In this regard, we will explain in Section~\ref{sec:antievaporation} how the $(C_1, C_{-1})$ directions have been incorrectly interpreted as the anti-evaporation channel once one-loop effects are included. For the moment,
however, as explained in Section~\ref{sec:classicalnearNariai}, we can use a convenient change of coordinates to set $C_1=C_{-1}=0$ and $C_0=1$ such that 
\be
\mathcal{S}_0(\t,\r) = \tanh \r,~~~\mathcal{R}_0(\t,\r) = \frac{2}{3} \tanh \r \,\log \cosh \r.
\ee
This choice matches the canonical near-Nariai geometry discussed in Section~\ref{sec:classicalnearNariai}. In this representative, the Killing horizons at the classical level sit at the null boundaries $\r\to\pm\infty$, as appropriate for the static patch. We keep $\mathcal{R}_0(\t,\r)$ coming from the quadratic correction to the dilaton potential throughout, as it encodes the leading geometric effect responsible for the horizon-temperature splitting.

At $O(\epsilon\hbar)$ the semiclassical equations for the one-loop correction $\mathcal{S}_1(\t,\r)$ can be arranged
into three linear equations coming from the metric EoMs whose homogeneous kernel is the same
SL$(2,\mathbb{R})$ zero-mode sector as at tree level, but with sources on the RHS:
\bea \label{eq:near_Nariai_tt}
\big[\partial^2_\r+\tanh \r\,\partial_\r+\sech^2 \r\big]\mathcal{S}_1&=&-\frac{N}{S_0} \,\bigg(\frac{\sech^2{\r} \tanh{\r}}{6 }+\frac{\tanh^3{\r}}{18 }+\frac{ \partial^2_{\r}\eta_1}{4 } \bigg),\\ \label{eq:near_Nariai_rr}
\big[\partial^2_\t+\tanh \r\,\partial_\r-\sech^2 \r\big]\mathcal{S}_1&=&-\frac{N}{S_0} \,\bigg(\frac{\tanh^3{\r}}{18 }+\frac{ \partial^2_{\t}\eta_1}{4 } \bigg),\\ \label{eq:near_Nariai_tr}
\big[\partial_\t \partial_\r +\tanh \r\, \partial_\t \big]\mathcal{S}_1&=&-\frac{N}{S_0} \, \frac{ \partial_{\t} \partial_{\r}\eta_1}{4 },
\eea
and similarly the dilaton EoM: 
\be \label{eq:oneloopdilatoneq}
(\partial_\r^2-\partial_\t^2)\mathcal{R}_1+2 \sech^2{\r} (\mathcal{R}_1-\mathcal{S}_1)=-\frac{N}{S_0} \, \frac{\sech^2{\r} \tanh{\r}}{24 }\Big( 24\ln{\sech{\r}}+23 \Big),
\ee
The corresponding 2d renormalized stress tensor is given by
\be \label{eq:Ttt}
\langle T_{\t \t} \rangle= -\frac{N \hbar}{24 \pi}\sech^2{\r}+\epsilon N \hbar \bigg[ \frac{\sech^2{\r} \tanh{\r}}{72 \pi }(2\cosh{2\r} -4 \ln{\cosh{\r}}-5) +\frac{\partial^2_\r \eta_1}{4 \pi }\bigg],
\ee
\be \label{eq:Trr}
\langle T_{\r \r} \rangle=\frac{N\hbar}{24 \pi}\sech^2{\r}+\epsilon N\hbar \bigg[ \frac{\sech^2{\r}\tanh{r}}{72 \pi } (2\cosh{2\r}+4 \ln{\cosh{\r}}+13) +\frac{\partial^2_\t \eta_1}{4 \pi} \bigg],
\ee
\be \label{eq:Ttr}
\langle T_{\t \r} \rangle= \epsilon N\hbar \frac{\partial_\t \partial_\r \eta_1}{4 \pi}.
\ee
where we have kept the $O(\hbar)$ quantum Nariai piece. We see that the state data $\eta_1(\t,\r)$ enters explicitly, and it again must satisfy the wave equation
\be
(\partial_\t^2-\partial_\r^2)\eta_1=0.
\ee
The general near-Nariai one-loop solution is naturally decomposed into particular and homogeneous pieces,
\be
\mathcal{S}_1=\mathcal{S}_{1p}+\mathcal{S}_{1h},\qquad
\mathcal{R}_1=\mathcal{R}_{1p}+\mathcal{R}_{1h}.
\ee
where the particular pair $(\mathcal{S}_{1p}, \mathcal{R}_{1p})$ solves the inhomogeneous equations for a chosen state deformation $\eta_1$, while the homogeneous pair $(\mathcal{S}_{1h}, \mathcal{R}_{1h})$ lies in the kernel of the same linearized operator encountered at tree level. In particular,
\be
\mathcal{S}_{1h}(\t,\r)=(c_1 e^\t+c_{-1} e^{-\t})\sech \r+c_{0}\tanh \r.
\ee
The analogous homogeneous sector in $\mathcal{R}_{1h}$ is likewise pure gauge, belonging to the residual diffeomorphism tower already discussed at tree level. Hence these homogeneous one-loop pieces do not correspond to new physical excitations; they can be absorbed into a redefinition of the tree-level constants, schematically $C_i\to C_i+\hbar c_i$, together with the corresponding homogeneous data in $\mathcal{R}_0$. We therefore fix this redundancy by setting $\mathcal{S}_{1h}=\mathcal{R}_{1h}=0$
so that the one-loop correction is represented entirely by the particular solution.

Since the horizon-tracing problem and the $\langle T_{ab} \rangle$ do not depend on
$\mathcal{R}_1$, it is natural to first study the three PDEs~\eqref{eq:near_Nariai_tt}--\eqref{eq:near_Nariai_tr} for
$(\mathcal{S}_1,\eta_1)$. The difference from the quantum Nariai case is that the PDE system now does
not determine a unique pair $(\mathcal{S}_1,\eta_1)$; it determines a family of solutions,
and further physical input is needed to select the state. 

Again, the appropriate requirement is that the state be smooth across each horizon in the appropriate Kruskal patches. This is sufficient to determine $\langle T_{ab} \rangle$. Since $\eta_1$ satisfies the wave equation, $\eta_1$ is chiral in $(u,v)$ with the general solution
\be
\eta_1=F(u)+G(v).
\ee
The relevant quantities are then $F''$ and $G''$, since these are exactly what
enter the $\langle T_{ab} \rangle$. 

The key point is that horizon regularity forces $F''$ and $G''$ to approach definite constants near
the relevant horizons, with exponentially decaying tails. Equivalently, $F$ and $G$ are
asymptotically quadratic. To see this, upon the appropriate coordinate transformations from \eqref{eq:Ttt}--\eqref{eq:Ttr}, one finds
\be
\langle T_{uu}\rangle
=
\frac{\epsilon N\hbar}{72\pi}\,\tanh \r\left(\sech^2 \r+2\right)
+\frac{\epsilon N\hbar}{4\pi}\,F''(u),
\ee
\be
\langle T_{vv}\rangle
=
\frac{\epsilon N\hbar}{72\pi}\,\tanh \r\left(\sech^2 \r+2\right)
+\frac{\epsilon N\hbar}{4\pi}\,G''(v).
\ee
Thus the state-dependent sector is purely chiral: $\langle T_{uu}\rangle$ depends only on
$F''$, while $\langle T_{vv}\rangle$ depends only on $G''$.

Now consider the future black-hole horizon $\mathcal H_b^+$, where $u\to+\infty$ and
$r\to-\infty$. Since
$
\tilde U_b=-e^{-\kappa_b u},$ and $
\tilde{V}_c =- e^{-\kappa_c v},
$
regularity of $\langle T_{\tilde U_b\tilde U_b}\rangle$ requires $\langle T_{uu}\rangle$ to vanish
at least as fast as $e^{-2\kappa_b u}$ for $\r\to-\infty$ with fixed $v$. Similarly for the cosmological horizon, regularity of $\langle T_{\tilde V_c\tilde V_c}\rangle$ requires $\langle T_{vv}\rangle$ to vanish
at least as fast as $e^{-2\kappa_c v}$ as $\r\to\infty$ with fixed $u$. This information is enough to determine the contribution from $\eta_1(\t,\r)$ to the stress tensor via
\begin{align}
\r\to -\infty:\qquad 
\langle T_{uu} \rangle 
&\approx \frac{\epsilon N \hbar}{4\pi}\left( - \frac{1}{9} + F''(u) \right)
&\Rightarrow\qquad 
F''(u) &= \frac{1}{9},
\\[0.3em]
\r\to \infty:\qquad 
\langle T_{vv} \rangle 
&\approx \frac{\epsilon N \hbar}{4\pi}\left( \frac{1}{9} + G''(v) \right)
&\Rightarrow\qquad 
G''(v) &= - \frac{1}{9}.
\end{align}
It then follows that the net flux is
\be \label{eq:nearNariaiTtrGF}
\langle T_{\t \r}\rangle
=
\langle T_{vv}\rangle-\langle T_{uu}\rangle
=
\frac{\epsilon N\hbar}{4\pi}\left(G''-F''\right)
=
-\frac{\epsilon N\hbar}{18\pi},
\ee
where the last equality holds in the late-time regime $\t\to\infty$ at fixed $\r$ in the static patch, as $F''$ and $G''$ approach their asymptotic near-horizon values. Thus the explicit horizon-regularity analysis directly gives a nonzero outward energy flux from the black-hole horizon toward the cosmological horizon.

It is also useful to state explicitly how the observer-adapted normal-ordering prescription affects the near-Nariai stress tensor at order $O(\epsilon\hbar)$. Similar to~\eqref{eq:Nariaishift}, we have
\be
\eta^{\text{(static)}}=\eta^{\text{(Kruskal)}}+\delta\eta_0+\epsilon\,\delta\eta_1.
\ee
It turns out that near the future black-hole and cosmological horizons, we have
\be
\partial_u^2(\delta\eta_0+\epsilon\,\delta\eta_1)=\frac{\kappa_b^2}{12},
\qquad
\partial_v^2(\delta\eta_0+\epsilon\,\delta\eta_1)=\frac{\kappa_c^2}{12}.
\ee
In physical units, the individual null components acquire the local thermal flux associated with the black-hole and cosmological horizons, 
\be
\Lambda \langle T_{uu} \rangle=\frac{\pi N \hbar}{12} T_b^2, \quad \Lambda \langle T_{vv} \rangle=\frac{\pi N \hbar}{12} T_c^2.
\ee
with $T_{b,c}$ defined by~\eqref{eq:Nariaitemperature}. Thus the observer-adapted prescription does affect the $O(\epsilon\hbar)$ null components individually: near the corresponding horizons they are no longer set to zero, but instead carry the linearized thermal baths dictated by $T_b$ and $T_c$. However, the transport component is unchanged, since the common exact-Nariai flux cancels in the difference. 
Indeed,
\be
\Lambda\langle T_{\t \r}\rangle
=
\Lambda\big(\langle T_{vv}\rangle-\langle T_{uu}\rangle\big)
=
\frac{\pi N\hbar}{12}\left(T_c^2-T_b^2\right)
=
-\frac{\epsilon N\hbar\Lambda}{18\pi},
\ee
in precise agreement with the direct late-time result~\eqref{eq:nearNariaiTtrGF}.\footnote{The quadratic correction to the dilaton potential discussed in
Section~\ref{sec:classicalnearNariai}, which goes beyond the strict JT approximation, is
essential for this result. In strict JT, the 2d metric remains rigid
dS$_2$ with $\mathcal{R}_0(\t,\r)=0$ even after the order-$\epsilon$ dilaton profile is turned on. Since the leading
matter sector only sees this rigid metric, the stress tensor contains no geometric
information distinguishing the black-hole and cosmological horizons. The only
$O(\epsilon\hbar)$ contribution then comes from the homogeneous sector of the auxiliary
field,
\be
\langle T_{uu}\rangle_{\rm JT}
=
\frac{\epsilon N\hbar}{4\pi}F''(u),
\qquad
\langle T_{vv}\rangle_{\rm JT}
=
\frac{\epsilon N\hbar}{4\pi}G''(v).
\ee
Horizon smoothness sets $F''=G''=0$ in the Kruskal normal-ordering prescription. Rewriting the same state in static-patch normal ordering gives only the common exact-Nariai thermal bath in \eqref{eq:Nariaiflux}, not separate $T_b^2$ and $T_c^2$ contributions. The reason is structural: the dilaton profile selects the relevant static patch, but a rigid dS$_2$ metric has only the single Nariai surface gravity. This is
consistent with analogous near-extremal black hole analyses in holography; see, for
example, \cite{Spradlin:1999bn, Almheiri:2019psf}.} We therefore recover, within the anomaly-induced one-loop effective formalism, the standard late-time thermal Hawking flux, in agreement with the well-established results of \cite{Markovic:1991ua,Tadaki:1990aa,Tadaki:1990cg}. From the 4d perspective, this corresponds to an $s$-wave energy flux density of order
$\Lambda^2\langle T_{\t \r}\rangle$, up to the trivial transverse-area factor of the $S^2$.

It is useful to go one step further and extract the corresponding late-time near-horizon
form of $\mathcal{S}_1$ implied by the thermal flux. We can plug the horizon-smooth thermal branch back into~\eqref{eq:near_Nariai_tt}-\eqref{eq:near_Nariai_tr}, and evaluate $\mathcal{S}_1$ near the horizons. Let us sketch the derivation for the black-hole horizon. Consider the limit of $\r\to-\infty$ with fixed $v$. Then $\partial_\t^2\eta_1\sim \partial_\r^2 \eta_1 \sim 0$ and $\partial_\t \partial_\r \eta_1 \sim -2/9$. Since $\sech^2\r \sim 0$ and $\tanh\r\sim -1$ the RHS of \eqref{eq:near_Nariai_tt}-\eqref{eq:near_Nariai_tr} becomes a constant. On the LHS we can assume that $\mathcal{S}_1(\r\to-\infty, v)$ is a smooth function of $v$. Therefore $\partial_\r \mathcal{S}_1 \sim \partial_\t \mathcal{S}_1 \sim \partial_v \mathcal{S}_1$ and \eqref{eq:near_Nariai_tt}-\eqref{eq:near_Nariai_tr} become identical ODE for the $v$-dependence of $\mathcal{S}_1(\r\to-\infty, v)$ which has a simple solution. A similar analysis holds near the cosmological horizon for $\r\to\infty$ and fixed $u$.

The results of the calculation we just outlined in the previous paragraph are
\be \label{eq:thermal_linear}
\mathcal{S}_1 \simeq - \frac{N}{18 S_0}\, v  \quad \text{(near $\mathcal{H}_b^+$)}, \qquad \mathcal{S}_1 \simeq  \frac{N}{18 S_0}\, u  \quad \text{(near $\mathcal{H}_c^+$)},
\ee
where as we argued, any homogeneous sectors to the PDEs have been redefined away at this order.  Thus, the horizon-smooth branch that reproduces the thermal Hawking flux also
induces a linear late-time drift of the near-horizon radius perturbation: linear in the
advanced null coordinate $v$ on $\mathcal{H}_b^+$ and linear in the retarded null coordinate $u$
on $\mathcal{H}_c^+$.  From the definition of $\phi$ we can infer the time evolution of the black-hole horizon area. In terms of the black hole (cosmological) horizon entropy $S_{b}$ ($S_{c}$), we get 
\be
\frac{\d S_{b}}{\d v} = - \frac{\pi N}{6 \sqrt{\Lambda}}(T_b-T_c) ,~~~~~~~~\frac{\d S_{c}}{\d u} = \frac{\pi N}{6 \sqrt{\Lambda}}(T_b-T_c).
\ee
Recall that in our conventions $v$ is dimensionless. The black hole entropy decreases and the cosmological horizon entropy increases as a function of time. The approximations used in this paper break down at very late times when $v, u \sim O(S_0/N)$. In particular, near the endpoint of black-hole evaporation, quantum-gravity effects are expected to become important.

The derivation above should be viewed as the explicit auxiliary-field realization of the
horizon-smooth branch within the one-loop effective description. At order $O(\epsilon\hbar)$,
however, the same state constraints admit a more invariant formulation in terms of the
Schwarzian transformation law of the null stress tensor. In that language, one defines the
state by null coordinates $(U,V)$, and horizon smoothness requires these state coordinates
to be asymptotically affine functions of the horizon-regular Kruskal variables. The
corresponding Schwarzian derivatives therefore approach constants, up to exponentially
decaying tails, which is precisely the condition needed to render the covariant stress tensor
finite at the horizons. This more elegant and more general formulation,
which makes the underlying Hadamard-state criterion manifest, will be given in Appendix~\ref{app:Schwarziantransform}.

\subsection{No-boundary condition and the fate of pair-created black holes}
\label{sec:noboundary}

As discussed in Section~\ref{sec:bigpicture}, near-Nariai black holes are most naturally produced through pair-creation rather than gravitational collapse. Since the near-Nariai regime corresponds to an SdS solution extremely close to the maximal-mass, degenerate-horizon limit, producing such a configuration by collapse requires a severe tuning of $\Lambda$ in an exponentially expanding background \cite{Bousso:1995cc, Bousso:1996au, Bousso:1997wi}. By contrast, Euclidean Nariai admits a smooth compact cap with $S^{2}\times S^{2}$ topology, and its analytic continuation provides a semiclassical preparation of a Lorentzian near-Nariai throat.

As we have shown in Section~\ref{sec:thermalHawking}, once one imposes horizon-smoothness, the near-Nariai black hole exhibits the standard thermal Hawking flux. A natural follow-up
question is then: \emph{how should a pair-created near-Nariai black hole evolve?} Since such
configurations are most naturally prepared by the Euclidean $S^2\times S^2$ cap, the answer
should be encoded in the corresponding no-boundary prescription~\cite{Hartle:1983ai}. This question was already addressed by Bousso and Hawking~\cite{Bousso:1997wi}, and they argued that the pair-created near-Nariai black hole evaporates at late times.  What was not made explicit, however, is whether this late-time evaporation is precisely the ordinary thermal Hawking process associated with the two static-patch horizons. We analyze the no-boundary preparation from the Euclidean $S^2\times S^2$ cap and compare its Lorentzian late-time behavior with the thermal flux derived in Section~\ref{sec:thermalHawking}.

Here, the Euclidean cap prepares a \emph{global} Lorentzian perturbation of the full
near-Nariai throat. By contrast, Hawking radiation is physically characterized only after one
restricts to a single static patch, where the black-hole and cosmological horizons are identified and the notion of Hawking flux is tied to the static-patch time. Thus, to make
contact with the Lorentzian problem, one must first identify the global profile selected by the
no-boundary construction and then project it onto a chosen static diamond. Only after this
projection does it become meaningful to ask whether the resulting state is compatible with a
smooth Hadamard behavior at the horizons and whether its late-time evolution matches the
thermal flux derived above.

With this interpretation in mind, we show that once the
global profile is projected properly into a static patch, the prescription does yield a late-time evolution compatible with the standard thermal Hawking flux.

\subsection*{Lorentzian global Nariai geometry}

We begin with the classical exact Nariai geometry in the global conformal coordinates $(\hat{t},x)$. A convenient way to introduce them is through embedding coordinates $X^I$, with $I=0,1,2$, via
\be
X_0=\sec \hat{t} \sin x,\qquad X_1=\tan \hat{t},\qquad  X_2=\sec \hat{t} \cos x.
\ee
For comparison, the static patch in the $(\t,\r)$ coordinates is obtained by
\be
X_0=\tanh \r,\qquad X_1=\sech \r \sinh \t,\qquad  X_2=\sech \r \cosh \t,
\ee
which only covers the $X_2>|X_1|$ region. Equating these two representations of the embedding coordinates can give a relation between $(\hat{t},x)$ and $(\t,\r)$. The metric in global coordinates becomes
\be\label{eq:quantum_nariai_global}
 \d s^{2}=\frac{1}{\Lambda \cos^{2}\hat{t}}\left(-\d \hat{t}^{2}+\d x^{2}\right)+\frac{1}{\Lambda}\d \Omega_{2}^{2},
\qquad
\hat{t}\in\Big(-\frac{\pi}{2},\,\frac{\pi}{2}\Big),\quad x\sim x+2\pi,
\ee
Here $x$ parametrizes the spatial $S^{1}$ of the
closed slicing, while $\hat{t}$ is the global conformal time that foliates the \emph{entire} dS$_2$ hyperboloid.  The endpoints $\hat{t}=\pm\pi/2$ are the conformal boundaries $\mathcal{I}^{\pm}$. The one-loop correction to this exact Nariai background only shifts the two radii by constants,
$e^{2\rho}\to \Lambda_1^{-1}\cos^{-2}\hat{t}$ and $e^{2\phi}\to \Lambda_2$, with $\Lambda_{1,2}=\Lambda+O(N/S_0)$, in agreement
with the quantum Nariai geometry in Section~\ref{sec:thermalHawking}. This chart is
well-suited for describing pair creation and no-boundary preparation, but it should be kept distinct from the static-patch
time: $\partial_{\hat{t}}$ is not a globally timelike Killing vector, so
stationarity and horizon regularity for a static-diamond observer will not be manifest in these variables.

We perturb the geometry by turning on $\epsilon$, similar to what we did in~\eqref{eq:linearperturbation}, in particular the dilaton profile
\be
\phi(\hat{t},x)=\phi_0-\epsilon \mathcal{S}(\hat{t},x).
\ee
At leading order, the linearized near-Nariai equations admit exactly the same three
$\mathrm{SL}(2,\mathbb R)$ modes as in the static-patch analysis $\mathcal{S}= C_I X^I$. The same solution written in global coordinates becomes
\be\label{eq:global_sl2_modes}
\mathcal{S}(t,x)
=(C_1+C_{-1})\,\sec \hat{t} \cos x+C_0\,\sec \hat{t} \sin x+ (C_1-C_{-1})\,\tan \hat{t}.
\ee
In light of the discussion
of Section~\ref{sec:anatomyNariai}, we should therefore choose the canonical representative in the static patch, namely the $C_0$ branch. That is, we use the residual gauge freedom to set $C_{1}=C_{-1}=0$. This choice might look arbitrary in the global coordinates but proves to be extremely convenient in the static patch.

\subsection*{No-boundary cap and the Euclidean solution}
\begin{figure}[t!]
\centering
\begin{tikzpicture}[scale=0.85, >=Latex]

    \draw[->, thick] (-0.6,0) -- (4.8,0) node[right] {$\mathrm{Re}(\tau)$};
    \draw[->, thick] (0,-0.8) -- (0,3.2) node[above] {$\mathrm{Im}(\tau)$};

    \fill[blue] (0,0) circle (2pt);
    \node[below left] at (0,0) {\small $0$};

    \fill[blue] (2.6,0) circle (2pt);
    \node[below, black] at (2.6,0) {\small $\frac{\pi}{2}$};

    \draw[blue, very thick, ->] (0,0) -- (1.4,0);
    \draw[blue, very thick, ] (1.4,0) -- (2.6,0);
    \draw[blue, very thick, ->] (2.6,0) -- (2.6,2.4);

    \node[black, right] at (2.6,1.2) {\small $\tau=\frac{\pi}{2}+i\tilde{\tau}$};
    \
\end{tikzpicture}
\caption{$\tau$ contour of the no-boundary geometry in our conventions.}
\label{fig:HHNBC}
\end{figure}
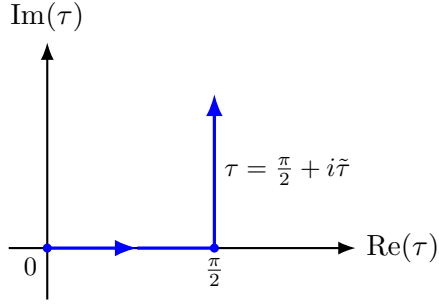

The no-boundary prescription prepares the Lorentzian pair-created configuration by specifying a smooth Euclidean ``cap" that
closes off the geometry in the past.  Concretely, one Wick-rotates the global conformal time,
\be
\hat{t}\ \to\ -\i \hat{t}_E,
\qquad
e^{2\rho}=\frac{1}{\Lambda_1\cosh^2 \hat{t}_E},
\ee
so that the dS$_2$ factor becomes a smooth $S^2$ (a hemisphere once we choose the nucleation slice).  It is convenient to
introduce a polar coordinate $\tau$ on this Euclidean $S^2$ by
\be\label{eq:u_def_rewrite}
\sin \tau=\frac{1}{\cosh \hat{t}_{E}},
\qquad
\tau\in\Big[0,\,\frac{\pi}{2}\Big],
\ee
which brings the Euclidean metric to 
\be\label{eq:Eucl_metric_u_rewrite}
\d s^{2}_{E}
=\frac{1}{\Lambda_{1}}\Bigl(\d \tau^{2}+\sin^{2}\tau\,\d x^{2}\Bigr)+\frac{1}{\Lambda_{2}}\,\d \Omega_{2}^{2}.
\ee
The point $\tau=0$ is the \emph{south pole} where the $x$-circle degenerates smoothly, and $\tau =\pi/2$ is the equatorial
nucleation slice across which we continue to Lorentzian signature.  The Lorentzian continuation is taken along
\be\label{eq:u_to_L_continuation}
\tau=\frac{\pi}{2}+\i \tilde{\tau},\qquad \tilde{\tau} \ge0,
\ee
so that $\sin{(\frac{\pi}{2}+\i \tilde{\tau})}=\cosh{\tilde{\tau}}$ and hence $\cosh{\tilde{\tau}}=\sec{\hat{t}}$ upon returning to the Lorentzian conformal time $\hat{t}$ used
in~\eqref{eq:quantum_nariai_global}. See Figure~\ref{fig:HHNBC}.

We then write the dilaton perturbation directly in the $\sin{x}$ harmonic sector, with a perturbative profile in $N/S_0 \ll1$,
\be
\phi_E(\tau,x)=\phi_0-\epsilon\, s(\tau)\sin x, \qquad s(\tau)=s_0(\tau)+\frac{N}{S_0} s_1(\tau),
\ee
Here we keep a general profile $s(\tau)$ for the following reason: at $\hbar=0$, this is just the Euclidean continuation of the classical global $\sec{\hat{t}}\sin{x}$-mode,
since $\sec(\i \hat{t}_E)=1/\cosh \hat{t}_E=\sin \tau$, so the classical profile is $s_0 (\tau) \propto \sin \tau$.
At one loop, however, we should not assume $s_1(\tau)$ is the Wick-rotated solution we found in Section~\ref{sec:thermalHawking}, because the no-boundary problem is precisely being used to select the state. Rather, the no-boundary prescription determines
the $O(N/S_0)$ correction by solving the Euclidean semiclassical equations in this chosen
harmonic sector, with the requirement that the solution reduces smoothly to the classical mode as $N/S_0\to0$.

The no-boundary geometry supplies an additional, genuinely Euclidean input: regularity at the south pole.
Since the $x$-circle pinches off at $\tau=0$, single-valuedness of the dilaton perturbation requires
\be\label{eq:southpole_bc_rewrite}
s(0)=0,
\ee
otherwise the dilaton would not extend smoothly through the tip of the cap.

Within this harmonic sector, the Euclidean $\hat{t}\hat{t}$- and $xx$-components of the
semiclassical equations, together with the Euclidean wave equation for the state data $\eta_1$, reduce
the problem to a single ODE for $s(\tau)$:
\be\label{eq:s_eq_u}
\sin^{2}\tau \, s''(\tau)
+ \sin \tau \cos \tau\, s'(\tau)
- \left(1 - \left(2 + \frac{N}{6 S_0}\right)\sin^{2}\tau \right)s(\tau)
= 0 \, .
\ee
This reduced equation should be interpreted with some care. In Lorentzian signature, the
mixed $\hat{t}x$-equation plays an essential role as a constraint and is needed to identify the allowed
static-patch branches. Here, by contrast, we are not classifying the full Lorentzian solution
space, but solving the Euclidean boundary-value problem in the already-selected global
harmonic sector. The additional input comes from smoothness of the Euclidean cap described above.

One can verify that indeed, at leading order, the unique solution obeying $s_0(0)=0$ is
\be
s_0(\tau)=C_0 \sin \tau,
\ee
with constant amplitude $C_0$, which is precisely the Euclidean continuation of the classical $C_0$-mode. We already determined in Section~\ref{sec:anatomyNariai} that in order to match with the classical near-Nariai solution, we should normalize $C_0=1$. This can be done without loss of generality since any other value can be absorbed in the normalization of $\epsilon$. At one loop, the solution that stays regular on $\tau\in[0,\pi/2]$ and still satisfies $s_1(0)=0$ is
\be\label{eq:s1_u_solution_revised}
s_1(\tau)=\frac{1}{18}\Bigl(\sin \tau\,\ln(1+\cos \tau)-\tan\frac{\tau}{2}\Bigr)+c_1\,\sin \tau,
\ee
The term proportional to $c_1$ is just the homogeneous solution and therefore amounts to
an $O(\hbar)$ renormalization of the overall amplitude. We may set $c_1=0$.

\subsection*{Lorentzian continuation and the late-time profile}

Now we continue off the nucleation slice to Lorentzian times via~\eqref{eq:u_to_L_continuation}. 
Along the Lorentzian segment, the profile $s(\tau)= s(\frac{\pi}{2}+\i \tilde{\tau})$ becomes
\be\label{eq:sigma_v_full}
s(\tilde{\tau})=  \cosh \tilde{\tau}
+\frac{N}{S_0}\Biggl[\frac{1}{18}\Bigl(\cosh \tilde{\tau}\ln(1-\i\sinh \tilde{\tau})-(\sech \tilde{\tau}+\i\tanh \tilde{\tau})\Bigr)
\Biggr]
\ee
and differentiating gives
\be\label{eq:sigma_v_derivative}
\partial_{\tilde{\tau}} s(\tilde{\tau})
=  \sinh \tilde{\tau}
+\frac{N}{S_0}\Biggl[\frac{1}{18}\Biggl(\sinh \tilde{\tau} \ln(1-\i\sinh \tilde{\tau})+\frac{3+\cosh(2 \tilde{\tau})}{2(\sinh \tilde{\tau}+\i)}\Biggr)
\Biggr].
\ee
At $\hbar=0$ the entire saddle (Euclidean cap plus Lorentzian continuation) can be chosen everywhere real. For $\hbar \neq 0$, the analytic continuation necessarily crosses the logarithmic branch cut, and a generic real Euclidean amplitude produces a complex Lorentzian history. This is not a pathology: in the no-boundary saddle-point approximation the relevant saddle is typically complex, and one instead imposes a late-time reality condition that the Lorentzian configuration approach the real section of phase space---namely, not only the field amplitude $s(\tilde{\tau})$ but also its conjugate momentum (equivalently $\partial_{\tilde{\tau}} s(\tilde{\tau})$) must become asymptotically real---so that the WKB phase generates a real trajectory.

To implement this, we take $\tilde{\tau} \gg 1$. Note that the late-time limit is uniform in the sense that taking large $\tilde{\tau}$ limit commutes with acting by $\partial_{\tilde{\tau}}$ on the asymptotic expansion. This indicates that the approach to the late-time regime is smooth and without a hidden rapidly varying phase, so the late-time profile is reached in a regular manner. The continued profile is generically complex at subleading order (and one can ensure reality if one specifically wants to make a stronger WKB reality claim).

For our purposes, however, only the universal late-time secular term
is needed. In the regime
\be
1\ll \tilde{\tau} \ll \frac{S_0}{N},
\ee
one finds
\be\label{eq:s_v_late}
s(\tilde{\tau})\simeq \frac{1}{2}e^{\tilde{\tau}}\left[1+\frac{N}{18S_0}\tilde{\tau}+O\left(\frac{N}{S_0}\right)\right],
\ee
Thus the no-boundary continuation produces a universal linear-in-$\tilde{\tau}$ correction to the
leading growth, independent of the subleading constant terms.

\subsection*{Static patch projection and the thermal Hawking flux}

Having established that the no-boundary cap yields a well-defined late-time Lorentzian continuation in global variables, we may
interpret it as preparing a (global) wavefunctional on the nucleation slice of the dS$_2\times S^2$ throat, i.e. a state on a
complete Cauchy slice of the Lorentzian spacetime. A static observer, however, has access only to the operator algebra within a
single causal diamond. The appropriate object to compare with our Lorentzian Hadamard analysis is therefore the
\emph{static-diamond restriction} of the globally prepared state, obtained by tracing out the complementary region. We will comment about the generic approach to global $\mathcal I^+$ in Section~\ref{sec:discussion}; here we focus only on the late-time limit relevant to a static observer. 

Translating \eqref{eq:s_v_late} back to global
Lorentzian time using $\cosh \tilde{\tau}=\sec \hat{t}$, or equivalently $\tilde{\tau}\sim \ln(2\sec \hat{t})$ as
$\hat{t}\to\pi/2$, gives
\be
\mathcal{S}(\hat{t},x)\simeq \sec \hat{t}\,\sin x\left[1+\frac{N}{18S_0}\ln(2\sec \hat{t})+\cdots\right],
\qquad \hat{t}\to\frac{\pi}{2}.
\ee
up to terms subleading in $N/S_0$. Using the exact global-to-static map $\sec \hat{t}\,\sin x=\tanh \r$, together with the late-time static-patch relation
\be
\ln(2\sec \hat{t})=\t+\ln\sech \r+O(e^{-2 \t}),
\ee
we obtain the static-diamond projection
\be\label{eq:static_late_t}
\mathcal{S}(\t,\r)\simeq
\tanh \r
\left[1+\frac{N}{18S_0}\bigl(\t+\ln\sech \r\bigr)+\cdots\right].
\ee
We may now evaluate the late-time horizon profile. Near the future black-hole horizon $\mathcal{H}_b^+$, we have
$\r\to-\infty$ at fixed $v=\t+\r$. Then one has $\tanh{\r} \to -1$ and $\t+\ln\sech \r \to v+\ln 2$. Therefore, in the large $v$ limit
\be\label{eq:Hb_profile} 
\mathcal{S}(\t,\r)\to
-1-\frac{N}{18S_0}v+\cdots, \qquad \text{(near $\mathcal{H}_b^+$)}.
\ee
\be\label{eq:Hb_profile} 
\partial_v \mathcal{S}(\t,\r)\to
-\frac{N}{18S_0}+\cdots, \qquad \text{(near $\mathcal{H}_b^+$)}.
\ee
Similarly, near the future cosmological horizon $\mathcal{H}_c^+$, $\r\to+\infty$ at fixed $u=\t -\r$, one finds $\tanh{\r} \to 1$ and $\t+\ln\sech \r \to u+\ln 2$. Hence in the large $u$ limit
\be\label{eq:Hc_profile}
\mathcal{S}(\t,\r)\to
1+\frac{N}{18S_0}u+\cdots, \qquad \text{(near $\mathcal{H}_c^+$)}.
\ee
Thus the static-diamond projection of the no-boundary state exhibits a linear late-time
drift along the horizon generators: linear in $v$ on $\mathcal{H}_b^+$ and linear in $u$ on $\mathcal{H}_c^+$.
This is exactly what we found in Section~\ref{sec:thermalHawking}, with the same slope given in~\eqref{eq:thermal_linear} from the
horizon-smooth Lorentzian analysis. The black-hole horizon is driven toward smaller area while the cosmological horizon is driven toward larger area.

This agreement is nontrivial beyond the exact Nariai limit. At exact Nariai, the thermal interpretation is closely tied to the fact that the static-patch time can be Wick rotated to a smooth Euclidean circle. At order $\epsilon$, however, the two derivations use rather different criteria: the no-boundary construction selects a global Lorentzian perturbation from the Euclidean cap, whereas the Lorentzian calculation fixes the state by requiring smoothness in the Kruskal frames of a chosen static diamond. The fact that these two prescriptions nevertheless lead to the same late-time thermal behavior, and in particular to the same net Hawking flux, is therefore a nontrivial check of the evaporation picture. We return to this point in Section~\ref{sec:discussion}.

\subsection{Anti-evaporation as a semiclassical pathology}
\label{sec:antievaporation}

We now turn
to the question: why did Bousso and Hawking~\cite{Bousso:1997wi}, working within the same
one-loop effective formalism, find the apparently opposite anti-evaporation behavior? We trace the origin of that surprising channel directly in the static-patch
description. We show that the would-be anti-evaporation channel arises due to a pathological choice of quantum state, rather than a genuine physical instability.

The calculation in \cite{Bousso:1997wi} was complicated by the fact that the authors did not use the $\mathrm{SL}(2,\mathbb R)$ freedom to set $C_{1}=C_{-1}=0$. Correspondingly, the source terms appearing in the semiclassical
equations \eqref{eq:near_Nariai_tt}--\eqref{eq:near_Nariai_tr} are modified by the explicit $C_I$-dependent
pieces in the background. However, two structural facts remain unchanged: neither the invariant
horizon-tracing condition \eqref{eq:gradPhi_null_22} nor the $\langle T_{ab} \rangle$ depends
on $\mathcal R_1$, so the entire question of evaporation versus anti-evaporation continues to
be controlled by the pair $(S_1,\eta_1)$ and by the anomaly-induced geometric sector. We shall show that the choice of matter state in \cite{Bousso:1997wi}, through a choice of homogeneous solution of $\eta_1$, is incompatible with globally smooth black-hole and cosmological horizons.

The analog of the choice of state in \cite{Bousso:1997wi} is to demand that $\eta_1$ satisfies
\be \label{eq:eta_1zero}
\partial_\t\partial_\r\eta_1=0,
\ee
together with the wave equation $(\partial_{\t}^2-\partial_{\r}^2)\eta_1=0$, which resembles the general solution of the exact Nariai branch in~\eqref{eq:Nariaieta0} such that $\eta_1=A(\t^2+\r^2)+\cdots$, where one finds $A=0$ instead. This is a perfectly consistent solution of the linearized semiclassical equations.\footnote{There is an ``evaporation" channel found by Bousso-Hawking (the analog of Eq.~(4.18) in~\cite{Bousso:1997wi} at the one-loop order), which is not a valid solution. It solves the reduced evolution equation but not the full constrained near-Nariai system: after imposing the mixed $\t \r$ constraint, the admissible $O(\epsilon)$ deformations are only the three $\mathrm{SL}(2,\mathbb R)$ modes, and this profile is not arising from one of them. Any attempt to construct it through the one-loop state sector would require promoting the $O(\epsilon\hbar)$ contribution from $\eta_1$ to $O(\epsilon)$, i.e. $\eta_1\sim 1/\hbar$, outside the controlled semiclassical expansion, representing a power-counting mismatch.} It should, however, be emphasized from the outset that this already represents a specific choice of state data. On this branch, the explicit $C_I$-dependence of $\langle T_{ab} \rangle$ is purely the geometric anomaly-driven sector:
\bea \label{eq:Ttt_1}
\langle T_{\t \t} \rangle&=& -\frac{N \hbar}{24 \pi}\sech^2{\r}+\epsilon N \hbar \bigg[(C_1 e^{\t} +C_{-1} e^{-\t}) \frac{\sech^3{\r}}{72 \pi} (\cosh{4 \r}-4 \ln{\cosh{r}}-6) 
\no\\
&\quad&+C_{0} \frac{\sech^2{\r} \tanh{\r}}{72 \pi }(2\cosh{2 \r} -4 \ln{\cosh{\r}}-5) \bigg],\\
\label{eq:Trr_1}
\langle T_{\r \r} \rangle&=&\frac{N\hbar}{24 \pi}\sech^2{\r}+\epsilon N\hbar \bigg[ (C_1 e^{\t} +C_{-1} e^{-\t}) \frac{\sech^3{\r}}{72 \pi} (8\cosh^4(\r)+4\ln{\cosh{\r}}+9)
\no\\
&\quad&+C_{0} \frac{\sech^2{\r}\tanh{\r}}{72 \pi } (\cosh{2\r}+2 \ln{\cosh{\r}}+13) \bigg],\\
\label{eq:Ttr_1}
\langle T_{\t \r} \rangle&=& \epsilon N\hbar \bigg[(C_1 e^{\t} -C_{-1} e^{-\t}) \frac{\sech{\r}\tanh{\r}}{18 \pi} (\cosh{2\r}+2) \bigg].
\eea
In particular, we see that the local energy flux $\langle T_{\t \r} \rangle$ need not vanish once $C_{\pm 1}$ are turned on.

\paragraph{$C_{\pm 1}=0$: Boulware-like state.}

Had we picked the gauge with $C_{\pm 1}=0$, then the restricted branch~\eqref{eq:eta_1zero} is the static-patch Boulware-like state: there is no local flux since $\langle T_{\t \r} \rangle=0$, the $\langle T_{uu} \rangle$ and $\langle T_{vv} \rangle$ approach bounded constants at the black-hole and cosmological horizons in the $(u, v)$ coordinates, but the state is in fact singular as it fails the horizon-smooth Kruskal test. This is consistent with what we found in Section~\ref{sec:thermalHawking}, where the horizon-smooth Hadamard state instead should satisfy
\be
\partial_\t\partial_\r\eta_1=-F''+G''=-\frac{2}{9}\neq 0,
\ee
near the horizons. The horizon-tracing analysis for this state choice would imply neither evaporation nor anti-evaporation, since the horizon areal radius is only $\r$-dependent and is perturbatively divergent at $O(\epsilon \hbar)$. We call it Boulware-like since we are in the static diamond rather than an asymptotically flat exterior~\cite{Boulware:1974dm}.

\paragraph{$C_{\pm 1} \neq 0$: anti-evaporation state.} If one keeps $C_{\pm 1} \neq 0$ while maintaining the same branch~\eqref{eq:eta_1zero}, one obtains a different Lorentzian state choice of the Bousso-Hawking type~\cite{Bousso:1997wi}. Note that the relevant horizons are then the shifted null generators discussed in Section~\ref{sec:anatomyNariai}, rather than the undeformed $C_{\pm 1}=0$ Killing horizons. By evaluating the horizon-tracing condition on this restricted branch, the one-loop correction at $O(\epsilon \hbar)$ to the traced horizon size acquires exponentially dominated late-time pieces.\footnote{A remark is that here ``late times" do not correspond to a physical static observer when $C_{\pm 1}\neq 0$, instead to the asymptotic ends of the horizon generators.} For the future black-hole branch, the area has the leading asymptotic behavior
\be
\mathcal{A}_b = \mathcal{A}_{0,b} +\frac{4 \epsilon N\sqrt{\pi} \sqrt{\mathcal{A}_0}}{9S_0} \frac{C_1^4}{(C_0+\sqrt{C_IC^I})^3 \sqrt{C_I C^I}} \frac{}{} e^{4\t}+\cdots, \qquad (\t \to \infty),
\ee
up to subleading tails. Here $\mathcal{A}_{0,b}$ is the time-independent background area, already incorporating the classical near-Nariai shift at $O(\epsilon)$ from Section~\ref{sec:classicalnearNariai}, and the static quantum Nariai correction at $O(\hbar)$ from Section~\ref{sec:thermalHawking}. We see that the coefficient of $e^{4 \t}$ is positive since the numerator $C_1^4>0$, while on the real-horizon branch one has $\sqrt{C_I C^I}>0$ and since  $\sqrt{C_IC^I}>|C_0|$, we have $C_0+\sqrt{C_IC^I}>0$. Thus the exponentially dominant term has a definite positive sign. A similar statement holds for the cosmological horizon branch, but now the time-dominant piece has an overall negative coefficient. By picking such a state with real-horizon branches, this behavior is gauge-invariant.

This reproduces the qualitative anti-evaporation behavior identified by Bousso and Hawking, indicating the growth of the black-hole branch together with the shrinkage of the cosmological branch (analog of Eq.~(4.17) in~\cite{Bousso:1997wi}, but now written in static patch variables and can be solved for asymptotic late times). However, the same exponential growth also signals that the semiclassical expansion eventually breaks down.

To identify the correct physical interpretation of such a state choice, we have to examine the regularity of the stress tensor. At a finite point on the horizon generator, one can always construct a local inertial frame (which is not of the Kruskal type when $C_{\pm 1}\neq0$). In this case, no extra pointwise divergence follows from the map: the transformation from $(u,v)$ to that local frame has a finite Jacobian, hence bounded $\langle T_{uu}\rangle$ and $\langle T_{vv}\rangle$ remain bounded locally. The real obstruction is global, when one follows the generators to their asymptotic ends: along $v=v_h$, one has $u \to \infty$, while along $u=u_h$, one has $v \to \infty$. In those limits, one finds schematically:
\be
\langle T_{uu}\rangle \sim \epsilon N \hbar\, C_1\, e^{u}, \qquad  \langle T_{vv}\rangle \sim \epsilon N \hbar\, C_{-1}\, e^{v},
\ee
so although the stress tensor can be pointwise regular on finite segments of the shifted horizons, it fails to remain bounded on the complete future generators. In this sense, this branch is not a globally smooth Hadamard state, consistent with the asymptotic breakdown of the horizon-tracing calculation. The correct interpretation is therefore not a new Hawking channel but a pathology of closing the semiclassical system with an inadmissible state choice.

This also clarifies in what sense $C_{\pm 1}$ modes remain ``pure gauge." At the classical level they lie on the SL(2,$\mathbb{R}$) orbit of the near-Nariai throat, but semiclassically the physical input is the geometry \emph{and} the quantum state. A change of representative in the geometric $C_I$ basis is physically innocuous only if the state data are transformed covariantly as well. If instead one turns on $C_{\pm 1}$ while keeping fixed the same state-defining data that we claim positive frequency with respect to, one has implicitly changed the quantum state, and $\langle T_{ab} \rangle$ need not be equivalent. From this viewpoint, the exponentially growing behavior that seeds anti-evaporation reflects an implicit, singular state choice rather than a physical instability in a smooth Hadamard state.

\section{Discussion and Outlook}
\label{sec:discussion}

In this paper, we sought to place the interpretation of the Nariai geometry as computing the black-hole nucleation rate in dS space on a firmer footing. The
main point is not simply that Euclidean Nariai is the unique smooth compact saddle in the SdS family, but that it computes the desired transition only once the question is formulated as an observer-refined process. In the presence of an observer witnessing the nucleation process, the gravitational path integral acquires the phase appropriate
to a transition rate. This also clarifies why the same Euclidean geometry can appear in different questions: the empty Nariai saddle is naturally associated with a norm or a state-counting observable, whereas the observer-dressed saddle is the object relevant for the black-hole nucleation in the static patch.

An equally central question is the subsequent Lorentzian fate of the nucleated black hole. The Euclidean saddle prepares the system at the Nariai point, but it is an unstable equilibrium where once the configuration is slightly perturbed, it enters the near-Nariai regime. We showed
that for a horizon-smooth state the black-hole horizon emits slightly more energy than
it absorbs and therefore shrinks, while the cosmological horizon grows. The black hole then undergoes standard thermal Hawking evaporation, until the geometry relaxes back to empty dS space. The timescale of this process is on the order of the Page time. This is consistent with interpreting the nucleation as a rare Boltzmann fluctuation of the dS vacuum rather than a vacuum decay, since the system eventually returns to its maximally entropic empty dS state \cite{Susskind:2021dfc}.

On the other hand, the details of the evaporation history of a near-Nariai black hole are subtle for several reasons. First, Bousso and Hawking found an anti-evaporation mode for near-Nariai black holes. Although they argued that the mode is either unexcited or only present initially, its mere existence is unexpected. We showed that their anti-evaporation mode is indeed pathological and arises from a state with singular horizons. This also justifies assuming standard thermal evaporation in other more recent applications~\cite{Montero:2019ekk, Chen:2024rpx}.\footnote{In particular, the picture we provide from the observer-dressed nucleation to the Lorentzian evaporation is then compatible with the von Neumann algebra perspective of \cite{Chen:2024rpx}, an evaporating SdS black hole provides the out-of-equilibrium ``clock" that makes it possible to define a nontrivial gauge-invariant algebra in a compact spacetime}

Furthermore, we showed that the no-boundary condition prepares a state whose late-time profile is compatible with the thermal Hawking radiation. In this case, the no-boundary cap prepares a global state of the near-Nariai throat, whereas Hawking radiation is an observer-dependent question formulated after restricting to a single static diamond. Hence, the relevant object for comparison with the Lorentzian analysis is not a global wavefunction, but its static-patch restriction, or equivalently the density matrix obtained after tracing out the complementary region~\cite{Ivo:2024ill}. In this sense, the situation is analogous to inflation, where regularity of the Euclidean cap selects the Bunch-Davies vacuum~\cite{Bunch:1978yq, Hartle:1983ai}, but here the subsequent projection to an observer's causal diamond is essential. Moreover, because the no-boundary saddle in the near-Nariai regime is generally complex, Euclidean smoothness alone does not automatically imply a KMS state with respect to the static Hamiltonian. The agreement we find with the horizon-smooth Lorentzian branch should therefore be viewed as a nontrivial consequence: the no-boundary state lies in the same late-time universality class as the thermal Hawking branch, producing the same linear horizon drift and the same net flux. 

Finally, quantum-gravity corrections become large in the near-extremal limit of black holes \cite{Ghosh:2019rcj,Iliesiu:2020qvm,Heydeman:2020hhw,Iliesiu:2022onk}, and one might worry that a similar situation occurs in the near-Nariai limit. This could strongly affect the Hawking rate \cite{Preskill:1991tb,Brown:2024ajk, Emparan:2025sao,Emparan:2025qqf,Biggs:2025nzs,Luo:2026epp}. We showed that the static patch of a near-Nariai black hole is well described by 2d JT gravity with a quadratic correction to the dilaton potential. Evaluating the sphere partition function suggests that this theory has no light modes analogous to the Schwarzian in the static patch. Such large corrections do appear when the SdS solution is interpreted as preparing a no-boundary wavefunction on $S^1 \times S^{D-2}$ \cite{Maldacena:2019cbz,Turiaci:2025xwi,Blacker:2025zca,Mariani:2025hee,Maulik:2025phe}, but we find here that this is not the case when SdS describes a physical black hole.\footnote{An intuitive picture for the absence of large quantum-gravity corrections in the static patch can
be understood from the difference between two inequivalent late-time limits of the same
near-Nariai geometry. The large Schwarzian enhancement found in the $S^1\times S^{D-2}$ no-boundary wavefunction~\cite{Turiaci:2025xwi} arises when one conditions on an expanding future slice with a large spatial circle. In that problem the classical answer becomes nearly independent of the modulus controlling the length of the $S^1$, and a sector of reparametrization modes develops an action suppressed by this large scale. By contrast, the static-diamond problem has no analogous large modulus: the observer remains within the region between the two horizons, and late time is reached by a correlated horizon-adjacent scaling rather than by taking the global $S^1$ to future infinity at fixed spatial position. Consequently there is no large spatial circle supporting a tower of parametrically light Schwarzian modes.} It is important to notice that although the geometry has the same form in both cases, their physical interpretation is completely different.

\medskip

We conclude with several open problems and future directions.

\paragraph{On the proliferation of dS spacetimes.} The Euclidean Nariai saddle has also been invoked in discussions of dS proliferation~\cite{Bousso:1998bn, Bousso:1999ms, Niemeyer:2000nq}. It was argued in the original Ginsparg-Perry picture~\cite{Ginsparg:1982rs} that the characteristic waiting time between nucleation events is parametrically longer than the evaporation time of the near-Nariai black hole, the ambient dS space effectively \emph{perdures}. 

The dS proliferation refers to a stronger global claim that sufficiently inhomogeneous perturbations along the $S^1$ of the nucleated $S^1\times S^2$ handle could create multiple pairs of apparent horizons. If the black-hole horizons then evaporate completely, the handle could pinch into several disconnected $S^3$ components, suggesting a branching global structure with multiple daughter dS regions.

Our analysis puts strong constraints on this scenario. The controlled linearized near-Nariai perturbation space is exhausted by the solution found in Section \ref{sec:anatomyNariai}, higher inhomogeneous perturbations are not freely specifiable $O(\epsilon)$ on-shell data. Trying to support such modes through one-loop state data would require parametrically large state dependence, outside the controlled semiclassical regime. Similarly, a no-boundary or complex-saddle construction must still solve the complexified equations of motion, hence it cannot generate new on-shell harmonics absent from the constraint-complete system.

If such higher-harmonic fragmentation exists, it must arise beyond linear order, from degrees of freedom outside the 2d model, or from genuinely quantum gravity effects. However, the sharpest tension would arise if one adopts a holographic viewpoint for dS space, in which the finite Gibbons-Hawking entropy $S_{\rm dS}$ reflects a finite number of microstates accessible to a single static patch. With this interpretation, a literal proliferation into $n$ disconnected asymptotic dS regions would suggest a tensor-product growth of Hilbert space dimension $\sim e^{nS_{\rm dS}}$, which is difficult to reconcile with a finite-entropy description of a single patch.

\paragraph{Charged dS black holes.} The charged case is an important extension because Reissner-Nordstr\"om-dS black holes have a richer extremal structure than the neutral SdS family. Besides the Nariai limit, where the outer black-hole horizon coincides with the cosmological horizon and the near-horizon geometry is of dS$_2\times S^2$ type, there are cold and ultracold limits associated with the degeneracy of the inner and outer black-hole horizons, or of all three horizons. These different limits need not have the same quantum dynamics. In particular, the cold near-extremal branch contains an AdS$_2$ throat and is therefore much closer to the usual near-extremal charged black-hole problem, where Schwarzian modes and large low-temperature quantum corrections can become important \cite{Iliesiu:2020qvm,Iliesiu:2022onk,Brown:2024ajk}. For related discussions of near-extremal black holes in dS, see \cite{Bhattacharjee:2025wfv, Aalsma:2025lcb}. By contrast, the charged Nariai branch is more directly connected to the nucleated black holes and proliferation scenarios~\cite{Bousso:1996pn,Bousso:1999ms}. This branch should also be distinguished from the lukewarm charged black holes that can realize the observer prescription discussed in Section~\ref{sec:Nariai-GPI-decay}: the lukewarm geometry already selects a thermal circle, whereas the exact Nariai geometry restores the enhanced $SO(3)\times SO(D-1)$ symmetry and therefore still requires an observer insertion, as explained in Appendix~\ref{sec:symmetries}. Thus the charged problem involves both the observer role of the lukewarm branch and
the nucleation/evaporation dynamics of the charged Nariai branch. It would therefore be valuable to revisit charged black-hole nucleation, proliferation, and evaporation in a unified framework. 

\paragraph{Anti-evaporation in modified gravity.} An analogous anti-evaporation behavior has been reported at the \emph{classical} level in certain modified-gravity theories (notably $f(R)$ models), where the Nariai solution admits a genuine \emph{classical instability} whose growing mode drives an increase of the apparent horizon radius~\cite{Nojiri:2013su,Addazi:2016prb} (see also~\cite{Nojiri:2014jqa} for the case of extremal Reissner-Nordstr\"om and~\cite{Addazi:2017cim} for a review in various modified gravity scenarios).

This is conceptually different from the one-loop story analyzed in this work: here the exponentially dominated channel that contaminates horizon tracing can be traced to an inadmissible (non-Hadamard) Lorentzian state choice in a sector tied to the $\mathrm{SL}(2,\mathbb{R})$ gauge orbit of the near-Nariai throat. By contrast, in $f(R)$ gravity ``anti-evaporation" is the classical evolution along an unstable mode of the modified field equations, and the key question is not the imposition of a Hadamard state but whether this unstable branch is realized for black holes produced by a well-defined nucleation/initial-data prescription, and how it competes with (or modifies) ordinary Hawking emission once quantum fields are included.

Existing attempts to include Hawking radiation in the $f(R)$ setting are largely kinematical~\cite{Addazi:2016prb}: rather than computing a renormalized $\langle T_{ab}\rangle$ in a horizon-smooth state on the dynamical (anti-evaporating) background and solving the coupled semiclassical equations, they argue that sufficiently rapid anti-evaporation can render the would-be emitting marginally trapped surface space-like on a timescale short compared to the effective emission time, thereby suppressing Hawking quanta. While suggestive, the domain of validity of this argument in a two-horizon, time-dependent geometry remains unclear. A systematic resolution therefore seems to require a genuinely semiclassical treatment.

\paragraph{Primordial black holes.} Our result also affects speculative applications to PBHs. Black-hole nucleation during inflation is one of the major physical motivations for treating the Nariai saddle as more than a formal Euclidean solution. As argued in~\cite{Bousso:1995cc, Bousso:1996au}, the nucleation rate is usually exponentially
suppressed, so an appreciable abundance would require either a sufficiently large inflationary Hubble scale (large effective $\Lambda$) or additional model-dependent enhancement. 

However, the scalar field setting discussed in Section~\ref{sec:HMplusNariai} suggests that this conclusion need not be universal: the competition between Hawking-Moss evolution and Nariai black-hole nucleation depends on the scalar potential, and in some regimes Nariai nucleation can become the dominant fluctuation channel. This is particularly interesting because PBHs produced sufficiently early during inflation are usually expected to be exponentially diluted by the subsequent expansion, unless the nucleation rate is large enough to compensate for this dilution.

This is also why the anti-evaporation scenario was potentially interesting: if the anti-evaporating branch were physical, near-Nariai black holes produced during inflation could survive much longer than in the standard Hawking picture, possibly making their late-time abundance sensitive to particle-physics or supersymmetric matter sectors~\cite{Elizalde:1999dw, Nojiri:1999vv}. Our result rules out this particular Einstein-gravity one-loop mechanism, where the anti-evaporation cannot be used to stabilize inflationary PBHs. This conclusion does not, however, exclude genuinely different mechanisms in modified gravity discussed above~\cite{Dialektopoulos:2017pgo}.  Whether such a mechanism can produce long-lived PBHs is a separate open problem, requiring a controlled nucleation prescription and a semiclassical treatment of Hawking emission in the modified theory.

\paragraph{Acknowledgements} It is a pleasure to thank Yiming Chen, Kristan Jensen, Mukund Rangamani, and Zhenbin Yang for interesting and useful discussions. This work was supported by the DOE Early Career Award DE-SC0026287 and by the University of Washington.

\begin{appendix}

\section{Why the Nariai black hole does not act as an observer}
\label{sec:symmetries}

The issue in using the Euclidean path integral of closed universes to extract information about the dS static patch has a kinematic origin: the Euclidean saddle often has a larger isometry group than the Lorentzian static patch whose states it is meant to count.
This mismatch indicates that the Euclidean saddle is not equivalent to the static patch physics by itself. Introducing an observer, or any equivalent defect/operator insertion, breaks the Euclidean symmetry down to the subgroup that preserves the chosen static patch. Thus including them in Euclidean path integral bridges to the static patch physics. In the rest of this Appendix we explain this fact, and use it to argue why the lukewarm black hole in dS (one whose event horizon is in equilibrium with the cosmological one) acts as an observer, while the Nariai black hole (when further the black hole and cosmological horizon area match) does not.

Lorentzian de Sitter space dS$_D$ has isometry group $SO(D,1)$. Choosing a static observer selects a causal diamond and reduces the manifest symmetry to
$$
    SO(1,1)\times SO(D-1),
$$
where $SO(1,1)$ is generated by translations of static patch time and $SO(D-1)$ rotates the transverse $S^{D-2}$. The remaining de Sitter generators move the observer to a different geodesic, and hence to a different static patch. The breaking
$$
    SO(D,1)
    \longrightarrow
    SO(1,1)\times SO(D-1)
$$
is the kinematic consequence of choosing an observer.

The same static-patch symmetry persists for SdS black holes. Although the black hole geometry is not maximally symmetric as dS, it preserves the static time translation and the rotations of the angular sphere. Hence the Lorentzian isometry group is again
$$
    SO(1,1)\times SO(D-1).
$$
This remains true throughout the SdS family, from small black holes up to the approach to the Nariai limit.

At the Nariai point the black-hole and cosmological horizons coincide, and the near-horizon geometry becomes
$
    {\rm dS}_2\times S^{D-2}.
$
The local symmetry is therefore enhanced to
$$
    SO(2,1)\times SO(D-1).
$$
However, this enhancement refers to the full dS$_2$ factor. A chosen static observer in the dS$_2$ factor still preserves only
$$
    SO(1,1)\times SO(D-1).
$$
Like empty dS, at the Nariai point there is a distinction between the symmetry of the full near-horizon geometry and the symmetry of a particular static patch.

We now compare this with the corresponding Euclidean saddles. The Euclidean continuation of global dS is the round sphere $S^D$, whose isometry group is
$$
    SO(D+1).
$$
This is the Euclidean real form of the Lorentzian de Sitter group $SO(D,1)$. By contrast, the Euclidean continuation of a single static patch preserves only
$$
    SO(2)\times SO(D-1),
$$
where the $SO(2)$ factor is the Euclidean thermal circle.

For a generic SdS black hole, the black-hole and cosmological horizons have different temperatures. Therefore a generic SdS black hole does not give a smooth compact Euclidean saddle without a conical defect. There are special exceptions, such as magnetically charged lukewarm black holes with $M=|Q|$, whose black hole and cosmological temperatures agree. In such cases one obtains a smooth Euclidean geometry of the warped-product form
$$
    \d s_E^2
    =
    \d s^2_{M_2}
    +
    r(x)^2 \d\Omega_{D-2}^2 ,
$$
where $M_2$ is the Euclidean $(\tau,x)$ space. Topologically $M_2$ is a two-sphere when both endpoints are smooth, but the radius of the transverse $S^{D-2}$ varies over $M_2$. Therefore the geometry is generically not a direct product. Its isometry group is
$$
    SO(2)\times SO(D-1),
$$
which agrees with the Euclidean continuation of the Lorentzian static-patch symmetry.

At the Nariai point, however, the Euclidean saddle becomes the direct product
$
    S^2\times S^{D-2},
$
with connected isometry group
$$
    SO(3)\times SO(D-1).
$$
This is the Euclidean counterpart of the Lorentzian near-horizon symmetry enhancement
$$
    SO(2,1)\times SO(D-1)
    \quad\longrightarrow\quad
    SO(3)\times SO(D-1).
$$
Again, this is larger than the symmetry of a chosen static patch.

For empty de Sitter and for Nariai, the corresponding Euclidean saddle has a larger symmetry than the Euclidean continuation of a single static patch isometries. Therefore, the necessary condition for the Euclidean path integral to be interpreted as computing a static-patch quantity is to break the excessive Euclidean symmetry.\footnote{We are currently exploring other ways to understand this and how it is related to the Hartle-Hawking wavefunction. } One natural choice is to include the Euclidean continuation of the static observer's worldline. In empty de Sitter this is modeled by a massive particle following a great circle on $S^D$, and it breaks \cite{Maldacena:2024spf}
$$
    SO(D+1)
    \longrightarrow
    SO(2)\times SO(D-1),
$$
In Nariai the observer follows a great circle on $S^2$ and for simplicity we consider it to be smeared over $S^{D-2}$. It breaks
$$
    SO(3)\times SO(D-1)
    \longrightarrow
    SO(2)\times SO(D-1).
$$
If the observer is localized on the transverse sphere then the rotational symmetry is further reduced to $SO(D-2)$. The observer thus selects the static patch whose Hilbert space or thermal trace the Euclidean path integral is meant to compute.

The lukewarm black hole is different. Because the black-hole geometry already singles out the Euclidean time circle and has nonconstant transverse-sphere radius over $M_2$, its Euclidean saddle has only
$$
    SO(2)\times SO(D-1)
$$
from the start. This already matches the Euclidean static-patch symmetry. In this sense the black hole itself plays the symmetry-breaking role of the observer insertion~\cite{Chen:2025jqm}. This distinction explains why the lukewarm black hole cannot be interpreted as an observer after it merges with the Nariai branch.

Finally, even within the branch of Nariai black holes of charge $Q$ there are two distinct behaviors. Consider conventions where a magnetic black hole has $F = Q \upepsilon_{D-2}$ and the action is $\frac{1}{16 \pi G_N} \int (R-2\Lambda - F^2)$. In 4D, when $Q<\sqrt{3/16\Lambda}$   the Nariai black hole is unstable both thermodynamically and in the Ginsparg-Perry sense of having an off-shell negative mode. The Nariai black hole with $Q>\sqrt{3/16\Lambda}$ is, on the other hand, thermodynamically stable~\cite{Bousso:1999ms}. It will not evaporate but at the same time it will not be produced by nucleation. The family of Nariai black holes terminates at the maximal charge $Q=\sqrt{1/2\Lambda}$.

\section{Observer trajectories on  \texorpdfstring{$S^2\times S^{D-2}$}{S2xS2}}
\label{app:ObserverandBetaIntegral}

We model the observer as a massive particle propagating on the Euclidean Nariai saddle $S^2\times S^{D-2}$ \cite{Maldacena:2024spf, Shi:2025amq, Ivo:2025yek}, with metric
\begin{equation}
\d s_E^2 = r_1^2\Big(\d\chi^2+\sin^2\chi\,\d\tau^2\Big) + r_2^2\Big(\d\theta^2+\sin^2\theta\,\d\phi^2+\cos^2\theta\,\d \Omega^2_{D-4} \Big),
\end{equation}
and periodicities $\tau\sim\tau+2\pi, ~ \phi\sim\phi+2\pi$. The first $S^2$ is the Euclidean continuation of the $\mathrm{dS}_2$ factor of Lorentzian Nariai, while the second $S^{D-2}$ is the spectator sphere.

In the semiclassical regime of large mass, the worldline path integral is dominated by closed geodesics, with action given by the proper length. Since the geometry is a direct product, a closed geodesic is obtained by choosing a closed geodesic on each factor and traversing the two factors with constant speeds. Up to the action of the isometry group, a convenient representative is
\begin{equation}
\chi=\frac{\pi}{2},\qquad \tau=n_1\sigma, \qquad \theta=\frac{\pi}{2}, \qquad \phi=n_2\sigma, \qquad
\sigma\sim\sigma+2\pi,
\end{equation}
with $n_1,n_2\in\mathbb Z$. The integers label the winding numbers on the two sphere factors. The sector $(0,0)$ is the constant loop and therefore corresponds to the absence of an observer, while each nontrivial sector $(n_1,n_2)\neq(0,0)$ describes a closed worldline which winds $|n_1|$ times on $S^2$ and $|n_2|$ times on $S^{D-2}$.

There are redundancies in this labeling. First, changing the sign of $n_1$ or $n_2$ reverses the orientation of the trajectory on the corresponding sphere, so $(n_1,n_2)$ and $(\pm n_1,\pm n_2)$ have the same classical action and the same local fluctuation spectrum. In practice one can include the corresponding degeneracy factor and restrict to $n_1,n_2\ge 0$. 
A further identification arises from Lorentzian picture. The saddles $(n_1,n_2)$ and $(mn_1,mn_2)$ with $m>1$ continue to the same geometric trajectory of the observer, since the common factor $m$ can be absorbed into a reparametrization of the noncompact worldline parameter. In particular, all $(n_1,0)$ describe the same timelike static observer at the center of the Nariai static patch. They are nevertheless distinct Euclidean saddles and have different action and fluctuation spectrum, because the Euclidean worldline is compact and they differ by the number of windings around the compact Euclidean circle. 

Another issue arises when we examine these Euclidean closed geodesics in the Lorentzian signature. After analytic continuation $\tau\to \i t$, the tangent
vector becomes
$$
u \propto n_1\,\partial_t+n_2\,\partial_\phi ,
$$
where $\phi$ denotes the affine angle along the chosen great circle in $S^{D-2}$. Its norm is
$$
u^2=-\,r_1^2 n_1^2+r_2^2 n_2^2 .
$$
Thus the Lorentzian avatar is timelike, null, or spacelike according as
$$
r_1|n_1|>r_2|n_2|,\qquad
r_1|n_1|=r_2|n_2|,\qquad
r_1|n_1|<r_2|n_2|,
$$
respectively. Thus, only saddles satisfying $r_1|n_1|>r_2|n_2|$ represent a realistic timelike observer.

For now we analyze the spectrum of general $(n_1, n_2)$ saddles. From the proper-length action
\begin{equation}
    I[x]=m\int_0^{2\pi} \d\sigma\,\sqrt{g_{\mu\nu}(x)\,\dot x^\mu \dot x^\nu},
\end{equation}
the on-shell worldline action for the $(n_1,n_2)$ saddle is given by
\begin{equation}\label{eq:I-n1n2-rewrite}
    I_{(n_1,n_2)} = m\oint \d s_E = 2\pi m \sqrt{r_1^2n_1^2+r_2^2n_2^2}.
\end{equation}
Expanding around classical solution, the transverse fluctuations on $S^2$ is $\chi(\sigma)=\frac{\pi}{2}+\eta(\sigma)$ while on $S^{D-2}$ we use local Cartesian coordinates $\vec y=(y^1,\dots,y^{D-3})$ transverse to the chosen great circle.\footnote{Writing the unit $S^{D-2}$ as
\begin{equation}
    X=
    \Big(
    \sqrt{1-\vec y^{\,2}}\cos\phi,\,
    \sqrt{1-\vec y^{\,2}}\sin\phi,\,
    y^1,\dots,y^{D-3}
    \Big),
\end{equation}
the metric becomes, to quadratic order,
\begin{equation}    \d\Omega_{D-2}^2
    =
    \d\vec y\cdot \d\vec y
    +
    (1-\vec y^{\,2})\,\d\phi^2
    +O(\text{cubic}).
\end{equation}
Hence along $\phi=n_2\sigma$,
\begin{equation}
    r_2^2\,\left(\frac{\d\Omega_{D-2}}{\d \sigma}\right)^2
    =
    r_2^2\Big[n_2^2+{\vec y}'^{\,2}-n_2^2\vec y^{\,2}\Big]+O(\text{cubic}).
\end{equation}}
The action expands as
\begin{equation}
    \delta^2I =\frac{m}{2\sqrt{r_1^2n_1^2+r_2^2n_2^2}}\int_0^{2\pi}\d\sigma\, \Big[r_1^2\big((\eta')^2-n_1^2\eta^2\big)+r_2^2\big(\vec y'^{\,2}-n_2^2\vec y^{\,2}\big)\Big].
\end{equation}
On Fourier modes $e^{\i k\sigma}$, $k\in\mathbb Z$, the eigenvalues are
\begin{equation}
    \lambda_k^{(1)}
    =
    \frac{m r_1^2}{\sqrt{r_1^2n_1^2+r_2^2n_2^2}}(k^2-n_1^2),
    \qquad
    \lambda_k^{(2)}
    =
    \frac{m r_2^2}{\sqrt{r_1^2n_1^2+r_2^2n_2^2}}(k^2-n_2^2),
\end{equation}
with multiplicities $1$ and $D-3$, respectively. We see that this action has some negative modes corresponding to $|k|<|n_{1,2}|$. In total we find
\begin{equation}
N_{\rm neg}(n_1,n_2)=
\begin{cases}
(2|n_1|-1)+(D-3)(2|n_2|-1), & n_1\neq 0,\; n_2\neq 0,\\[4pt]
(D-3)(2|n_2|-1), & n_1=0,\; n_2\neq 0,\\[4pt]
2|n_1|-1, & n_1\neq 0,\; n_2=0,\\[4pt]
0, & n_1=n_2=0.
\end{cases}
\label{eq:NegativeModesParticle}
\end{equation}
Therefore the phase for the observer sector is given by
\begin{equation}
    \frac{Z_{\rm particle}}{|Z_{\rm particle}|}=(-\ii)^{N_{\rm Neg}}
\end{equation}
There are also zero modes for $|k|=|n_{1,2}|$. They generate rotations of the great circle within the equator plane and correspond to the isometries spontaneously broken by the choice of path. They give finite factors in the path integral since the subgroups of the isometries are all compact.

In particular, the first non-trivial observer saddle $(1,0)$ has one negative mode and acquires one factor of $-\i$ in the one-loop determinant.

\section{Greybody factors in the Nariai limit}
\label{app:greybody}
In this appendix we compute the greybody factor for a probe scalar field in the Nariai geometry and compare the result with the zero-frequency limit discussed in \cite{Crispino:2013pya}. Since some of the formulas in \cite{Crispino:2013pya} are presented in a regime corresponding to small SdS black holes, it is useful to repeat the computation directly in the Nariai background, where the analysis simplifies considerably.

We begin with the Nariai metric written as
$
\d s^2 = \frac{1}{\Lambda}(\frac{-\d\t^2+\d\r^2}{\cosh^2 \r} + \d\Omega_2^2).
$
This describes the static patch with the black-hole horizon at $\r\to -\infty$ and the cosmological horizon at $\r\to +\infty$, as explained in Section~\ref{sec:anatomyNariai}. Consider a probe scalar field ${\sf f}$ of mass $m$, obeying
$
\nabla^2 {\sf f} - m^2 {\sf f} = 0.
$
Expanding in spherical harmonics and Fourier modes,
\begin{equation}
{\sf f}(\t,\r,\Omega) = \sum_{\omega, \ell,m} e^{-\i\omega \t}\,\tilde{{\sf f}}(\r)\,Y_{\ell m}(\Omega),
\end{equation}
the wave equation reduces to
\begin{equation}
\cosh^2 \r\,\tilde{{\sf f}}''(\r)+\omega^2\cosh^2 \r\,\tilde{{\sf f}}(\r)-\left(\frac{m^2}{\Lambda}+\ell(\ell+1)\right)\tilde{{\sf f}}(\r)=0.
\end{equation}
It is convenient to define
$
\mu_\ell^2 \equiv \frac{m^2}{\Lambda}+\ell(\ell+1),
$
and
$
\Delta_\ell \equiv -\frac12+\sqrt{\frac14-\mu_\ell^2}.
$
The general solution is then
\begin{equation}
\tilde{{\sf f}}(\r)
=
C_1\,P_{\Delta_\ell}^{\,\i\omega}(\tanh \r)
+
C_2\,P_{\Delta_\ell}^{\,-\i\omega}(\tanh \r).
\end{equation}

We now impose the standard scattering boundary conditions. Near the black-hole horizon $\r\to -\infty$, we require a purely ingoing transmitted wave,
\begin{equation}
\tilde{{\sf f}}(\r)\sim A_{\rm tr}\,e^{-\i\omega \r},
\qquad \r\to -\infty,
\end{equation}
while near the cosmological horizon $\r\to +\infty$ we write
\begin{equation}
\tilde{{\sf f}}(\r)\sim A_{\rm in}\,e^{-\i\omega \r}+A_{\rm out}\,e^{\i\omega \r},
\qquad \r\to +\infty.
\end{equation}
The greybody factor is
\begin{equation}
\Gamma_\omega=\left|\frac{A_{\rm tr}}{A_{\rm in}}\right|^2.
\end{equation}

Expanding the solution near $\r\to -\infty$, one finds
\begin{align}
\tilde{{\sf f}}
&\sim
e^{\i\omega \r}
\frac{\i\pi\,\mathrm{csch}(\pi\omega)}{\Gamma(1+\i\omega)}
\left(
-\frac{C_2\sin(\pi\Delta_\ell)}{\pi}
+\frac{C_1}{\Gamma(-\Delta_\ell-\i\omega)\Gamma(1+\Delta_\ell-\i\omega)}
\right)
\nonumber\\
&\quad
+
e^{-\i\omega \r}
\frac{\mathrm{csch}(\pi\omega)\big(\pi C_2-C_1\sin(\pi\Delta_\ell)\Gamma(\i\omega-\Delta_\ell)\Gamma(1+\Delta_\ell+\i\omega)\big)}
{\omega\,\Gamma(-\i\omega)\Gamma(\i\omega-\Delta_\ell)\Gamma(1+\Delta_\ell+\i\omega)}.
\end{align}
The absence of an outgoing mode at the black-hole horizon implies that the coefficient of $e^{\i\omega \r}$ must vanish. This fixes
\begin{equation}
C_2
=
C_1\,
\frac{\pi\,\csc(\pi\Delta_\ell)}
{\Gamma(-\Delta_\ell-\i\omega)\Gamma(1+\Delta_\ell-\i\omega)}.
\end{equation}

Choosing $C_1=1$ as a normalization, the near-horizon behavior becomes
\begin{equation}
\tilde{{\sf f}}(\r)\sim A_{\rm tr}\,e^{-\i\omega \r},
\qquad
A_{\rm tr}
=
\frac{\csc(\pi\Delta_\ell)\sinh^2(\pi\omega)\Gamma(\i\omega)}{\pi},
\qquad \r\to -\infty.
\end{equation}
Similarly, expanding near the cosmological horizon gives
$
\tilde{{\sf f}}(\r)\sim
A_{\rm in}\,e^{-\i\omega \r}
+
A_{\rm out}\,e^{\i\omega \r},
$
with
\begin{equation}
A_{\rm in}
=
\frac{\pi\,\csc(\pi\Delta_\ell)}
{\Gamma(1+\i\omega)\Gamma(-\Delta_\ell-\i\omega)\Gamma(1+\Delta_\ell-\i\omega)},
\qquad
A_{\rm out}
=
\frac{1}{\Gamma(1-\i\omega)}.
\end{equation}

Combining these expressions, the greybody factor is
\begin{equation}
\Gamma_\omega
=
\frac{2\sinh^2(\pi\omega)}
{\cos\big(\pi\sqrt{1-4\mu_\ell^2}\big)+\cosh(2\pi\omega)},
\qquad
\mu_\ell^2=\frac{m^2}{\Lambda}+\ell(\ell+1).
\end{equation}
This has the familiar form of an SL(2,$\mathbb R$) Plancherel measure, suggesting there is a more direct derivation based on the enhanced symmetry of the Nariai geometry.

It is useful to consider a few limits. For either a massive scalar, or a higher spherical harmonic of a massless scalar, the low-frequency behavior is
\begin{equation}
\Gamma_\omega
=
\frac{2\pi^2}{1+\cos\big(\pi\sqrt{1-4\mu_\ell^2}\big)}\,\omega^2
+O(\omega^4).
\end{equation}
Thus the greybody factor vanishes at low frequency.

By contrast, for the $\ell=0$ mode of a massless scalar one has $\mu_0^2=0$, and therefore
\begin{equation}
\Gamma_\omega
=
\frac{2\sinh^2(\pi\omega)}{\cosh(2\pi\omega)-1}
=1.
\end{equation}
This is the expected result: in two dimensions the left- and right-moving modes of a massless scalar decouple, so the s-wave experiences no potential barrier. In particular, the low-frequency limit remains finite. Relative to the $s$-wave massless case, the emission of massive modes or higher harmonics is therefore suppressed by
\begin{equation}
\frac{\Gamma_\omega(m\neq 0 \ \text{or}\ \ell>0)}{\Gamma_\omega(m=0,\ell=0)}
\sim \omega^2.
\end{equation}

Finally, according to \cite{Crispino:2013pya}, the zero-frequency greybody factor for a massless scalar with $\ell=0$ in SdS is
\begin{equation}
\Gamma_0
=
\frac{4r_c^2r_b^2}{(r_c^2+r_b^2)^2}.
\end{equation}
In the Nariai limit one has $r_c=r_b=r_N$, and therefore
\begin{equation}
\Gamma_0=1,
\end{equation}
in agreement with the exact result above.

One may similarly ask whether, when both horizons emit, there is a distinct greybody factor associated with each horizon, multiplying the corresponding Planckian spectrum. In the Nariai limit this turns out not to be an additional complication: the absorption coefficient associated with the cosmological horizon is equal to the one found above for the black-hole horizon. This is ultimately a consequence of the fact that in the exact Nariai geometry, dS$_2\times S^2$ in the static patch, there is no intrinsic distinction between the two horizons. To meaningfully distinguish them one must move slightly away from the Nariai limit, or equivalently leave the exact static patch description.

To see this explicitly, consider the behavior of the two independent solutions near the cosmological horizon. Expanding the associated Legendre functions for $\r\to +\infty$, one finds that $P_{\Delta_\ell}^{\,\i\omega}(\tanh\r)$ is purely incoming, while $P_{\Delta_\ell}^{\,-\i\omega}(\tanh\r)$ is purely outgoing towards the cosmological horizon. This is in contrast with the behavior near the black-hole horizon, where each solution contains both components. It is therefore sufficient to consider, up to an overall normalization, the mode
\begin{equation}
\tilde{{\sf f}}(\r)=P_{\Delta_\ell}^{\,\i\omega}(\tanh\r).
\end{equation}
Expanding this solution near the two horizons gives
\begin{align}
\tilde{{\sf f}}(\r)
&\sim
\underbrace{\frac{1}{\Gamma(1-\i\omega)}}_{A_{\rm tr}}\,e^{\i\omega \r},
\qquad \r\to +\infty,
\\
\tilde{{\sf f}}(\r)
&\sim
\underbrace{
\frac{\pi\,e^{\i\omega \r}\,\mathrm{csch}(\pi\omega)}
{\omega\,\Gamma(\i\omega)\Gamma(-\Delta_\ell-\i\omega)\Gamma(1+\Delta_\ell-\i\omega)}
}_{A_{\rm in}}
-\frac{\sin(\pi\Delta_\ell)\,e^{-\i\omega \r}\,\mathrm{csch}(\pi\omega)}
{\omega\,\Gamma(-\i\omega)},
\qquad \r\to -\infty.
\end{align}
The absorption coefficient associated with the cosmological horizon is then obtained from the ratio of the transmitted and incoming amplitudes. One finds
\begin{equation}
\Gamma_\omega
=
\frac{2\sinh^2(\pi\omega)}
{\cos\big(\pi\sqrt{1-4\mu_\ell^2}\big)+\cosh(2\pi\omega)},
\qquad
\mu_\ell^2=\frac{m^2}{\Lambda}+\ell(\ell+1).
\end{equation}
This is exactly the same expression obtained above for the black-hole horizon. Thus, in the Nariai limit, the effect of the greybody factor on the total emission rate is simply an overall multiplicative factor which to leading order is one for $s$-waves. Compared to the $s$-wave, the full radiation spectrum for massive modes or higher harmonics is suppressed by an overall factor of $\omega^2$ at low frequency.

\section{Fermions in the presence of a charged dS black hole}
\label{app:fermions}

In the main text, we made the simplifying assumption that the number of massless fields in the 2d dilaton-gravity description was large $N\gg1$. If we discuss scalar fields, this requires a corresponding large number of massless fields in higher dimensions such that the 2d fields are their $s$-wave modes, which seems unrealistic. 

It is possible to construct a more realistic model that avoids this shortcoming. A solution proposed in \cite{Maldacena:2018gjk} was to consider magnetically charged black holes in the presence of electrically charged fermions. For a macroscopic black hole with a large magnetic charge, there will be a large number of lowest Landau levels which appear as massless fermions in 2d.   

\smallskip

Consider Einstein gravity with $\Lambda>0$ coupled to a $\U(1)$ Maxwell field $F_{\mu\nu}$, together with a charged black hole with magnetic charge $F = P \epsilon_{S^2}$, where $\epsilon_{S^2}$ is the $S^2$ volume element. The geometry has the same form as SdS but with a redshift factor 
\beq
f(r) = 1 - \frac{2m}{r} + \frac{P^2}{4 r^2} - \frac{\Lambda r^2}{3}.
\eeq
The black-hole and cosmological horizons have the same radius $r_N$ when $f(r_N)=f'(r_N)=0$ which leads to
\beq\label{eq:RNFP}
4 r_N^2 (1-\Lambda r_N^2) = P^2.
\eeq
The geometry between the two horizons in this regime becomes 
\beq
\d s^2 = \frac{1}{\Lambda_1}  \frac{-\d \t^2 + \d \r^2}{\cosh^2 \r} + \frac{1}{\Lambda_2} \d \Omega_2^2,
\eeq
where $\Lambda_2 = 1/r_N^2$ is a function of $P$ obtained by solving \eqref{eq:RNFP} and the field equations imply that $\Lambda_1 + \Lambda_2 = 2\Lambda$, from which we can extract $\Lambda_1$ and therefore the dS$_2$ radius as a function of $\Lambda$ and $P$. See \cite{Bousso:1996pn} for a thorough study of the space of solutions with magnetic charge.

\smallskip

Other than the difference between the curvature of the dS$_2$ and $S^2$ factors the solution close to the Nariai limit is structurally the same as the neutral case. We can again consider small deviations from Nariai and formulate the 4d problem as a 2d dilaton-gravity theory coupled to matter. In order to do this we still write 
\be \label{eq:spherical_ansatz_gauged_app}
\d s^2_{4\mathrm{D}}=\frac{1}{\Lambda}\left[\Phi^{-1/2}(x) \, g_{ab}(x) \d x^a \d x^b + \Phi(x) \d \Omega_2^2 \right],
\ee
where for simplicity we turned off the $\SU(2)$ gauge fields. Consider an electrically-charged fermion $\Psi$ in 4d of charge $q$. Its action is
\beq
S = \i \int \d^4 x\,\sqrt{-g^{(4)}}\, \bar{\Psi}  \Gamma^\mu D_\mu \Psi
\eeq
where the covariant derivative is $D_\mu=\nabla_\mu - \i q A_\mu$ and $A_\mu = - P \cos \theta \d \varphi$ is the potential corresponding to the magnetic charge $P$. To rewrite the 4d fermion in terms of its 2d Landau levels we can write a representation of the 4d Clifford algebra $\Gamma^a= \gamma^a \otimes I_2$ and $\Gamma^i = \gamma^3 \otimes \tilde{\gamma}^i$, where $a=\t,\r$ and $i=\theta,\varphi$. $\gamma^a$ are the $1+1$ gamma matrices and $\tilde{\gamma}^i$ the Euclidean 2d matrices. Before integrating over the sphere, redefine $\Psi =\Lambda^{3/2} \Phi^{-3/8} \psi $. The 4d action can then be written as
\beq
S = \i \int \d^4 x \, \sqrt{-g} \sqrt{g_{S^2}}\, \bar{\psi} (\slashed{\nabla}_x+ e^{3 \phi/2} \gamma_3 \otimes \slashed{D}_{y})\psi
\eeq
where $\nabla_x$ is the 2d derivative along $x^a$ in the metric $g_{ab}$, and $D_y$ is the covariant derivative including the $\U(1)$ contribution. We can expand $\psi(x,y)$ in spin harmonics $\chi(y)$ on the transverse sphere, such that
$\i \slashed{D} \chi(y) = \lambda \chi(y)$. The mass of the 2d fermion is therefore proportional to the eigenvalue of the $S^2$ harmonic. Since we are after massless modes on dS$_2$ we can restrict to the zero modes
\beq
\psi(x,y) = \sum_{m=1}^{N_f} \psi_m(x) \otimes \chi_m(y) + \ldots,~~~~N_f=|2 q P|
\eeq
where $\chi_m$ are the $N_f$ zero-modes of $\i\slashed{D}$ and the dots represent contributions from higher harmonics with $\lambda \neq 0$. To derive that the number of zero-modes is $|2 q P|$ one can use the Atiyah-Bott theorem, or explicitly by constructing the harmonics, see \cite{Maldacena:2018gjk}. The resulting matter sector of the 2d theory is
\beq
S = \sum_{m=1}^{N_f} \i \int \d^2x \, \sqrt{-g} \, \bar{\psi}_m \slashed{\nabla} \psi_m + \ldots,
\eeq
with the dots representing massive modes. For a macroscopic black hole with $P\gg1$, there is correspondingly a large number of massless 2d matter fields $N_f \gg 1$ with a single massless charged fermion in 4d. Moreover, the action of the 2d fermions also exactly decouples from the dilaton. This is precisely the setup considered in Section~\ref{sec:Antievaporation}, up to only small changes in the coefficients of the effective action.

\section{Smooth Hadamard states and the Schwarzian transformation law}
\label{app:Schwarziantransform}

In Section~\ref{sec:thermalHawking} we derived the physical thermal Hawking flux directly
from the one-loop auxiliary-field stress tensor by imposing horizon smoothness in Kruskal
coordinates. The purpose of this appendix is to recast the same conclusion in a more
invariant language. The Schwarzian formulation is complementary but more powerful: it
makes the Hadamard-state criterion manifest, and shows directly which choices of state are compatible
with smoothness across the static-patch horizons.

At order $O(\epsilon\hbar)$, the state dependence in the one-loop effective description sits
in the homogeneous auxiliary-field mode $\eta_1$. In null coordinates $x^\pm$ one may
write
\be
\eta_1(x^\pm)=F(x^+)+G(x^-).
\ee
Under a large conformal reparametrization $x^\pm\mapsto y^\pm(x^\pm)$, the renormalized chiral stress tensor transforms with a Schwarzian shift,
\be \label{eq:Schwarzian_recap}
\langle T_{\pm\pm}(y^\pm)\rangle
=\Big(\frac{\d x^\pm}{\d y^\pm}\Big)^{\!2}\langle T_{\pm\pm}(x^\pm)\rangle
-\frac{N \hbar}{24\pi}\{x^\pm,y^\pm\},
\ee
In conformal gauge one may equivalently write (as confirmed by covariant point-splitting)
\be
\langle \Psi|T_{\pm\pm}(x^\pm)|\Psi\rangle
=\langle \Psi|:T_{\pm\pm}(x^\pm):|\Psi\rangle
-\frac{N\hbar}{12\pi}\big(\partial_\pm\rho\,\partial_\pm\rho-\partial_\pm^2\rho\big).
\ee
The second term is \emph{state-independent} and vanishes in flat space ($\rho\to0$).  If one chooses the
reference vacuum $|x^\pm\rangle$ defined by positive frequency in $x^\pm$, then
$\langle x^\pm|:T_{\pm\pm}(x^\pm):|x^\pm\rangle=0$ and the full covariant tensor equals the geometric
piece
\be \label{eq:tensorgeo_recap}
\langle x^\pm|T_{\pm\pm}(x^\pm)|x^\pm\rangle
=-\frac{N\hbar}{12\pi}\big(\partial_\pm\rho\,\partial_\pm\rho-\partial_\pm^2\rho\big).
\ee
Changing the physical state from $|x^\pm\rangle$ to $|\tilde x^\pm\rangle$ shifts only the normal-ordered
part and hence shifts the full covariant stress tensor by the \emph{state} Schwarzian:
\be
\langle \tilde x^\pm|T_{\pm\pm}(x^\pm)|\tilde x^\pm\rangle
=\langle x^\pm|T_{\pm\pm}(x^\pm)|x^\pm\rangle
-\frac{N\hbar}{24\pi}\{\tilde x^\pm,x^\pm\}.
\ee
Thus, if one moves along the $\mathrm{SL}(2,\mathbb{R})$ orbit of the geometry while \emph{holding fixed}
a vacuum notion defined by chosen positive-frequency coordinates, one has implicitly changed the
physical state; the resulting $\langle T_{ab}\rangle$ can then develop flux even though the underlying
geometric deformation is gauge. This is the key reason behind the anti-evaporation mode, as we have emphasized in Section~\ref{sec:antievaporation}. 

We now define the quantum state by choosing null coordinates, where $(u, v)$ are the Eddington-Finkelstein coordinates~\cite{Markovic:1991ua}
\be \label{eq:statecoords_app}
U=\int^u h_u(u')\,\d u',\qquad V=\int^v h_v(v')\,\d v',
\ee
and declaring positive frequency with respect to $(U,V)$. The corresponding state
dependence enters the stress tensor through
\be \label{eq:schwarzian_def_app}
\{U,u\}=\frac{h_u''}{h_u}-\frac{3}{2}\frac{h_u'^2}{h_u^2},
\qquad
\{V,v\}=\frac{h_v''}{h_v}-\frac{3}{2}\frac{h_v'^2}{h_v^2}.
\ee
A horizon-smooth Hadamard state is precisely one for which the state-defining coordinates
extend as \emph{regular locally inertial null coordinates} across each horizon patch. Concretely,
near $\mathcal H_b^+$ this means
\be \label{eq:asymp_affine_app}
U=f(\tilde U_b),
\qquad
f\in C^3 \text{ near }\tilde U_b=0,
\qquad
f'(0)\neq 0,
\ee
and similarly for $V$ near the other horizons, where $(\tilde{U}, \tilde{V})$ are the Kruskal patches defined in~\eqref{eq:KruskalBH} and~\eqref{eq:KruskalCH}. The condition $f'(0)\neq 0$ is local
invertibility (so the $U$-frequency notion is locally Minkowskian at the horizon), while $f\in C^3$ is the minimal regularity needed for the Schwarzian
and hence for $\langle T_{ab}\rangle$ to be a well-defined local observable.

\paragraph{Horizon smoothness implies constant Schwarzian tails.} The key technical fact is that a horizon-smooth state can contribute only a constant plus
exponentially decaying tails to the chiral stress tensor near each horizon. To see this,
consider $\mathcal H_b^+$, where $u\to+\infty$ and $\tilde U_b\to0$. Writing
$U(u)=f(\tilde U_b(u))$, the Schwarzian chain rule gives
\be \label{eq:chainrule_step_app}
\{U,u\}=(\partial_u\tilde U_b)^2\,\{f,\tilde U_b\}+\{\tilde U_b,u\}.
\ee
Since
\be
(\partial_u\tilde U_b)^2=\kappa_b^2 e^{-2\kappa_b u},
\qquad
\{\tilde U_b,u\}=-\frac{\kappa_b^2}{2},
\ee
and $\{f,\tilde U_b\}$ is finite at $\tilde U_b=0$ whenever $f\in C^3$ and $f'(0)\neq0$,
one obtains
\be \label{eq:schwarzian_affine_U}
\{U,u\}
=
-\frac{\kappa_b^2}{2}
+
O(e^{-2\kappa_b u}),
\qquad (u\to+\infty).
\ee
The same reasoning gives
\be \label{eq:schwarzian_affine_V}
\{V,v\}
=
-\frac{\kappa_c^2}{2}
+
O(e^{-2\kappa_c v}),
\qquad (v\to+\infty)
\ee
near the future cosmological horizon, with analogous statements on the past horizons.\footnote{For dilaton-coupled $s$-wave matter, the exact anomalous transformation law of the normal-ordered null stress tensor is not purely Schwarzian. In addition to the Schwarzian derivative, it contains local terms proportional to $(x^{\pm\prime\prime}/x^{\pm\prime})\,\partial_{y^\pm}\phi$ and $\ln[(dx^+/dy^+)(dx^-/dy^-)](\partial_{y^\pm}\phi)^2$ \cite{Fabbri:2003vy}. Equivalently, the quantum state is not encoded by chiral functions alone: in this theory the state data are described by functions $t_\pm(x^+,x^-)$ and $t(x^+,x^-)$ obeying modified conservation laws, and $t_\pm$ are no longer chiral \cite{Fabbri:2003vy}. Nevertheless, for the horizon-smoothness criterion used here, these extra dilaton terms are subleading. Near a smooth non-extremal horizon one has $\{U,u\}=-\kappa_h^2/2+O(e^{-2\kappa_h u})$, while in the near-Nariai throat the dilaton approaches a constant at the horizon, implying $\partial_u\phi=O(e^{-\kappa_h u})$. Hence the additional dilaton terms contribute only decaying tails, $(U''/U')\partial_u\phi=O(e^{-\kappa_h u})$ and $\ln(U'V')(\partial_u\phi)^2=O(u e^{-2\kappa_h u})$, and do not modify the leading constant term that enforces regularity in Kruskal components.}

Indeed, this is what we need to cancel the universal divergence induced by the
state-independent geometric term when converting to Kruskal coordinates.  Near a smooth
non-extremal horizon one has the general expression
\be
-\frac{N \hbar}{12\pi}\big(\partial_\pm\rho\,\partial_\pm\rho-\partial_\pm^2\rho\big)
\approx -\frac{N \hbar}{48\pi}\kappa_h^2,
\ee
with $\partial_u\rho\to-\kappa_h/2$, $\partial_v\rho\to+\kappa_h/2$ and $\partial_u^2\rho,\partial_v^2\rho\to0$.
To cancel the Jacobian divergence in Kruskal components, the Schwarzian must supply a \emph{leading}
constant term as dictated by~\eqref{eq:schwarzian_affine_U} and~\eqref{eq:schwarzian_affine_V}.

Equivalently, the state coordinates must be asymptotically affine functions of the horizon-regular
Kruskal null coordinates.  The minimal choice is
\be
U(u)=a+b\,\tilde U_b(u)=a-be^{-\kappa_b u},\qquad
V(v)=\bar a+\bar b\,\tilde V_c(v)=\bar a-\bar b\,e^{-\kappa_c v},
\ee
so that $h_u(u)=U'(u)=b\kappa_b e^{-\kappa_b u}$ and similarly for $h_v$. On the other hand, any attempt to engineer a Schwarzian with qualitatively different asymptotic behavior would necessarily require $f(\tilde{U}_b)$ to cease being asymptotically affine. That is, one may either have $f'(0)=0$ or a divergence in $f''$ or $f'''$ as $\tilde{U}_b \to 0$. Such a choice lies outside the horizon-smooth Hadamard class.

Thus horizon smoothness forces the Schwarzian to be a constant plus exponentially decaying
tails. In the auxiliary-field language of Section~\ref{sec:thermalHawking}, the same
state-dependent null stress tensor was encoded by
\be
\eta_1=F(u)+G(v),
\qquad
\langle T_{uu} \rangle_{\rm state}=\frac{\epsilon N\hbar}{4\pi}F''(u),
\qquad
\langle T_{vv} \rangle_{\rm state}=\frac{\epsilon N\hbar}{4\pi}G''(v).
\ee
Matching the auxiliary-field and Schwarzian descriptions shows that the asymptotic affinity
of the state-defining coordinates is equivalent to asymptotically constant $F''$ and $G''$,
and therefore to asymptotically quadratic $F$ and $G$. This is the invariant counterpart of
the explicit $F''$, $G''$ derivation given in Section~\ref{sec:thermalHawking}, which reproduces the thermal Hawking flux.

\end{appendix}
\bibliographystyle{utphys2}
{\bibliography{bibliography}{}}

@article{Marolf:2022ybi,
    author = "Marolf, Donald",
    title = "{Gravitational thermodynamics without the conformal factor problem: partition functions and Euclidean saddles from Lorentzian path integrals}",
    eprint = "2203.07421",
    archivePrefix = "arXiv",
    primaryClass = "hep-th",
    doi = "10.1007/JHEP07(2022)108",
    journal = "JHEP",
    volume = "07",
    pages = "108",
    year = "2022"
}

@article{Ivo:2025yek,
    author = "Ivo, Victor and Maldacena, Juan and Sun, Zimo",
    title = "{Physical instabilities and the phase of the Euclidean path integral}",
    eprint = "2504.00920",
    archivePrefix = "arXiv",
    primaryClass = "hep-th",
    month = "4",
    year = "2025"
}

@article{Crispino:2013pya,
    author = "Crispino, Lu{\'\i}s C. B. and Higuchi, Atsushi and Oliveira, Ednilton S. and Rocha, Jorge V.",
    title = "{Greybody factors for nonminimally coupled scalar fields in Schwarzschild{\textendash}de Sitter spacetime}",
    eprint = "1304.0467",
    archivePrefix = "arXiv",
    primaryClass = "gr-qc",
    doi = "10.1103/PhysRevD.87.104034",
    journal = "Phys. Rev. D",
    volume = "87",
    pages = "104034",
    year = "2013"
}

@article{Mariani:2025hee,
    author = "Mariani, Francesca and Toldo, Chiara",
    title = "{Gravitational dynamics of near-extreme Kerr (Anti-)de Sitter black holes}",
    eprint = "2505.02674",
    archivePrefix = "arXiv",
    primaryClass = "hep-th",
    doi = "10.1007/JHEP02(2026)052",
    journal = "JHEP",
    volume = "02",
    pages = "052",
    year = "2026"
}

@article{Luo:2026epp,
    author = "Luo, Shu and Pando Zayas, Leopoldo A.",
    title = "{Quantum-Corrected Evaporation and Absorption Cross-Section of Near-Extremal Rotating Black Holes}",
    eprint = "2601.06720",
    archivePrefix = "arXiv",
    primaryClass = "hep-th",
    reportNumber = "LITP-25-14",
    month = "1",
    year = "2026"
}

@article{Emparan:2025qqf,
    author = "Emparan, Roberto and Trezzi, Stefano",
    title = "{Quantum transparency of near-extremal black holes}",
    eprint = "2507.03398",
    archivePrefix = "arXiv",
    primaryClass = "hep-th",
    doi = "10.1007/JHEP10(2025)023",
    journal = "JHEP",
    volume = "10",
    pages = "023",
    year = "2025"
}

@article{Castro:2022cuo,
    author = "Castro, Alejandra and Mariani, Francesca and Toldo, Chiara",
    title = "{Near-extremal limits of de Sitter black holes}",
    eprint = "2212.14356",
    archivePrefix = "arXiv",
    primaryClass = "hep-th",
    doi = "10.1007/JHEP07(2023)131",
    journal = "JHEP",
    volume = "07",
    pages = "131",
    year = "2023"
}

@article{Gibbons:1976ue,
    author = "Gibbons, G. W. and Hawking, S. W.",
    title = "{Action Integrals and Partition Functions in Quantum Gravity}",
    reportNumber = "PRINT-76-0995 (CAMBRIDGE)",
    doi = "10.1103/PhysRevD.15.2752",
    journal = "Phys. Rev. D",
    volume = "15",
    pages = "2752--2756",
    year = "1977"
}

@article{Christensen:1977jc,
    author = "Christensen, S. M. and Fulling, S. A.",
    title = "{Trace Anomalies and the Hawking Effect}",
    doi = "10.1103/PhysRevD.15.2088",
    journal = "Phys. Rev. D",
    volume = "15",
    pages = "2088--2104",
    year = "1977"
}

@article{Kay:1988mu,
    author = "Kay, Bernard S. and Wald, Robert M.",
    title = "{Theorems on the Uniqueness and Thermal Properties of Stationary, Nonsingular, Quasifree States on Space-Times with a Bifurcate Killing Horizon}",
    reportNumber = "PRINT-88-0840 (CHICAGO)",
    doi = "10.1016/0370-1573(91)90015-E",
    journal = "Phys. Rept.",
    volume = "207",
    pages = "49--136",
    year = "1991"
}

@article{Radzikowski:1996pa,
    author = "Radzikowski, M. J.",
    title = "{Micro-local approach to the Hadamard condition in quantum field theory on curved space-time}",
    doi = "10.1007/BF02100096",
    journal = "Commun. Math. Phys.",
    volume = "179",
    pages = "529--553",
    year = "1996"
}

@article{Conti:2014uda,
      author         = "Conti, Gabriele and Hertog, Thomas",
      title          = "{Two wave functions and dS/CFT on S$^{1}\times$S$^{2}$}",
      journal        = "JHEP",
      volume         = "06",
      year           = "2015",
      pages          = "101",
      doi            = "10.1007/JHEP06(2015)101",
      eprint         = "1412.3728",
      archivePrefix  = "arXiv",
      primaryClass   = "hep-th",
      SLACcitation   = "%%CITATION = ARXIV:1412.3728;%%"
}

@article{Iliesiu:2020qvm,
      author         = "Iliesiu, Luca V. and Turiaci, Gustavo J.",
      title          = "{The statistical mechanics of near-extremal black holes}",
      year           = "2020",
      eprint         = "2003.02860",
      archivePrefix  = "arXiv",
      primaryClass   = "hep-th",
      SLACcitation   = "%%CITATION = ARXIV:2003.02860;%%"
}

@article{Cotler:2019nbi,
    author = "Cotler, Jordan and Jensen, Kristan and Maloney, Alexander",
    title = "{Low-dimensional de Sitter quantum gravity}",
    eprint = "1905.03780",
    archivePrefix = "arXiv",
    primaryClass = "hep-th",
    doi = "10.1007/JHEP06(2020)048",
    journal = "JHEP",
    volume = "06",
    pages = "048",
    year = "2020"
}

@article{Bousso:1996au,
	Archiveprefix = {arXiv},
	Author = {Bousso, Raphael and Hawking, Stephen W.},
	Doi = {10.1103/PhysRevD.54.6312},
	Eprint = {gr-qc/9606052},
	Journal = {Phys. Rev.},
	Pages = {6312-6322},
	Primaryclass = {gr-qc},
	Reportnumber = {DAMTP-R-96-33},
	Slaccitation = {%%CITATION = GR-QC/9606052;%%},
	Title = {{Pair creation of black holes during inflation}},
	Volume = {D54},
	Year = {1996}}

@article{Bousso:1995cc,
	Archiveprefix = {arXiv},
	Author = {Bousso, R. and Hawking, S. W.},
	Doi = {10.1103/PhysRevD.52.5659},
	Eprint = {gr-qc/9506047},
	Journal = {Phys. Rev.},
	Pages = {5659-5664},
	Primaryclass = {gr-qc},
	Reportnumber = {DAMTP-R-95-33},
	Slaccitation = {%%CITATION = GR-QC/9506047;%%},
	Title = {{The probability for primordial black holes}},
	Volume = {D52},
	Year = {1995}}

@article{Bousso:1998na,
	Archiveprefix = {arXiv},
	Author = {Bousso, Raphael and Hawking, Stephen W.},
	Doi = {10.1103/PhysRevD.59.103501, 10.1103/PhysRevD.60.109903},
	Eprint = {hep-th/9807148},
	Journal = {Phys. Rev.},
	Note = {[Erratum: Phys.  Rev. D 60, 109903 (1999)]},
	Pages = {103501},
	Primaryclass = {hep-th},
	Reportnumber = {SU-ITP-98-26, DAMTP-98-87, DAMTP-1998-87},
	Slaccitation = {%%CITATION = HEP-TH/9807148;%%},
	Title = {{Lorentzian condition in quantum gravity}},
	Volume = {D59},
	Year = {1999}}

@article{Nariai,
	Adsurl = {http://adsabs.harvard.edu/abs/1950SRToh..34..160N},
	Author = {{Nariai}, H.},
	Journal = {Sci.~Rep.~Tohoku Univ.~Eighth Ser.},
	Title = {{On some static solutions of Einstein's gravitational field equations in a spherically symmetric case}},
	Volume = 34,
	Year = 1950}

@article{Ginsparg:1982rs,
	Author = {Ginsparg, Paul H. and Perry, Malcolm J.},
	Doi = {10.1016/0550-3213(83)90636-3},
	Journal = {Nucl. Phys.},
	Pages = {245-268},
	Reportnumber = {HUTP-82/A035},
	Slaccitation = {%%CITATION = NUPHA,B222,245;%%},
	Title = {{Semiclassical Perdurance of de Sitter Space}},
	Volume = {B222},
	Year = {1983}}

@article{Hartle:1983ai,
	Author = {Hartle, J. B. and Hawking, S. W.},
	Doi = {10.1103/PhysRevD.28.2960},
	Journal = {Phys. Rev.},
	Note = {[Adv. Ser. Astrophys. Cosmol.3,174(1987)]},
	Pages = {2960-2975},
	Reportnumber = {PRINT-83-0937 (CAMBRIDGE)},
	Slaccitation = {%%CITATION = PHRVA,D28,2960;%%},
	Title = {{Wave Function of the Universe}},
	Volume = {D28},
	Year = {1983}}

@article{Jackiw:1984je,
	Author = {Jackiw, R.},
	Booktitle = {{1984 Meeting of the Division of Particles and Fields of the APS Santa Fe, New Mexico, October 31-November 3, 1984}},
	Doi = {10.1016/0550-3213(85)90448-1},
	Journal = {Nucl. Phys.},
	Pages = {343-356},
	Reportnumber = {MIT-CTP-1203, C84-06-04, C84-10-31},
	Slaccitation = {%%CITATION = NUPHA,B252,343;%%},
	Title = {{Lower Dimensional Gravity}},
	Volume = {B252},
	Year = {1985}}

@article{Teitelboim:1983ux,
	Author = {Teitelboim, C.},
	Doi = {10.1016/0370-2693(83)90012-6},
	Journal = {Phys. Lett.},
	Pages = {41-45},
	Slaccitation = {%%CITATION = PHLTA,B126,41;%%},
	Title = {{Gravitation and Hamiltonian Structure in Two Space-Time Dimensions}},
	Volume = {B126},
	Year = {1983}}

@article{Engelsoy:2016xyb,
	Archiveprefix = {arXiv},
	Author = {Engels{\"o}y, Julius and Mertens, Thomas G. and Verlinde, Herman},
	Doi = {10.1007/JHEP07(2016)139},
	Eprint = {1606.03438},
	Journal = {JHEP},
	Pages = {139},
	Primaryclass = {hep-th},
	Slaccitation = {%%CITATION = ARXIV:1606.03438;%%},
	Title = {{An investigation of AdS$_{2}$ backreaction and holography}},
	Volume = {07},
	Year = {2016}}

@article{Jensen:2016pah,
	Archiveprefix = {arXiv},
	Author = {Jensen, Kristan},
	Eprint = {1605.06098},
	Primaryclass = {hep-th},
	Slaccitation = {%%CITATION = ARXIV:1605.06098;%%},
	Title = {{Chaos and hydrodynamics near AdS$_2$}},
	Year = {2016}}

@article{Page:1993df,
	Archiveprefix = {arXiv},
	Author = {Page, Don N.},
	Doi = {10.1103/PhysRevLett.71.1291},
	Eprint = {gr-qc/9305007},
	Journal = {Phys. Rev. Lett.},
	Pages = {1291-1294},
	Primaryclass = {gr-qc},
	Reportnumber = {ALBERTA-THY-22-93},
	Slaccitation = {%%CITATION = GR-QC/9305007;%%},
	Title = {{Average entropy of a subsystem}},
	Volume = {71},
	Year = {1993}}

@article{Anninos:2012ft,
	Archiveprefix = {arXiv},
	Author = {Anninos, Dionysios and Denef, Frederik and Harlow, Daniel},
	Doi = {10.1103/PhysRevD.88.084049},
	Eprint = {1207.5517},
	Journal = {Phys. Rev.},
	Number = {8},
	Pages = {084049},
	Primaryclass = {hep-th},
	Reportnumber = {SU-ITP-12-19},
	Slaccitation = {%%CITATION = ARXIV:1207.5517;%%},
	Title = {{Wave function of Vasiliev's universe: A few slices thereof}},
	Volume = {D88},
	Year = {2013}}

@article{Maldacena:2016upp,
	Archiveprefix = {arXiv},
	Author = {Maldacena, Juan and Stanford, Douglas and Yang, Zhenbin},
	Doi = {10.1093/ptep/ptw124},
	Eprint = {1606.01857},
	Journal = {PTEP},
	Number = {12},
	Pages = {12C104},
	Primaryclass = {hep-th},
	Slaccitation = {%%CITATION = ARXIV:1606.01857;%%},
	Title = {{Conformal symmetry and its breaking in two dimensional Nearly Anti-de-Sitter space}},
	Volume = {2016},
	Year = {2016}}

@article{Banks:2000fe,
    author = "Banks, Tom",
    editor = "Duff, Michael J. and Liu, J. T. and Lu, J.",
    title = "{Cosmological breaking of supersymmetry?}",
    eprint = "hep-th/0007146",
    archivePrefix = "arXiv",
    reportNumber = "RUNHETC-2000-24, SCIPP-00-23",
    doi = "10.1142/S0217751X01003998",
    journal = "Int. J. Mod. Phys. A",
    volume = "16",
    pages = "910--921",
    year = "2001"
}

@article{Susskind:2021omt,
    author = "Susskind, Leonard",
    title = "{De Sitter Holography: Fluctuations, Anomalous Symmetry, and Wormholes}",
    eprint = "2106.03964",
    archivePrefix = "arXiv",
    primaryClass = "hep-th",
    doi = "10.3390/universe7120464",
    journal = "Universe",
    volume = "7",
    number = "12",
    pages = "464",
    year = "2021"
}

@article{Banks:2005bm,
    author = "Banks, T.",
    title = "{Some thoughts on the quantum theory of stable de Sitter space}",
    eprint = "hep-th/0503066",
    archivePrefix = "arXiv",
    reportNumber = "SCIPP-05-01",
    month = "3",
    year = "2005"
}

@article{Gibbons:1977mu,
    author = "Gibbons, G. W. and Hawking, S. W.",
    title = "{Cosmological Event Horizons, Thermodynamics, and Particle Creation}",
    doi = "10.1103/PhysRevD.15.2738",
    journal = "Phys. Rev. D",
    volume = "15",
    pages = "2738--2751",
    year = "1977"
}

@book{coleman_1985,
	Author = {Coleman, Sidney},
	Doi = {10.1017/CBO9780511565045},
	Place = {Cambridge},
	Publisher = {Cambridge University Press},
	Title = {Aspects of Symmetry: Selected Erice Lectures},
	Year = {1985,~Chapter 7}}

@article{Cotler:2024xzz,
    author = "Cotler, Jordan and Jensen, Kristan",
    title = "{Non-perturbative de Sitter Jackiw-Teitelboim gravity}",
    eprint = "2401.01925",
    archivePrefix = "arXiv",
    primaryClass = "hep-th",
    doi = "10.1007/JHEP12(2024)016",
    journal = "JHEP",
    volume = "12",
    pages = "016",
    year = "2024"
}

@article{Maldacena:2024spf,
    author = "Maldacena, Juan",
    title = "{Real observers solving imaginary problems}",
    eprint = "2412.14014",
    archivePrefix = "arXiv",
    primaryClass = "hep-th",
    month = "12",
    year = "2024"
}

@article{Polchinski:1988ua,
    author = "Polchinski, Joseph",
    title = "{The phase of the sum over spheres}",
    reportNumber = "UTTG-31-88",
    doi = "10.1016/0370-2693(89)90387-0",
    journal = "Phys. Lett. B",
    volume = "219",
    pages = "251--257",
    year = "1989"
}

@article{Heydeman:2020hhw,
    author = "Heydeman, Matthew and Iliesiu, Luca V. and Turiaci, Gustavo J. and Zhao, Wenli",
    title = "{The statistical mechanics of near-BPS black holes}",
    eprint = "2011.01953",
    archivePrefix = "arXiv",
    primaryClass = "hep-th",
    reportNumber = "PUPT-2621",
    doi= "10.1088/1751-8121/ac3be9",
    journal = "J. Phys. A",
    volume = "55",
    number = "1",
    pages = "014004",
    year = "2022"
}

@article{Preskill:1991tb,
      author         = "Preskill, John and Schwarz, Patricia and Shapere, Alfred
                        D. and Trivedi, Sandip and Wilczek, Frank",
      title          = "{Limitations on the statistical description of black
                        holes}",
      journal        = "Mod. Phys. Lett.",
      volume         = "A6",
      year           = "1991",
      pages          = "2353-2362",
      doi           = "10.1142/S0217732391002773",
      reportNumber   = "IASSNS-HEP-91-34, CALT-68-1730",
      SLACcitation   = "%%CITATION = MPLAE,A6,2353;%%"
}

@article{Hartle:1976tp,
    author = "Hartle, J. B. and Hawking, S. W.",
    title = "{Path Integral Derivation of Black Hole Radiance}",
    doi = "10.1103/PhysRevD.13.2188",
    journal = "Phys. Rev. D",
    volume = "13",
    pages = "2188--2203",
    year = "1976"
}

@article{Maldacena:2018gjk,
    author = "Maldacena, Juan and Milekhin, Alexey and Popov, Fedor",
    title = "{Traversable wormholes in four dimensions}",
    eprint = "1807.04726",
    archivePrefix = "arXiv",
    primaryClass = "hep-th",
    month = "7",
    year = "2018"
}

@article{Blacker:2025zca,
    author = "Blacker, Matthew J. and Castro, Alejandra and Sybesma, Watse and Toldo, Chiara",
    title = "{Quantum corrections to the path integral of near extremal de Sitter black holes}",
    eprint = "2503.14623",
    archivePrefix = "arXiv",
    primaryClass = "hep-th",
    reportNumber = "NORDITA 2025-016",
    month = "3",
    year = "2025"
}

@article{Iliesiu:2022onk,
    author = "Iliesiu, Luca V. and Murthy, Sameer and Turiaci, Gustavo J.",
    title = "{Revisiting the Logarithmic Corrections to the Black Hole Entropy}",
    eprint = "2209.13608",
    archivePrefix = "arXiv",
    primaryClass = "hep-th",
    month = "9",
    year = "2022"
}

@article{Maulik:2025phe,
    author = "Maulik, Sabyasachi and Mitra, Arpita and Mukherjee, Debangshu and Ray, Augniva",
    title = "{Logarithmic corrections to near-extremal entropy of charged de Sitter black holes}",
    eprint = "2503.08617",
    archivePrefix = "arXiv",
    primaryClass = "hep-th",
    month = "3",
    year = "2025"
}

@article{Shi:2025amq,
    author = "Shi, Xiaoyi and Turiaci, Gustavo J.",
    title = "{The phase of the gravitational path integral}",
    eprint = "2504.00900",
    archivePrefix = "arXiv",
    primaryClass = "hep-th",
    doi = "10.1007/JHEP07(2025)047",
    journal = "JHEP",
    volume = "07",
    pages = "047",
    year = "2025"
}

@article{Volkov:2000ih,
    author = "Volkov, Mikhail S. and Wipf, Andreas",
    title = "{Black hole pair creation in de Sitter space: A complete one-loop analysis}",
    eprint = "hep-th/0003081",
    archivePrefix = "arXiv",
    reportNumber = "FSUJ-TPI-00-03",
    doi = "10.1016/S0550-3213(00)00287-X",
    journal = "Nucl. Phys. B",
    volume = "582",
    pages = "313--362",
    year = "2000"
}

@misc{WIP,
    author = "{M. Rangamani, X. Shi, and G. J. Turiaci, \textit{work in progress}}"
}

@article{Polyakov:1981rd,
    author = "Polyakov, Alexander M.",
    editor = "Khalatnikov, I. M. and Mineev, V. P.",
    title = "{Quantum Geometry of Bosonic Strings}",
    reportNumber = "Print-81-0351 (LANDAU INST)",
    doi = "10.1016/0370-2693(81)90743-7",
    journal = "Phys. Lett. B",
    volume = "103",
    pages = "207--210",
    year = "1981"
}

@article{Grumiller:2002nm,
      author         = "Grumiller, D. and Kummer, W. and Vassilevich, D. V.",
      title          = "{Dilaton gravity in two-dimensions}",
      journal        = "Phys. Rept.",
      volume         = "369",
      year           = "2002",
      pages          = "327-430",
      doi           = "10.1016/S0370-1573(02)00267-3",
      eprint         = "hep-th/0204253",
      archivePrefix  = "arXiv",
      primaryClass   = "hep-th",
      reportNumber   = "TUW-02-01",
      SLACcitation   = "%%CITATION = HEP-TH/0204253;%%"
}

@article{Maldacena:2019cbz,
    author = "Maldacena, Juan and Turiaci, Gustavo J. and Yang, Zhenbin",
    title = "{Two dimensional Nearly de Sitter gravity}",
    eprint = "1904.01911",
    archivePrefix = "arXiv",
    primaryClass = "hep-th",
    doi = "10.1007/JHEP01(2021)139",
    journal = "JHEP",
    volume = "01",
    pages = "139",
    year = "2021"
}

@phdthesis{Laflamme:1986bc,
  author = {Laflamme, R.},
  title  = {Time and Quantum Cosmology},
  school = {University of Cambridge},
  type   = {{Ph.D. thesis}},
  year   = {(1986)}
}

@book{Fabbri:2005mw,
    author = "Fabbri, A. and Navarro-Salas, J.",
    title = "{Modeling black hole evaporation}",
    doi = "10.1142/p378",
    isbn = "978-1-86094-527-4, 978-1-86094-722-3, 978-1-78326-038-6",
    publisher = "World Scientific",
    address = "Singapore",
    year = "2005"
}

@article{Ghosh:2019rcj,
      author         = "Ghosh, Animik and Maxfield, Henry and Turiaci, Gustavo
                        J.",
      title          = "{A universal Schwarzian sector in two-dimensional
                        conformal field theories}",
      year           = "2019",
      eprint         = "1912.07654",
      archivePrefix  = "arXiv",
      primaryClass   = "hep-th",
      SLACcitation   = "%%CITATION = ARXIV:1912.07654;%%"
}

@article{Spradlin:1999bn,
    author = "Spradlin, Marcus and Strominger, Andrew",
    title = "{Vacuum states for AdS(2) black holes}",
    eprint = "hep-th/9904143",
    archivePrefix = "arXiv",
    reportNumber = "HUTP-99-A014",
    doi= "10.1088/1126-6708/1999/11/021",
    journal = "JHEP",
    volume = "11",
    pages = "021",
    year = "1999"
}

@article{Almheiri:2019psf,
    author = "Almheiri, Ahmed and Engelhardt, Netta and Marolf, Donald and Maxfield, Henry",
    title = "{The entropy of bulk quantum fields and the entanglement wedge of an evaporating black hole}",
    eprint = "1905.08762",
    archivePrefix = "arXiv",
    primaryClass = "hep-th",
    doi = "10.1007/JHEP12(2019)063",
    journal = "JHEP",
    volume = "12",
    pages = "063",
    year = "2019"
}

@article{Chandrasekaran:2022cip,
    author = "Chandrasekaran, Venkatesa and Longo, Roberto and Penington, Geoff and Witten, Edward",
    title = "{An algebra of observables for de Sitter space}",
    eprint = "2206.10780",
    archivePrefix = "arXiv",
    primaryClass = "hep-th",
    doi = "10.1007/JHEP02(2023)082",
    journal = "JHEP",
    volume = "02",
    pages = "082",
    year = "2023"
}

@article{Mahajan:2021nsd,
    author = "Mahajan, Raghu and Stanford, Douglas and Yan, Cynthia",
    title = "{Sphere and disk partition functions in Liouville and in matrix integrals}",
    eprint = "2107.01172",
    archivePrefix = "arXiv",
    primaryClass = "hep-th",
    doi = "10.1007/JHEP07(2022)132",
    journal = "JHEP",
    volume = "07",
    pages = "132",
    year = "2022"
}

@article{Biggs:2025nzs,
    author = "Biggs, Anna",
    title = "{Following the state of an evaporating charged black hole into the quantum gravity regime}",
    eprint = "2503.02051",
    archivePrefix = "arXiv",
    primaryClass = "hep-th",
    month = "3",
    year = "2025"
}

@article{Emparan:2025sao,
    author = "Emparan, Roberto",
    title = "{Quantum Cross-section of Near-extremal Black Holes}",
    eprint = "2501.17470",
    archivePrefix = "arXiv",
    primaryClass = "hep-th",
    month = "1",
    year = "2025"
}

@article{Mertens:2022irh,
    author = "Mertens, Thomas G. and Turiaci, Gustavo J.",
    title = "{Solvable models of quantum black holes: a review on Jackiw\textendash{}Teitelboim gravity}",
    eprint = "2210.10846",
    archivePrefix = "arXiv",
    primaryClass = "hep-th",
    doi = "10.1007/s41114-023-00046-1",
    journal = "Living Rev. Rel.",
    volume = "26",
    number = "1",
    pages = "4",
    year = "2023"
}

@article{Hawking:1975vcx,
    author = "Hawking, S. W.",
    editor = "Gibbons, G. W. and Hawking, S. W.",
    title = "{Particle Creation by Black Holes}",
    doi = "10.1007/BF02345020",
    journal = "Commun. Math. Phys.",
    volume = "43",
    pages = "199--220",
    year = "1975",
    note = "[Erratum: Commun. Math. Phys. 46, 206 (1976)]"
}

@article{Boulware:1974dm,
    author = "Boulware, David G.",
    title = "{Quantum Field Theory in Schwarzschild and Rindler Spaces}",
    reportNumber = "RLO-1388-683",
    doi = "10.1103/PhysRevD.11.1404",
    journal = "Phys. Rev. D",
    volume = "11",
    pages = "1404",
    year = "1975"
}

@article{Page:2013dx,
      author         = "Page, Don N.",
      title          = "{Time Dependence of Hawking Radiation Entropy}",
      journal        = "JCAP",
      volume         = "1309",
      year           = "2013",
      pages          = "028",
      doi            = "10.1088/1475-7516/2013/09/028",
      eprint         = "1301.4995",
      archivePrefix  = "arXiv",
      primaryClass   = "hep-th",
      SLACcitation   = "%%CITATION = ARXIV:1301.4995;%%"
}

@article{Page:1993wv,
      author         = "Page, Don N.",
      title          = "{Information in black hole radiation}",
      journal        = "Phys. Rev. Lett.",
      volume         = "71",
      year           = "1993",
      pages          = "3743-3746",
      doi            = "10.1103/PhysRevLett.71.3743",
      eprint         = "hep-th/9306083",
      archivePrefix  = "arXiv",
      primaryClass   = "hep-th",
      reportNumber   = "ALBERTA-THY-24-93",
      SLACcitation   = "%%CITATION = HEP-TH/9306083;%%"
}

@article{Wu:2023uyb,
    author = "Wu, Chih-Hung and Xu, Jiuci",
    title = "{Islands in non-minimal dilaton gravity: exploring effective theories for black hole evaporation}",
    eprint = "2303.03410",
    archivePrefix = "arXiv",
    primaryClass = "hep-th",
    doi = "10.1007/JHEP10(2023)094",
    journal = "JHEP",
    volume = "10",
    pages = "094",
    year = "2023"
}

@article{Mukhanov:1994ax,
    author = "Mukhanov, Viatcheslav F. and Wipf, A. and Zelnikov, A.",
    title = "{On 4-D Hawking radiation from effective action}",
    eprint = "hep-th/9403018",
    archivePrefix = "arXiv",
    reportNumber = "ETH-TH-94-08",
    doi = "10.1016/0370-2693(94)91255-6",
    journal = "Phys. Lett. B",
    volume = "332",
    pages = "283--291",
    year = "1994"
}

@article{Bousso:1997cg,
    author = "Bousso, Raphael and Hawking, Stephen W.",
    title = "{Trace anomaly of dilaton coupled scalars in two-dimensions}",
    eprint = "hep-th/9705236",
    archivePrefix = "arXiv",
    reportNumber = "DAMTP-R-97-25",
    doi = "10.1103/PhysRevD.56.7788",
    journal = "Phys. Rev. D",
    volume = "56",
    pages = "7788--7791",
    year = "1997"
}

@article{Nojiri:1999vv,
    author = "Nojiri, S. and Obregon, O. and Odintsov, S.D.",
    title = "{Unified approach to study quantum properties of primordial black holes, wormholes and of quantum cosmology}",
    eprint = "gr-qc/9907008",
    archivePrefix = "arXiv",
    doi = "10.1142/S0217732399001401",
    journal = "Mod. Phys. Lett. A",
    volume = "14",
    pages = "1309--1316",
    year = "1999"
}

@article{Wald:1978pj,
    author = "Wald, Robert M.",
    title = "{Trace Anomaly of a Conformally Invariant Quantum Field in Curved Space-Time}",
    doi = "10.1103/PhysRevD.17.1477",
    journal = "Phys. Rev. D",
    volume = "17",
    pages = "1477--1484",
    year = "1978"
}

@article{Kummer:1998dc,
    author = "Kummer, W. and Vassilevich, D. V.",
    title = "{Effective action and Hawking radiation for dilaton coupled scalars in two-dimensions}",
    eprint = "hep-th/9811092",
    archivePrefix = "arXiv",
    reportNumber = "TUW-98-22, TUW--98--22",
    doi = "10.1103/PhysRevD.60.084021",
    journal = "Phys. Rev. D",
    volume = "60",
    pages = "084021",
    year = "1999"
}

@article{Balbinot:2000iy,
    author = "Balbinot, R. and Fabbri, A. and Frolov, Valeri P. and Nicolini, Piero and Sutton, P. and Zelnikov, A.",
    title = "{Vacuum polarization in the Schwarzschild space-time and dimensional reduction}",
    eprint = "hep-th/0012048",
    archivePrefix = "arXiv",
    doi = "10.1103/PhysRevD.63.084029",
    journal = "Phys. Rev. D",
    volume = "63",
    pages = "084029",
    year = "2001"
}

@article{Kummer:1999zy,
    author = "Kummer, W. and Vassilevich, D. V.",
    title = "{Hawking radiation from dilaton gravity in (1+1)-dimensions: A Pedagogical review}",
    eprint = "gr-qc/9907041",
    archivePrefix = "arXiv",
    reportNumber = "TUW-99-14",
    doi = "10.1002/(SICI)1521-3889(199912)8:10<801::AID-ANDP801>3.0.CO;2-O",
    journal = "Annalen Phys.",
    volume = "8",
    pages = "801--827",
    year = "1999"
}

@article{Fabbri:2003vy,
    author = "Fabbri, A. and Farese, S. and Navarro-Salas, J.",
    title = "{Generalized Virasoro anomaly and stress tensor for dilaton coupled theories}",
    eprint = "hep-th/0309160",
    archivePrefix = "arXiv",
    doi = "10.1016/j.physletb.2003.09.012",
    journal = "Phys. Lett. B",
    volume = "574",
    pages = "309--318",
    year = "2003"
}

@article{Hofmann:2004kk,
    author = "Hofmann, D. and Kummer, W.",
    title = "{Effective action and Hawking flux from covariant perturbation theory}",
    eprint = "gr-qc/0408088",
    archivePrefix = "arXiv",
    reportNumber = "TUW-04-20",
    doi = "10.1140/epjc/s2005-02129-9",
    journal = "Eur. Phys. J. C",
    volume = "40",
    pages = "275--286",
    year = "2005"
}

@article{Hofmann:2005yv,
    author = "Hofmann, D. and Kummer, W.",
    title = "{IR renormalisation of general effective actions and Hawking flux in 2-D gravity theories}",
    eprint = "gr-qc/0512163",
    archivePrefix = "arXiv",
    reportNumber = "TUW-05-18",
    doi = "10.1140/epjc/s2006-02553-3",
    journal = "Eur. Phys. J. C",
    volume = "48",
    pages = "291--301",
    year = "2006"
}

@article{Karakhanian:1994gs,
    author = "Karakhanian, D. R. and Manvelyan, R. P. and Mkrtchian, R. L.",
    title = "{Area preserving structure of 2-d gravity}",
    eprint = "hep-th/9401031",
    archivePrefix = "arXiv",
    doi = "10.1016/0370-2693(94)90758-7",
    journal = "Phys. Lett. B",
    volume = "329",
    pages = "185--188",
    year = "1994"
}

@article{Jackiw:1995qh,
    author = "Jackiw, R.",
    title = "{Another view on massless matter - gravity fields in two-dimensions}",
    eprint = "hep-th/9501016",
    archivePrefix = "arXiv",
    reportNumber = "MIT-CTP-2377",
    month = "1",
    year = "1995"
}

@article{Navarro-Salas:1995lmi,
    author = "Navarro-Salas, J. and Navarro, M. and Talavera, C. F.",
    title = "{Weyl invariance and black hole evaporation}",
    eprint = "hep-th/9505139",
    archivePrefix = "arXiv",
    reportNumber = "FTUV-95-17, IFIC-95-17, IMPERIAL-TP-94-95-31",
    doi = "10.1016/0370-2693(95)00848-F",
    journal = "Phys. Lett. B",
    volume = "356",
    pages = "217--222",
    year = "1995"
}

@article{Bunch:1978yq,
    author = "Bunch, T. S. and Davies, P. C. W.",
    title = "{Quantum Field Theory in de Sitter Space: Renormalization by Point Splitting}",
    doi = "10.1098/rspa.1978.0060",
    journal = "Proc. Roy. Soc. Lond. A",
    volume = "360",
    pages = "117--134",
    year = "1978"
}

@article{Bousso:1997wi,
    author = "Bousso, Raphael and Hawking, Stephen W.",
    title = "{(Anti)evaporation of Schwarzschild-de Sitter black holes}",
    eprint = "hep-th/9709224",
    archivePrefix = "arXiv",
    reportNumber = "DAMTP-R-97-26",
    doi = "10.1103/PhysRevD.57.2436",
    journal = "Phys. Rev. D",
    volume = "57",
    pages = "2436--2442",
    year = "1998"
}

@article{Nojiri:1998ue,
    author = "Nojiri, Shin'ichi and Odintsov, Sergei D.",
    title = "{Effective action for conformal scalars and anti-evaporation of black holes}",
    eprint = "hep-th/9802160",
    archivePrefix = "arXiv",
    reportNumber = "NDA-FP-44",
    doi = "10.1142/S0217751X9900066X",
    journal = "Int. J. Mod. Phys. A",
    volume = "14",
    pages = "1293--1304",
    year = "1999"
}

@article{Nojiri:1998ph,
    author = "Nojiri, Shin'ichi and Odintsov, Sergei D.",
    title = "{Quantum evolution of Schwarzschild-de Sitter (Nariai) black holes}",
    eprint = "hep-th/9804033",
    archivePrefix = "arXiv",
    reportNumber = "NDA-FP-46",
    doi = "10.1103/PhysRevD.59.044026",
    journal = "Phys. Rev. D",
    volume = "59",
    pages = "044026",
    year = "1999"
}

@article{Brown:2024ajk,
    author = "Brown, Adam R. and Iliesiu, Luca V. and Penington, Geoff and Usatyuk, Mykhaylo",
    title = "{The evaporation of charged black holes}",
    eprint = "2411.03447",
    archivePrefix = "arXiv",
    primaryClass = "hep-th",
    month = "11",
    year = "2024"
}

@article{Elizalde:1999dw,
    author = "Elizalde, E. and Nojiri, S. and Odintsov, S. D.",
    title = "{Possible quantum instability of primordial black holes}",
    eprint = "hep-th/9901026",
    archivePrefix = "arXiv",
    reportNumber = "IEEC-98-81, NDA-FP-52",
    doi = "10.1103/PhysRevD.59.061501",
    journal = "Phys. Rev. D",
    volume = "59",
    pages = "061501",
    year = "1999"
}

@article{Anninos:2020hfj,
    author = "Anninos, Dionysios and Denef, Frederik and Law, Y. T. Albert and Sun, Zimo",
    title = "{Quantum de Sitter horizon entropy from quasicanonical bulk, edge, sphere and topological string partition functions}",
    eprint = "2009.12464",
    archivePrefix = "arXiv",
    primaryClass = "hep-th",
    doi = "10.1007/JHEP01(2022)088",
    journal = "JHEP",
    volume = "01",
    pages = "088",
    year = "2022"
}

@article{Wald:1977up,
    author = "Wald, Robert M.",
    title = "{The Back Reaction Effect in Particle Creation in Curved Space-Time}",
    doi = "10.1007/BF01609833",
    journal = "Commun. Math. Phys.",
    volume = "54",
    pages = "1--19",
    year = "1977"
}

@article{Wald:1978ce,
    author = "Wald, Robert M.",
    title = "{Axiomatic Renormalization of the Stress Tensor of a Conformally Invariant Field in Conformally Flat Space-Times}",
    doi = "10.1016/0003-4916(78)90040-4",
    journal = "Annals Phys.",
    volume = "110",
    pages = "472--486",
    year = "1978"
}

@article{Bousso:1998bn,
    author = "Bousso, Raphael",
    title = "{Proliferation of de Sitter space}",
    eprint = "hep-th/9805081",
    archivePrefix = "arXiv",
    reportNumber = "SU-ITP-98-24",
    doi = "10.1103/PhysRevD.58.083511",
    journal = "Phys. Rev. D",
    volume = "58",
    pages = "083511",
    year = "1998"
}

@article{Bousso:1999ms,
    author = "Bousso, Raphael",
    title = "{Quantum global structure of de Sitter space}",
    eprint = "hep-th/9902183",
    archivePrefix = "arXiv",
    reportNumber = "SU-ITP-99-7",
    doi = "10.1103/PhysRevD.60.063503",
    journal = "Phys. Rev. D",
    volume = "60",
    pages = "063503",
    year = "1999"
}

@article{Niemeyer:2000nq,
    author = "Niemeyer, Jens C. and Bousso, Raphael",
    title = "{The Nonlinear evolution of de Sitter space instabilities}",
    eprint = "gr-qc/0004004",
    archivePrefix = "arXiv",
    doi = "10.1103/PhysRevD.62.023503",
    journal = "Phys. Rev. D",
    volume = "62",
    pages = "023503",
    year = "2000"
}

@article{Markovic:1991ua,
    author = "Markovic, D. and Unruh, W. G.",
    title = "{Vacuum for a massless scalar field outside a collapsing body in de Sitter space-time}",
    doi = "10.1103/PhysRevD.43.332",
    journal = "Phys. Rev. D",
    volume = "43",
    pages = "332--339",
    year = "1991"
}

@article{Addazi:2017cim,
    author = "Addazi, Andrea and Marciano, Antonino",
    title = "{Evaporation and Antievaporation instabilities}",
    eprint = "1710.07962",
    archivePrefix = "arXiv",
    primaryClass = "gr-qc",
    doi = "10.3390/sym9110249",
    journal = "Symmetry",
    volume = "9",
    number = "11",
    pages = "249",
    year = "2017"
}

@article{Nojiri:2013su,
    author = "Nojiri, Shin'ichi and Odintsov, Sergei D.",
    title = "{Anti-Evaporation of Schwarzschild-de Sitter Black Holes in $F(R)$ gravity}",
    eprint = "1301.2775",
    archivePrefix = "arXiv",
    primaryClass = "hep-th",
    doi = "10.1088/0264-9381/30/12/125003",
    journal = "Class. Quant. Grav.",
    volume = "30",
    pages = "125003",
    year = "2013"
}

@article{Addazi:2016prb,
    author = "Addazi, Andrea and Capozziello, Salvatore",
    title = "{The fate of Schwarzschild-de Sitter Black Holes in $f(R)$ gravity}",
    eprint = "1602.00485",
    archivePrefix = "arXiv",
    primaryClass = "gr-qc",
    doi = "10.1142/S0217732316500541",
    journal = "Mod. Phys. Lett. A",
    volume = "31",
    number = "09",
    pages = "1650054",
    year = "2016"
}

@article{Nojiri:2014jqa,
    author = "Nojiri, Shin'ichi and Odintsov, Sergei D.",
    title = {{Instabilities and anti-evaporation of Reissner{\textendash}Nordstr{\"o}m black holes in modified $F(R)$ gravity}},
    eprint = "1405.2439",
    archivePrefix = "arXiv",
    primaryClass = "gr-qc",
    doi = "10.1016/j.physletb.2014.06.070",
    journal = "Phys. Lett. B",
    volume = "735",
    pages = "376--382",
    year = "2014"
}

@article{Susskind:2021dfc,
    author = "Susskind, Leonard",
    title = "{Black Holes Hint towards De Sitter Matrix Theory}",
    eprint = "2109.01322",
    archivePrefix = "arXiv",
    primaryClass = "hep-th",
    doi = "10.3390/universe9080368",
    journal = "Universe",
    volume = "9",
    number = "8",
    pages = "368",
    year = "2023"
}

@article{Easson:2025ekn,
    author = "Easson, Damien A.",
    title = "{The fate of Schwarzschild--de Sitter black holes: nonequilibrium evaporation}",
    eprint = "2511.11873",
    archivePrefix = "arXiv",
    primaryClass = "hep-th",
    month = "11",
    year = "2025"
}

@article{Bhattacharjee:2025wfv,
    author = "Bhattacharjee, Arindam and Saha, Muktajyoti",
    title = "{Quantum evolution of de Sitter black holes near extremality}",
    eprint = "2510.18035",
    archivePrefix = "arXiv",
    primaryClass = "hep-th",
    month = "10",
    year = "2025"
}

@article{Aalsma:2025lcb,
    author = "Aalsma, Lars and Lin, Puxin and van der Schaar, Jan Pieter and Shiu, Gary and Sybesma, Watse",
    title = "{Limits on the Statistical Description of Charged de Sitter Black Holes}",
    eprint = "2511.03867",
    archivePrefix = "arXiv",
    primaryClass = "hep-th",
    month = "11",
    year = "2025"
}

@article{Montero:2019ekk,
    author = "Montero, Miguel and Van Riet, Thomas and Venken, Victoria",
    title = "{Festina Lente: EFT Constraints from Charged Black Hole Evaporation in de Sitter}",
    eprint = "1910.01648",
    archivePrefix = "arXiv",
    primaryClass = "hep-th",
    doi = "10.1007/JHEP01(2020)039",
    journal = "JHEP",
    volume = "01",
    pages = "039",
    year = "2020"
}

@article{deCesare:2025ccs,
    author = "de Cesare, Marco and Miranda, Marcello and Porfyriadis, Achilleas P.",
    title = "{Scalar field scattering in a Schwarzschild-de Sitter geometry}",
    eprint = "2511.09168",
    archivePrefix = "arXiv",
    primaryClass = "gr-qc",
    reportNumber = "CCTP-2025-12, ITCP-IPP 2025/12",
    month = "11",
    year = "2025"
}

@article{Chen:2025jqm,
    author = "Chen, Yiming and Stanford, Douglas and Tang, Haifeng and Yang, Zhenbin",
    title = "{On the phase of the de Sitter density of states}",
    eprint = "2511.01400",
    archivePrefix = "arXiv",
    primaryClass = "hep-th",
    month = "11",
    year = "2025"
}

@article{Ivo:2025xek,
    author = "Ivo, Victor and Sun, Zimo",
    title = "{The phase of charged Nariai solutions}",
    eprint = "2511.06604",
    archivePrefix = "arXiv",
    primaryClass = "hep-th",
    month = "11",
    year = "2025"
}

@article{Ivo:2025fwe,
    author = "Ivo, Victor",
    title = "{One loop aspects of Coleman de Luccia instantons at small backreaction}",
    eprint = "2509.18651",
    archivePrefix = "arXiv",
    primaryClass = "hep-th",
    month = "9",
    year = "2025"
}

@article{Tadaki:1990aa,
    author = "Tadaki, Shin-Ichi and Takagi, Shin",
    title = "{Quantum Field Theory in Two-dimensional Schwarzschild-de Sitter Space-time 1. Empty Space}",
    reportNumber = "KUNS-999",
    doi = "10.1143/PTP.83.941",
    journal = "Prog. Theor. Phys.",
    volume = "83",
    pages = "941--952",
    year = "1990"
}

@article{Tadaki:1990cg,
    author = "Tadaki, S. and Takagi, S.",
    title = "{Quantum field theory in two-dimensional Schwarzschild-de Sitter space-time. 2: Space with a collapsing star}",
    doi = "10.1143/PTP.83.1126",
    journal = "Prog. Theor. Phys.",
    volume = "83",
    pages = "1126--1139",
    year = "1990"
}

@article{Kolanowski:2019xpq,
    author = "Kolanowski, Maciej",
    title = "{(Anti-)evaporation of Schwarzschild-de Sitter black holes revisited}",
    eprint = "1908.01716",
    archivePrefix = "arXiv",
    primaryClass = "gr-qc",
    month = "8",
    year = "2019"
}

@article{Turiaci:2024cad,
    author = "Turiaci, Gustavo J.",
    title = "{Les Houches lectures on two-dimensional gravity and holography}",
    eprint = "2412.09537",
    archivePrefix = "arXiv",
    primaryClass = "hep-th",
    month = "12",
    year = "2024"
}

@article{Turiaci:2025xwi,
    author = "Turiaci, Gustavo J. and Wu, Chih-Hung",
    title = "{The wavefunction of a quantum S$^{1}$ {\texttimes} S$^{2}$ universe}",
    eprint = "2503.14639",
    archivePrefix = "arXiv",
    primaryClass = "hep-th",
    doi = "10.1007/JHEP07(2025)158",
    journal = "JHEP",
    volume = "07",
    pages = "158",
    year = "2025"
}

@article{Weinberg:2006pc,
    author = "Weinberg, Erick J.",
    title = "{Hawking-Moss bounces and vacuum decay rates}",
    eprint = "hep-th/0612146",
    archivePrefix = "arXiv",
    reportNumber = "CU-TP-1171, KIAS-P06062",
    doi = "10.1103/PhysRevLett.98.251303",
    journal = "Phys. Rev. Lett.",
    volume = "98",
    pages = "251303",
    year = "2007"
}

@article{Hawking:1981fz,
    author = "Hawking, S. W. and Moss, I. G.",
    title = "{Supercooled Phase Transitions in the Very Early Universe}",
    reportNumber = "Print-82-0181 (CAMBRIDGE)",
    doi = "10.1016/0370-2693(82)90946-7",
    journal = "Phys. Lett. B",
    volume = "110",
    pages = "35--38",
    year = "1982"
}

@article{Tolman:1930ona,
    author = "Tolman, Richard and Ehrenfest, Paul",
    title = "{Temperature Equilibrium in a Static Gravitational Field}",
    doi = "10.1103/PhysRev.36.1791",
    journal = "Phys. Rev.",
    volume = "36",
    number = "12",
    pages = "1791--1798",
    year = "1930"
}

@article{Tolman:1930zza,
    author = "Tolman, Richard C.",
    title = "{On the Weight of Heat and Thermal Equilibrium in General Relativity}",
    doi = "10.1103/PhysRev.35.904",
    journal = "Phys. Rev.",
    volume = "35",
    pages = "904--924",
    year = "1930"
}

@article{Choudhury:2004ph,
    author = "Choudhury, T. Roy and Padmanabhan, T.",
    title = "{Concept of temperature in multi-horizon spacetimes: Analysis of Schwarzschild-de Sitter metric}",
    eprint = "gr-qc/0404091",
    archivePrefix = "arXiv",
    doi = "10.1007/s10714-007-0489-0",
    journal = "Gen. Rel. Grav.",
    volume = "39",
    pages = "1789--1811",
    year = "2007"
}

@article{Kanti:2014dxa,
    author = "Kanti, Panagiota and Pappas, Thomas and Pappas, Nikolaos",
    title = "{Greybody factors for scalar fields emitted by a higher-dimensional Schwarzschild{\textendash}de Sitter black hole}",
    eprint = "1409.8664",
    archivePrefix = "arXiv",
    primaryClass = "hep-th",
    doi = "10.1103/PhysRevD.90.124077",
    journal = "Phys. Rev. D",
    volume = "90",
    number = "12",
    pages = "124077",
    year = "2014"
}

@article{Brady:1996za,
    author = "Brady, Patrick R. and Chambers, Chris M. and Krivan, William and Laguna, Pablo",
    title = "{Telling tails in the presence of a cosmological constant}",
    eprint = "gr-qc/9611056",
    archivePrefix = "arXiv",
    reportNumber = "CGPG-96-11-3",
    doi = "10.1103/PhysRevD.55.7538",
    journal = "Phys. Rev. D",
    volume = "55",
    pages = "7538--7545",
    year = "1997"
}

@article{Kanti:2005ja,
    author = "Kanti, P. and Grain, Julien and Barrau, A.",
    title = "{Bulk and brane decay of a (4+n)-dimensional Schwarzschild-de-Sitter black hole: Scalar radiation}",
    eprint = "hep-th/0501148",
    archivePrefix = "arXiv",
    reportNumber = "DCPT-05-01",
    doi = "10.1103/PhysRevD.71.104002",
    journal = "Phys. Rev. D",
    volume = "71",
    pages = "104002",
    year = "2005"
}

@article{Chen:2024rpx,
    author = "Chen, Chang-Han and Penington, Geoff",
    title = "{A clock is just a way to tell the time: gravitational algebras in cosmological spacetimes}",
    eprint = "2406.02116",
    archivePrefix = "arXiv",
    primaryClass = "hep-th",
    month = "6",
    year = "2024"
}

@article{Draper:2022xzl,
    author = "Draper, Patrick and Farkas, Szilard",
    title = "{de Sitter black holes as constrained states in the Euclidean path integral}",
    eprint = "2203.02426",
    archivePrefix = "arXiv",
    primaryClass = "hep-th",
    doi = "10.1103/PhysRevD.105.126022",
    journal = "Phys. Rev. D",
    volume = "105",
    number = "12",
    pages = "126022",
    year = "2022"
}

@article{Morvan:2022ybp,
    author = "Morvan, Edward K. and van der Schaar, Jan Pieter and Visser, Manus R.",
    title = "{On the Euclidean action of de Sitter black holes and constrained instantons}",
    eprint = "2203.06155",
    archivePrefix = "arXiv",
    primaryClass = "hep-th",
    doi = "10.21468/SciPostPhys.14.2.022",
    journal = "SciPost Phys.",
    volume = "14",
    number = "2",
    pages = "022",
    year = "2023"
}

@article{Bousso:1996pn,
    author = "Bousso, Raphael",
    title = "{Charged Nariai black holes with a dilaton}",
    eprint = "gr-qc/9608053",
    archivePrefix = "arXiv",
    reportNumber = "DAMTP-R-96-39",
    doi = "10.1103/PhysRevD.55.3614",
    journal = "Phys. Rev. D",
    volume = "55",
    pages = "3614--3621",
    year = "1997"
}

@article{Coleman:1980aw,
    author = "Coleman, Sidney R. and De Luccia, Frank",
    title = "{Gravitational Effects on and of Vacuum Decay}",
    reportNumber = "SLAC-PUB-2463",
    doi = "10.1103/PhysRevD.21.3305",
    journal = "Phys. Rev. D",
    volume = "21",
    pages = "3305",
    year = "1980"
}

@article{Jensen:1983ac,
    author = "Jensen, Lars Gerhard and Steinhardt, Paul Joseph",
    title = "{Bubble Nucleation and the {Coleman-Weinberg} Model}",
    reportNumber = "UPR-0231T",
    doi = "10.1016/0550-3213(84)90021-X",
    journal = "Nucl. Phys. B",
    volume = "237",
    pages = "176--188",
    year = "1984"
}

@article{Tomasevic:2025clf,
    author = "Toma{\v{s}}evi{\'c}, Marija and Wu, Chih-Hung",
    title = "{Unveiling horizons in quantum critical collapse}",
    eprint = "2509.03584",
    archivePrefix = "arXiv",
    primaryClass = "gr-qc",
    month = "9",
    year = "2025"
}

@article{Ivo:2024ill,
    author = "Ivo, Victor and Li, Yue-Zhou and Maldacena, Juan",
    title = "{The no boundary density matrix}",
    eprint = "2409.14218",
    archivePrefix = "arXiv",
    primaryClass = "hep-th",
    doi = "10.1007/JHEP02(2025)124",
    journal = "JHEP",
    volume = "02",
    pages = "124",
    year = "2025"
}

@article{Dialektopoulos:2017pgo,
    author = "Dialektopoulos, Konstantinos F. and Nathanail, Antonios and Tzikas, Athanasios G.",
    title = "{Cosmological production of black holes: a way to constrain alternative theories of gravity}",
    eprint = "1712.10177",
    archivePrefix = "arXiv",
    primaryClass = "gr-qc",
    doi = "10.1103/PhysRevD.97.124059",
    journal = "Phys. Rev. D",
    volume = "97",
    number = "12",
    pages = "124059",
    year = "2018"
}

@article{Tomasevic:2025kqy,
    author = "Toma{\v{s}}evi{\'c}, Marija and Wu, Chih-Hung",
    title = "{Quantum Critical Collapse Abhors a Naked Singularity}",
    eprint = "2509.03587",
    archivePrefix = "arXiv",
    primaryClass = "gr-qc",
    month = "9",
    year = "2025"
}

@article{Faruk:2023uzs,
    author = "Faruk, Mir Mehedi and Morvan, Edward and van der Schaar, Jan Pieter",
    title = "{Static sphere observers and geodesics in Schwarzschild-de Sitter spacetime}",
    eprint = "2312.06878",
    archivePrefix = "arXiv",
    primaryClass = "gr-qc",
    doi = "10.1088/1475-7516/2024/05/118",
    journal = "JCAP",
    volume = "05",
    pages = "118",
    year = "2024"
}

@article{Aalsma:2026kqj,
    author = "Aalsma, Lars and Faruk, Mir Mehedi",
    title = "{Complex Geodesics in the Nariai Geometry}",
    eprint = "2604.26662",
    archivePrefix = "arXiv",
    primaryClass = "hep-th",
    month = "4",
    year = "2026"
}

\end{document}